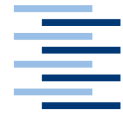

Hochschule für Angewandte Wissenschaften Hamburg
*Hamburg University of Applied Sciences*

# Bachelorarbeit

Philipp Grulich

Skalierbare Echtzeitverarbeitung mit Spark Streaming: Realisierung und Konzeption eines Car Information Systems

*Fakultät Technik und Informatik
Department Informatik*

*Faculty of Engineering and Computer Science
Department of Computer Science*

# Philipp Grulich

Skalierbare Echtzeitverarbeitung mit Spark Streaming: Realisierung und Konzeption eines Car Information Systems

**Philipp Grulich**

**Thema der Arbeit**
Skalierbare Echtzeitverarbeitung mit Spark Streaming: Realisierung und Konzeption eines Car Information Systems

**Stichworte**
Echtzeitverarbeitung, Spark, Spark Streaming, Car Information System, Event Processing, Skalierbarkeit, Fehlertoleranz

**Kurzzusammenfassung**
Stream Verarbeitung ist mittlerweile eines der relevantesten Bereiche im Rahmen der Big Data Analyse, da es die Verarbeitung einer Vielzahl an Events innerhalb einer kurzen Latenz erlaubt. Eines der momentan am häufigsten genutzten Stream Verarbeitungssysteme ist Spark Streaming. Dieses wird im Rahmen dieser Arbeit anhand der Konzeption und Realisierung eines Car Information Systems demonstriert und diskutiert, wobei viel Wert auf das Erzeugen einer möglichst generischen Anwendungsarchitektur gelegt wird. Abschließend wird sowohl das CIS als auch Spark Streaming mittels dem *Goal-Question-Metric*-Modell evaluiert. Hierbei zeigt sich, dass sich Spark Streaming zwar für die Realisierung eines skalierbaren und ausfallsicheren Systems eignet, aber durch das schnelle Voranschreiten der Weiterentwicklung von Spark verschiedene Probleme bei der Entwicklung eines auf Spark basierten Systems entstehen können.

**Philipp Grulich**

**Title of the paper**
Scalable real-time processing with Spark Streaming: implementation and design of a Car Information System

**Keywords**
real-time processing, Spark, Spark Streaming, Car Information System, event processing, scalability, fault tolerant

**Abstract**
Streaming data processing is a hot topic in big data these days, because it made it possible to process a huge amount of events within a low latency. One of the most common used open-source stream processing platforms is Spark Streaming, which is demonstrated and discussed based on a real-world use-case in this paper. The use-case is about a Car Information System, which is an example for a classic stream processing system. First the System is designed and engineered, whereby the application architecture is created carefully, because it should be adaptable for similar use-cases. At the end of this paper the CIS and Spark Streaming is evaluated by the use of the Goal Question Metric model. The evaluation proves that Spark Streaming is capable to create stream processing in a scalable and fault tolerant manner. But it also shows that Spark is a very fast moving project, which could cause problems during the development and maintenance of a software project.

# Inhaltsverzeichnis

# 1  Einleitung

Der Begriff Big Data hat in den letzten Jahren immer mehr an Bedeutung gewonnen und beschreibt Daten die aufgrund ihrer Schnelllebigkeit (Velocity), ihrer Größe (Volume) oder ihrer Heterogenität (Variety) nur schwer mit traditionellen Datenbanksystemen verarbeitbar sind. Des Weiteren wird Big Data oft in einen Zusammenhang mit Datenanalyse gebracht. So ermöglichen Big Data Analysen beispielsweise Vorhersagen oder Simulationen des Kundenverhaltens, wobei hier oft Techniken des maschinellen Lernens eingesetzt werden [Plat13]. Google war eine der ersten Firmen, die über solche Daten verfügte und diese analysieren wollte. Um das Arbeiten mit solchen Daten zu vereinfachen und gleichzeitig eine hohe Ausfallsicherheit und Skalierbarkeit zu erreichen, entwickelten die Google Mitarbeiter Jeffrey Dean und Sanjay Ghemawat den Map-Reduce Algorithmus, welchen sie im Jahr 2004 veröffentlichten. Durch Map-Reduce war es nun möglich Datenverarbeitungen über eine große Anzahl an Rechnerknoten zu verteilen und parallel auszuführen [DeGh04]. Über die darauffolgenden Jahre wurde der Bedarf für Big Data Analysen auch in anderen Unternehmen und Organisationen immer größer, sodass verschiedene Open-Source Lösungen entstanden, die den Map-Reduce Ansatz implementieren. Hadoop gilt als eine der bekanntesten.

Des Weiteren wurden immer mehr Machine-Learning Algorithmen für die Analyse von Big Data genutzt, wodurch ein Bedarf für eine möglichst effiziente Umsetzung iterativer Algorithmen entstand, welche nur schwer mittels Hadoop realisiert werden konnten [Piat15]. Dies war einer der Gründe für die Entwicklung von Spark, einer Plattform für verteilte Datenverarbeitung, welche seit 2009 am AMPLab der UC Berkeley entwickelt wurde und seit 2014 als Top Level Projekt von der Apache Foundation verwaltet wird [Apac16a]. Es baut zum Teil auf durch Hadoop bekannte Techniken auf, legt aber bei der Abarbeitung von Aufgaben den Fokus auf In-Memory-Berechnungen, wodurch eine erhebliche Leistungssteigerung gegenüber Hadoop erreicht werden konnte [ZCDD12]. Mittlerweile ist Spark das aktivste Open-Source Big Data Projekt und verfügt über ein vielfältiges Ökosystem [Apac16b, Data15]. Neben der klassischen Batchanalyse ergeben sich durch die In-Memory-Arbeitsweise von Spark im Bereich der Big Data-Verarbeitung einige komplett neue Anwendungsbereiche. So ermöglicht das Spark-Streaming Projekt Eventverarbeitung nahezu in Echtzeit. Dies wird beispielsweise für das schnelle Reagieren auf besondere Ereignisse benötigt. So können beispielsweise E-Commerce-Unternehmen die Aktionen eines Kunden verarbeiten, während dieser noch deren Onlineshop besucht [Akid15]. Da es sich bei Spark Streaming um ein relativ junges Projekt handelt, ist die Evaluation anhand eines relevanten Anwendungsfalls besonders interessant. So kann hier analysiert werden, wie sich Spark Streaming während der Umsetzung verhält und welche Probleme während der Nutzung von Spark Streaming entstehen können.

## 1.1 Zielsetzung

Das Ziel dieser Arbeit ist es, Spark Streaming anhand eines praxisnahen Beispiels zu demonstrieren und zu diskutieren. Im theoretischen Teil werden die grundlegenden Techniken von Spark und Spark Streaming dargestellt. Hierbei werden auch alternative Systeme zum Vergleich vorgestellt, um so die spätere Diskussion zu ermöglichen. Im Hauptteil dieser Arbeit wird ein Car Information System (CIS) konzipiert und mit Spark Streaming umgesetzt. Mittels des CIS können verschiedene, während einer Autofahrt gewonnene Daten verarbeitet und analysiert werden. Die Analyse kann verschiedene Use-Cases abdecken und beispielsweise den Fahrer während der Fahrt über bestimmte Ereignisse benachrichtigen. Im Rahmen der prototypischen Umsetzung des CIS werden zunächst verschiedene User-Stories dargestellt und anschließend eine für die Realisierung ausgewählt. Zum Abschluss der Arbeit wird Spark Streaming anhand des geschilderten Anwendungsfalls diskutiert. Hierbei werden zunächst Bewertungskriterien definiert und in der Diskussion herangezogen. Die Diskussion konzentriert sich vor allem auf das Spark Streaming-Projekt und die architektonische Realisierung des CIS.

## 1.2 Aufbau der Arbeit

Die Arbeit ist in sechs Kapitel gegliedert, die folgendermaßen aufgeteilt sind:

> Kapitel 1 – *Einleitung* – Gibt einen Überblick über die Arbeit und stellt die Relevanz des Themas dar.
>
> Kapitel 2 – *Spark* – Hier wird eine Einführung in Spark gegeben und die technisch relevanten Konzepte werden erläutert.
>
> Kapitel 3 – *Spark Streaming* – Es wird das Projekt Spark Streaming vorgestellt und die technischen Besonderheiten wie das Micro Batch-Verfahren vorgestellt.
>
> Kapitel 4 – *Alternative Streaming Systeme* – Hier werden verschiedene Streaming Systeme vorgestellt und es wird auf deren Unterschiede zu Spark Streaming eingegangen.
>
> Kapitel 5 – *Praxisbeispiel* – In diesem Kapitel wird das Car Information System konzipiert und realisiert.
>
> Kapitel 6 – *Diskussion* – In diesem Kapitel wird zunächst das CIS und anschließend Spark Streaming mittels des GQM Modells evaluiert.
>
> Kapitel 7 – *Fazit und Ausblick* – Hier wird eine abschließende Zusammenfassung der Arbeit und werden hier zukünftige Fragestellungen bezüglich des Anwendungsfalls sowie bezüglich Spark Streaming im Allgemeinen dargestellt.

# 2 Spark

Spark verfolgt vor allem zwei Ziele. Zum einen erhöht es durch In-Memory-Berechnungen die Effizienz des durch Hadoop bekannten Map-Reduce Verarbeitungsverfahrens, wodurch große Datenmengen in kürzerer Zeit analysiert werden können. Zum anderen abstrahiert es Map-Reduce und eignet sich so für eine größere Anzahl an Datenverarbeitungsanwendungsfällen. So bietet Spark mittlerweile Bibliotheken für Machine-Learning, Stream-Verarbeitung, Graphenverarbeitung, sowie für Datenverarbeitung mittels einer SQL-Abstraktion an. Spark selbst besteht zum einen aus einer Anwendungsbibliothek, über welche die jeweiligen Verarbeitungsverfahren definiert werden und aus einer Cluster-Komponente, welche einzelne Tasks auf einem Rechencluster verteilt und verwaltet. [KKWZ15]

Im Folgenden wird auf die jeweiligen Aspekte von Spark eingegangen. Hierfür wird zunächst die grundlegende Spark-Technolgie *Resilient Distibuted Dataset* vorgestellt. Darauf aufbauend wird in Punkt 2.2 das Verteilen bzw. das Scheduling der einzelnen Datenverarbeitungsaufgaben auf dem Rechner-Cluster näher erläutert. In Punkt 2.3 wird die Clusterverwaltung und das Deployment von Spark-Anwendungen auf verschiedenen Plattformen dargestellt. In Punkt 2.4 wird auf zustandsbehaftete Spark-Operationen und deren Umsetzung eingegangen. Im letzten Punkt werden verschiedene Projekte des Spark-Ökosystems vorgestellt.

## 2.1 Resilient Distributed Dataset

Die *Resilient Distributed Datasets* kurz RDDs bilden die grundlegende Datenabstraktion von Spark und ermöglichen fehlertolerante und verteilte In-Memory-Berechnungen, wodurch unter anderem interaktive Anwendungen und das effiziente Wiederverwenden von Zwischenergebnissen ermöglicht werden. Von letzterem profitieren unter anderem viele iterative Machine-Learning-Algorithmen. [ZCDD12]

Aus der Sicht eines Spark Nutzers sind RDDs unveränderbare Datensammlungen, welche über mehrere Rechnerknoten verteilt sind. Diese Verteilung wird auch Partitionierung genannt, eine RDD besteht also immer aus einer Partition oder mehreren Partitionen.

In Spark können RDDs nur auf zwei Arten erzeugt werden:
- Durch das Einlesen aus einer persistenten Datenquelle, z.B. einer Datei aus dem Hadoop Distributed File System (HDFS)
- Durch bestimmte *Transformation*-Operationen auf einem schon existierenden RDD.

Ein RDD wird innerhalb von Spark durch mehrere Attribute beschrieben:
- Eine Liste mit Abhängigkeiten, entweder zu den Vorgänger-RDDs (auch *Parent* RDDs genannt) oder zu der Ursprungsdatenquelle. Dies ist abhängig davon, ob die RDD durch eine Transformation oder durch das Einlesen einer Datei erstellt wurde.
- Eine Liste der jeweiligen Partitionen einer RDD.
- Eine Funktion, welche eine Operation beschreibt, die parallel auf den einzelnen Partitionen ausgeführt wird.

Des Weiteren kann ein RDD über eine spezifische Partitionierungsfunktion verfügen, über welche bestimmt werden kann, nach welchem Schema die Daten über das Cluster verteilt werden. Außerdem kann die RDD Informationen über bestimmte Rechner verfügen, um beispielsweise die Berechnungen möglichst auf den Rechnern durchzuführen, welche auch die jeweiligen benötigten Daten speichern.

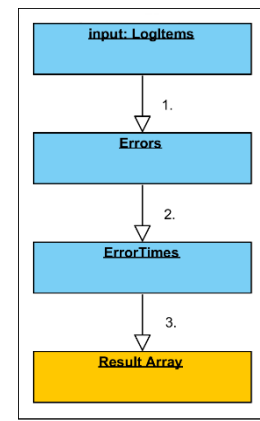

Um mit den RDDs zu arbeiten, nutzt Spark sogenannte *coarse-grained transformations*. Hierbei werden Änderungen auf einem Datensatz nicht wie in einer traditionellen Datenbank direkt auf jedem Element vollzogen, sondern über eine sogenannte *lineage* (etwa Abstammungsreihenfolge) zusammengetragen. (Abbildung 1) Hierdurch ist es möglich, mehrere aufeinander aufbauende Verarbeitungsschritte als ein Gesamtes zu betrachten. Diese Folge von Verarbeitungsschritten wird daraufhin als gerichteter azyklischer Graph (DAG) weiterverarbeitet. Spark definiert für die Nutzung von RDDs zwei Arten von Operationen. *Transformations, z.B. map() und filter(),* werden *lazy* ausgeführt. Das heißt, dass nicht direkt ein konkretes Ergebnis erzeugt wird, sondern neue RDDs mit einer veränderten *lineage* erzeugt werden. Hingegen sind *Actions, z.B. collect(), count() und forEach(),* Output-Operationen und liefern konkrete Werte an die Anwendung zurück.

**Abbildung 1: Darstellung der lineage anhand des Codes von Listing 1**

Folgendes Beispiel verdeutlicht diese Arbeitsweise:

```
1.  var input = readLogItemsAsRDD()
2.  var errors = input.filter(_.isError())
3.  errors.persist()
4.  var errorTimes= errors.map((item)=> item.getTime())
5.  var resultArray = errorTimes.collect()
```

**Listing 1: Einfache Batchverarbeitung mit Spark - Beispiel** [ZCDD12]

In Listing 1 wird eine Spark-RDD-Verarbeitung definiert, in der zunächst *logItems* als RDD eingelesen werden (Zeile 1) und anschließend nach Fehlermeldungen gefiltert werden (Zeile 2). Mittels der *persist()* Funktion in Zeile 3 speichert Spark die gefilterten *logItems* nach einer ersten Berechnung zwischen, um diese später erneut verwenden zu können. In Zeile 4 und 5 werden die einzelnen Zeitpunkte der Fehlermeldungen ausgegeben. In dieser Verarbeitung wird bis einschließlich Punkt 4 keine Operation auf den Daten ausgeführt. Erst die Action *collect()* in Zeile 5 führt dazu, dass die einzelnen Verarbeitungsschritte wirklich durchgeführt werden. Durch diese Art der Verarbeitung kann Spark die eigentliche Abarbeitung optimieren. Das vorgestellte Beispiel kann beispielsweise wie folgt erweitert werden:

```
6. var mysqlErrors = errors.filter((item) => item.message.contains("MySQL")))
7. mysqlErrors.count()
```

Listing 2: Erweiterung von Listing 1

In Listing 2 wird erneut auf die in Listing 1 erstellte Variable *errors* zugegriffen und noch weiter auf „MySQL Fehlermeldungen" eingeschränkt. Hier kann Spark nun auf die in Zeile 3 mittels *persist()* gespeicherten Zwischenergebnisse zurückgreifen und so auf das Filtern nach Errors auf dem gesamten Datenbestand verzichten. Die DAG-Repräsentation bildet auch die Grundlage der Fehlertoleranz von RDDs. Anhand der Informationen über die jeweiligen Abhängigkeiten einer RDD und der *Transformation* Funktion kann die jeweilige RDD jederzeit, z.B. im Fall des Ausfalls eines Workers, erneut berechnet werden. Diese erneute Berechnung muss für alle RDDs bis zum Originaldatensatz, im Beispiel oben das Lesen der LogItems, oder bis zu einem mit *persist()* gespeicherten Zwischenergebnis geschehen. Durch diese Art der erneuten Berechnung kann auf das redundante Speichern von Zwischenergebnissen verzichtet werden, da nur das verlässliche Speichern des DAGs nötig ist, um in jeder Situation die Berechnung auf einem anderen Worker zu wiederholen.

## 2.2 Scheduling

Damit Spark die einzelnen Transformationen auf einer RDD möglichst effizient verarbeiten und parallelisieren kann, wird nach dem Aufrufen einer *Action*, zunächst der DAG durch den *DAGScheduler* optimiert. Hierfür unterteilt Spark die Abhängigkeit zwischen einzelnen Transformationen in *narrow dependencies* und *wide dependencies*.

Bei einer *narrow dependency* wird jede Partition der Parent RDD für maximal eine Partition der Child RDD benötigt. Dies ist z.B. immer der Fall bei *map()* und *filter()* Transformationen und bedeutet, dass diese Transformationen lokal auf einem Worker ausgeführt werden können. Das Verteilen von Zwischenergebnissen zwischen mehreren Workern (Shuffle) entfällt.

Bei einer *wide dependency* hingegen nutzen mehrere Child Partitionen dieselbe Parent Partition, wodurch ein Shuffle nötig wird. Dies ist z.B. bei einfachen *join()* Operationen nötig, da hier ein kartesisches Produkt entsteht und alle Elemente einer Menge mit allen Elementen einer anderen Menge vereinigt werden. Der

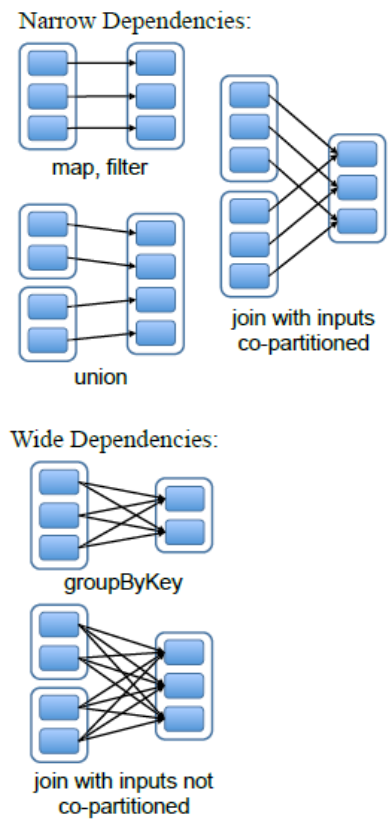

Abbildung 2: Narrow Dependencies & Wide Dependencies

*DAGScheduler* versucht nun möglichst viele *narrow dependencies* zu einer *stage* zusammenzufassen, um so möglichst viel Arbeit lokal auf den einzelnen Workern zu erledigen und die Anzahl an nötigen Shuffle Operationen zu minimieren. In Abbildung 3 ist dies dargestellt. So ist zu sehen, dass in der Stage 2 mehrere *narrow dependencies* zusammengefasst wurden (*C, D, E, F*). Des Weiteren bilden die *wide dependencies* hier immer den Übergang zwischen den einzelnen Stages. Nach der Optimierung des DAGs erstellt der *DAGScheduler* für jede Stage einzelne Tasks und übermittelt diese an den *TaskScheduler*. Dieser verwaltet nun die Abarbeitung der Tasks im Cluster. [Khoa15a, Khoa15b]

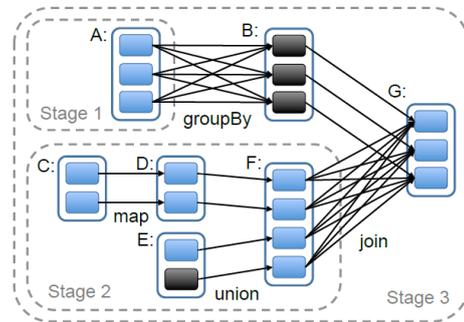

Abbildung 3 Darstellung eines komplexen DAG mit mehreren Stages

## 2.3 Deployment

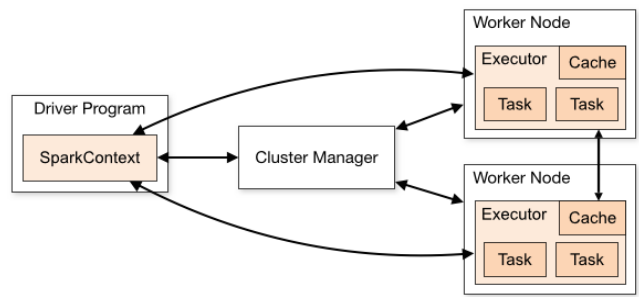

**Abbildung 4: Darstellung der Spark Cluster Komponenten und deren Kommunikation** [Apac16a]

Ein Spark Deployment im Cluster besteht immer aus folgenden Komponenten:

**Driver Programm:**
Der Driver ist der zentrale Koordinator des Spark Systems [KKWZ15]. Hier ist außerdem die eigentliche Anwendungslogik implementiert und Spark wird über den SparkContext angesprochen. Das Driver Programm ist somit auch für die Verwaltung der RDDs und deren Scheduling zuständig.

**Worker/Slave/Executor:**
Bei der Benennung dieser Komponente wird zwischen dem Worker, dem Slave und dem Executor unterschieden. Der Worker ist der jeweilige Rechner, somit die physikalische Hardware. Der Slave benennt den Teil der Cluster-Software, welche auf dem Worker ausgeführt wird und mit dem Cluster-Manager kommuniziert. So läuft auf jedem Worker genau ein Slave. Der Executor wird im Rahmen der jeweiligen Spark Anwendung gestartet und ist die Software

Komponente zum Ausführen einzelner Tasks auf einem Slave. Jeder Executor startet eine eigene Java VM, wobei es möglich ist, mehrere Executor auf einem Slave zu starten. Hauptsächlich sind die Executor für die Abarbeitung der vom Driver erstellten Task zuständig und liefern diesem die Ergebnisse zurück. Des Weiteren können die Executors RDDs zwischenspeichern, sodass mehrere Tasks mit denselben Daten arbeiten können. [KKWZ15]

**Cluster Manager:**
Der Cluster Manager ist für das initiale Starten der Executor auf den im Cluster verfügbaren Slaves zuständig. Außerdem ist es auch möglich das Driver Programm über den Cluster Manager zu starten. Dies ermöglicht eine Fehlertoleranz gegenüber Ausfällen des Drivers, da hier der Cluster Manager die Laufzeit des Drivers überwacht. Der Cluster Manager selbst lässt sich über verschiedene Möglichkeiten realisieren. Diese werden nun kurz beschrieben:

- **Standalone Cluster Manager:**

Dies ist ein einfacher direkt mit Spark mitgelieferter Cluster Manager. Er besteht aus mindestens einem Masterknoten und mehreren Worker-Knoten. In der Standardkonfiguration bietet er eine Fehlertoleranz gegenüber dem Ausfall von Worker-Knoten. Um den Ausfall des Masters selbst kompensieren zu können, muss dieser zunächst auf mehreren Rechnern redundant gestartet werden und sich über das System ZooKeepers mit den anderen Mastern synchronisieren. [Apac16b]

- **YARN**:

Bei YARN handelt es sich um den Resource Manager von Hadoop. So kann eine Spark Anwendung auf demselben Cluster betrieben werden, auf dem eine Hadoop Anwendung läuft und so die verfügbaren Ressourcen ohne Mehraufwand mitbenutzen. Hierfür kommuniziert der Spark-Driver direkt mit dem YARN-Master. Außerdem wird auf den meisten YARN-Worker-Knoten auch HDFS genutzt, sodass Spark die Daten auf denselben Knoten verarbeiten kann, auf denen sie auch gespeichert sind. Dies reduziert den Netzwerk Overhead und steigert die Verarbeitungsgeschwindigkeit. [KKWZ15]

- **Mesos**:

Mesos verwaltet einen Cluster als einen einzigen Pool von Ressourcen und bietet verschiedene Schnittstellen an, sodass verschiedenste Anwendungen parallel auf dem Cluster ausgeführt und gescheduled werden können [Apac16c]. Es ist z.B. möglich auf demselben Cluster eine Spark-Anwendung und DockerDockerDocker-Container mit einer Webanwendung ausfallsicher zu betreiben. [KKWZ15]

## 2.4 Gemeinsame Variablen

Wie zuvor beschreiben, ist es möglich auf einem RDD verschiedene Operationen auszuführen, welche jeweils parallel auf dem Cluster ausgeführt werden. Die einzelne Operation wird meist über eine Funktion definiert, die der jeweiligen RDD-Operation als Parameter übergeben wird. Diese Funktionen werden aus der Sicht von Spark als Closures behandelt. Ein einfaches Beispiel stellt hierfür Listing 3 dar.

```
errors.filter((item)=> item.message.contains("MySQL")))
```

**Listing 3: Spark Beispiel mit Filterfunktion ohne Variable im Closure**

```
var filterTerm = "MySQL"
errors.filter((item)=> item.message.contains(filterTerm)))
```

**Listing 4: Spark Beispiel mit Filterfunktion mit Variable im Closure**

Listing 4 stellt hingegen eine Funktion dar, bei der in der inneren Funktion auf eine äußere Variable zugegriffen wird. Um die Verteilung der Operation auf dem Cluster zu ermöglichen, muss Spark somit nicht nur die übergebende Funktion serialisieren, sondern auch noch die benötigten Eigenschaften des Closure-Scopes. Aus diesem Grund ist es somit nur möglich serialisierbare Objekte in der inneren Funktion zu nutzen. Spark verfügt zudem über Möglichkeiten den jeweiligen Closure zu analysieren, um nicht benötigte bzw. nicht serialisierbare Eigenschaften aus dem Closure zu entfernen. Bei der Verteilung der Closure-Variablen im Cluster entstehen auf den ersten Blick gemeinsame Variablen, auf denen jeweils der Driver und die Executor arbeiten können. Hierbei ist zu beachten, dass Spark keine Wertänderungen zwischen den Variablen synchronisiert. Des Weiteren werden die Closure-Variablen für jede Operation erneut auf die Executor kopiert, wodurch bei größeren Objekten ein signifikanter Overhead entstehen kann [KKWZ15].

Als Alternative zu diesem Vorgehen bietet Spark zwei Konzepte an:

- **Accumulators:**
  Hierdurch wird es möglich, dass die Executor Änderungen an einer gemeinsamen Variable durchführen können und diese auch auf den Driver synchronisiert wird. Zum Beispiel lässt sich so eine Zählvariable realisieren, welche in jeder verteilten Transformation inkrementiert wird. Für die Worker ist es aber nicht möglich, auf den jeweils aktuellen Wert zuzugreifen, dies kann nur der Spark Driver.

- **Broadcast variables:**
  Diese Funktion ermöglicht es, große Objekte sehr effizient auf alle Worker zu verteilen. Dies eignet sich z.B. gut für Lookup-Tabellen, die während des Verarbeitungsprozesses gebraucht werden. Eine Einschränkung der Broadcast Variablen ist aber erneut, dass die geteilten Objekte auf den Workern nicht veränderbar sind.

## 2.5 Spark Ökosystem

Das Spark Projekt verfügt mittlerweile über mehrere Sub-Projekte, welche jeweils verschiedene Anwendungsfälle abdecken. Im Folgenden werden die nicht weiter in dieser Arbeit behandelten Projekte kurz vorgestellt:

**Spark SQL:**
Spark SQL erleichtert die Arbeit mit strukturierten und semistrukturierten Daten. So bietet es Möglichkeiten Daten aus verschieden Datenquellen direkt zu laden, wie z.B. JSON, Hive oder Prequed und diese mittels einem SQL-Dialekt zu verarbeiten. Außerdem lassen sich auch mittels JDBC klassische Datenbanken anbinden und direkt aus Spark heraus abfragen. Um diese Art der Verarbeitung zu ermöglichen, abstrahiert Spark SQL die RDDs in sogenannte DataFrames bzw. Datasets. Durch die Abstraktion verfügt Spark SQL über zusätzliche Schema-Informationen der jeweiligen Daten und kann zusätzliche Optimierungen nutzen, sodass hier die Ausführungsgeschwindigkeit in manchen Fällen gegenüber normalen RDDs steigen kann [XiAL15]. Die DataFrame-Abstraktion ist vergleichbar mit den Tabellen in relationalen Datenbanken und kann z.B. automatisch anhand von einem JSON-String erzeugt werden. Datasets hingegen sind stark typisiert und dadurch vergleichbar mit den RDDs. Sie profitieren aber zugleich von den erweiterten Optimierungsmöglichkeiten des Spark SQL Projekts. [Apac16d, KKWZ15]

**Spark MLlib**:
MLlib hat sich als Ziel gesetzt, Machine Learning einfacher und skalierbarer zu machen. Hierfür bietet es zum einen Algorithmen für unteranderem Klassifizierung, Regression und Clustering an. Zum anderen bietet es eine API, welche auf den aus Spark SQL bekannten DataFrames aufbaut und es ermöglicht, Machine Learning Pipelines zu erstellen. [Apac16e, KKWZ15]

**GraphX:**
GraphX ist ein Projekt, welches die API von Spark um Graph Verarbeitung erweitert. Hierfür führt es zunächst eine RDD-Abstraktion ein, welche einen gerichteten Property Graphen darstellt. Um Berechnungen auf dem Graphen durchzuführen, implementiert GraphX zum einen einige Low-Level-Graph-Operationen sowie die Pregel-API. Außerdem bietet es einige fertige Algorithmen wie z.B. PageRank an [Apac16f, KKWZ15].

# 3 Spark Streaming

Spark Streaming ist ein Projekt, welches Spark um Fähigkeiten zur Verarbeitung von Daten in „Echtzeit" erweitert. Dies ist für einen großen Bereich von Big Data-Anwendungsfällen besonders wichtig. So muss z.B. eine Betrugserkennung bei einer Kreditkartenzahlung möglichst sofort während der jeweiligen Transaktion geschehen, um diese eventuell noch abbrechen zu können. Sobald die Datenanalyse aber nur in langwierigen Batchprozessen erfolgen kann, ist der Informationsgewinn in diesem Fall wertlos, da das Geld schon überwiesen wurde. Spark Streaming hat sich als Ziel gesetzt, Analysen zu ermöglichen, die zum einen parallel auf einer Vielzahl von Rechnern lauffähig sind und zum anderen maximal eine Latenz im Sekundenbereich tolerieren [ZDLH13]. Anwendungsfälle mit noch geringeren Latenzanforderungen, z.B. Hochfrequenz-Aktienhandel, gehören nicht zu der Zielgruppe von Spark Streaming [ZDLH13]. Ein weiteres Ziel ist es, den Ausfall von Rechnern im Cluster möglichst gut und schnell zu kompensieren. In Spark Streaming werden die einzelnen Datenverarbeitungsschritte auf kontinuierlichen Datenströmen definiert, sogenannten Streams. Diese können aus verschiedenen Quellen stammen, z.B. aus einem TCP-Socket, wobei hier jede empfangene Nachricht ein Element im Stream darstellt. Um die Verarbeitung von Streams zu ermöglichen, führt Spark das Konzept der Discretized Streams (DStreams) ein, welche eine Abstraktion der bekannten RDDs darstellen. Im Folgenden wird auf die Eigenschaften dieser Abstraktion eingegangen.

## 3.1 Micro-Batch Verarbeitung

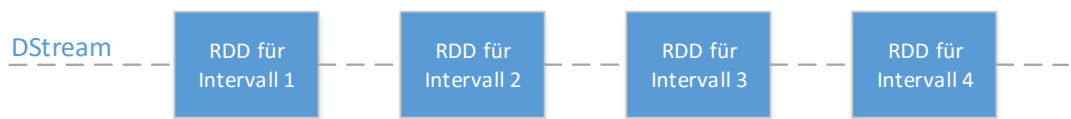

**Abbildung 5: Darstellung eines DStreams mit 4 RDDs für die jeweiligen Zeitintervalle**

In Spark Streaming werden die einzelnen Nachrichten in dem sogenannten Micro-Batch Verfahren verarbeitet. Hierbei werden mehrere Nachrichten aus einem gemeinsamen Zeitintervall in ein RDD zusammengefasst und anschließend jeweils wie eine bei Spark übliche Batch-Verarbeitung behandelt [ZDLS12]. Dieses Verfahren unterscheidet sich stark von dem z.B. bei Apache Storm üblichen one-record-at-a-time Verfahren, bei dem jedes eintreffende Element direkt und einzeln weiterverarbeitet wird. Die Länge des Batch-Intervalls wird beim Erzeugen eines DStreams als Parameter übergeben und kann zwischen 100ms und einigen Sekunden liegen.

## 3.2 DStream Operationen

Da die DStreams intern auf RDDs basieren, bieten sie zum Teil ähnliche Operationen an. So werden diese erneut in *Transformations* und *Actions* unterteilt. Hierbei erzeugen die *Transformations* jeweils neue DStreams. Als Transformationen existieren zum einen die von RDDs bekannten *map()* und *filter()* Operationen, welche direkt auf den einzelnen Daten eines Streams arbeiten. Zum anderen bietet Spark Streaming auch eine *transform()* Funktion an, welche eine vom Benutzer definierte Funktion auf jedem RDD ausführt und somit sehr generisch ist. Die *Actions* umfassen erneut Output-Operationen, z.B. *forEachRDD()*. Des Weiteren lassen sich auch verschiedene DStreams über die Operation *join()*, bzw. *union()* zusammenfassen. Dies ist auch möglich, wenn die Ursprungsstreams aus verschiedenen Quellen stammen. Eine vollständige Liste aller angebotenen Operationen ist im Anhang enthalten.

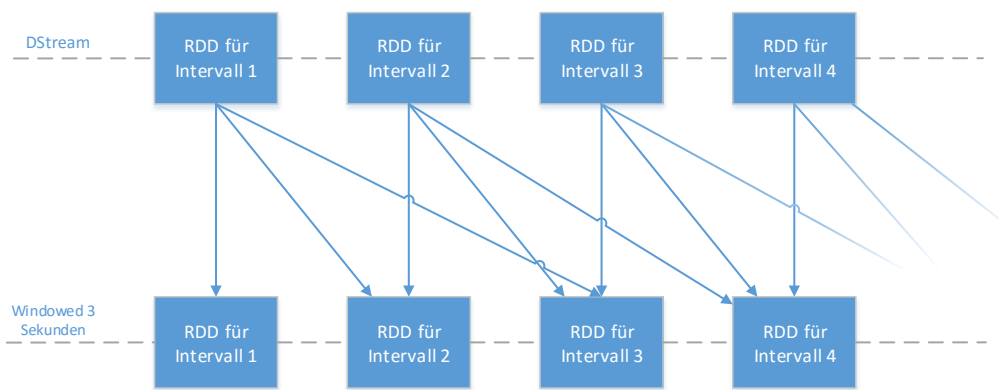

**Abbildung 6: Windowed DStream mit einem Window von 3 Sekunden**

Darüber hinaus bieten DStreams sogenannte Window Operationen an. Hierbei werden mehrere Zeitintervalle bzw. mehrere RDDs zusammengefasst, wodurch sich mehrere Datenintervalle zusammen verarbeiten lassen. Hierbei wird zwischen der Window-Länge und dem Sliding-Intervall unterschieden. Ein rein über die Window-Länge manipulierter DStream ist in Abbildung 6 dargestellt. Hierbei enthält beispielsweise das *Intervall 3* im neuen DStream alle Daten aus den Intervallen 1,2,3 des ursprünglichen DStreams.

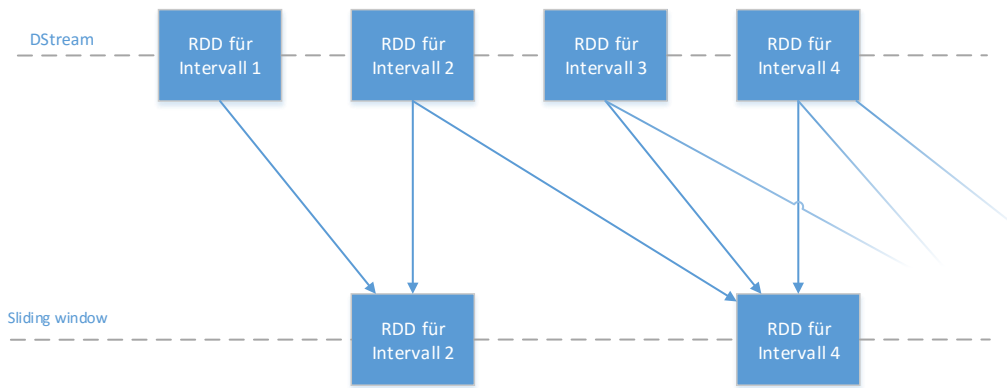

**Abbildung 7: Sliding Window mit einer Window Länge von 3 Sekunden und einem Sliding-Intervall von 2 Sekunden**

In Abbildung 7 ist die Veränderung des Sliding Intervalls dargestellt, mittels welchem die Zeit, die zwischen den einzelnen RDDs liegt, verändert werden kann. Dieser Wert muss zwingend größer, als das bei der Erzeugung des initialen DStreams angegebene Intervall, sein.
Des Weiteren bietet Spark Streaming sogenannte Stateful-Transformations an. Hierüber lässt sich ein Zustand über mehre DStream-Intervalle hinweg festhalten und manipulieren. Wichtig ist, dass diese Operationen nur auf PairedDStreams angeboten werden, die zugrundeliegenden Datensätze also als Key-Value-Paar vorliegen müssen. Der jeweilige Zustand wird immer zu den jeweiligen Keys zugeordnet. Durch die Beschränkung auf Key-Value Paare ist es möglich, dass Spark die einzelnen Datensätze nach ihrem Key partitionieren kann und so den Zustand nur lokal auf den Workern verwalten muss, sodass auf das Synchronisieren von Änderungen der Zustände zwischen den einzelnen Workern verzichtet werden kann.

## 3.3 Datenquellen

Zur Erzeugung von DStreams werden von Spark Streaming verschiedene Datenquellen unterstützt, welche über Receiver angesprochen werden können. So bietet Spark Streaming unter anderem Receiver für Apache Kafka, Apache Flume, Twitter, Sockets oder HDFS Dateien an. Außerdem können eigene Datenquellen in Spark integriert werden, wofür lediglich eine von der Klasse Receiver erbende Klasse erstellt werden muss. Der Receiver empfängt zunächst die Nachricht von der Datenquelle und übermittelt dann die empfangenen Nachrichten, durch den Aufruf der *store()* Funktion an Spark. Hierbei werden die Daten zum einen über das Spark-Cluster verteilt, aber auch repliziert, um eine sichere Aufbewahrung zu garantieren. Zur Laufzeit der Spark Anwendung wird der Receiver auf einem zufälligen Worker gestartet und benötigt zwingend einen eigenen CPU-Kern. Spark Streaming kann also nicht auf einem System lokal ausgeführt werden, welches nur über einen CPU-Kern verfügt, da so keine Kapazitäten für die eigentliche Datenverarbeitung zur Verfügung stehen [KKWZ15].

## 3.4 Fehlertoleranz

Spark Streaming läuft oft im sogenannten 24/7 Betrieb und bietet eine exactly once-Fehlersemantik über die komplette Verarbeitungskette hinweg. Dies bedeutet, dass jede von dem System empfangene Nachricht genau einmal verarbeitet wird, auch im Fall des Ausfalls eines Knotens, wodurch Spark eine hohe Fehlertoleranz bieten muss. Andere Streaming Systeme, wie z.B. Apache Storm verwenden oft eine Replikation oder einen Upstream-Backup-Mechanismus, um eine Toleranz gegenüber Fehlerfällen zu bieten. Beide Verfahren sind allerdings nicht sehr effizient. So werden bei einer Replikation die Daten immer auf mehreren Rechnern bereitgehalten, wodurch zwar schnell auf Fehler reagiert werden kann, aber zusätzliche Kosten entstehen. Beim Upstream-Backup hingegen hält der sendende Rechner einen Puffer vor und kann die Daten im Fehlerfall erneut einspielen, was zum Teil sehr viel Zeit in Anspruch nehmen kann. Eine kurze Wiederherstellungszeit ist aber gerade bei der Streamverarbeitung besonders wichtig, da nur so alle Daten mit den nötigen Latenzzeiten verarbeitet werden können. Die Fehlertoleranz von DStreams basiert vor allem auf der Möglichkeit die zu Grunde liegenden RDDs jederzeit erneut berechnen zu können. Des Weiteren führte Spark Streaming das sogenannte Checkpointing ein, bei dem periodisch Daten auf ein zuverlässiges externes System, z.B. HDFS, geschrieben werden. Hierdurch wird zum einen erreicht, dass nur alle Operationen bis zum letzten Checkpoint wiederholt werden müssen und zum anderen, dass eine Ausfallsicherheit des Drivers gewährleistet werden kann, da so ein neuer Driver die Arbeit an einem vorherigen Checkpoint wiederaufnehmen kann.
Im Folgenden wird auf die Fehlertoleranz der einzelnen Spark Streaming Komponenten eingegangen:

**Executer**:
Hierfür werden die Wiederherstellungsmechanismen der RDDs genutzt.

**Receiver**:
Falls der Knoten des Receivers ausfällt, wird Spark diesen automatisch erneut auf einem anderen Knoten starten. Hierbei könnte es jedoch unter Umständen zu einem Datenverlust kommen, da Daten eventuell schon von der Datenquelle empfangen wurden, diese aber noch nicht sicher im Spark System gespeichert wurden. Die Datenquelle würde die Daten auch nicht erneut senden, da sie davon ausgeht, dass alle Daten korrekt empfangen werden konnten. Da dieses gegen die exactly-once Semantik von Spark Streaming spricht, werden die einzelnen Receiver in *reliable* und *unreliable* unterteilt. Reliable Receiver können die Datenquelle z.B. über ein *Acknowledgement* darüber benachrichtigen, dass die Daten vom System korrekt entgegengenommen wurden und somit sicher sind. So kann, im Fall eines Ausfalls des Receivers, die Datenquelle die jeweils noch nicht bestätigten Daten erneut an Spark senden. Unreliable Receiver können den Empfang der Daten hingegen nicht bestätigen, sodass es zu dem Verlust von Daten kommen kann und hierdurch nur noch at-most-once Fehlersemantik geboten werden kann.

**Driver:**

Bei dem Ausfall des Drivers wird die komplette Anwendung beendet und alle im Speicher gehaltenen Zustände werden unbrauchbar. Um hier trotzdem eine Fehlertoleranz zu bieten, wird der zuvor angesprochene Checkpoint Mechanismus verwendet. Außerdem muss beim Empfangen der Daten ein Write-Ahead-Log verwendet werden, da die Datenquelle die schon bestätigten Nachrichten nicht erneut senden wird. Zudem muss sichergestellt werden, dass der Driver auch neugestartet wird. Hierfür kann z.B. der Standalone-Cluster-Manager die Überwachung des Drivers übernehmen und diesen auf einen beliebigen anderen Worker-Knoten neu starten.

# 4   Alternative Streaming Systeme

Im folgenden Kapitel werden die Streamingplattformen Storm und Flink vorgestellt, die zum Teil in Konkurrenz zu Spark Streaming stehen. So ist Storm eine reine Streaminglösung, Flink beherrscht sowohl Batch- als auch Stream-Verarbeitung

## 4.1   Storm

Storm ist eine verteilte Plattform zur Echtzeit-Stream-Verarbeitung. Unter anderem ermöglicht es Realtime Analytics, online-machine-learning und verteilte RPCs und sorgt für eine hohe Ausfallsicherheit und Skalierbarkeit [Apac15a]. Storm wurde 2011 von Twitter veröffentlicht und ist seit 2014 ein Apache Top Level Projekt [Goet14, Marz11a]. Storm verarbeitet Streams nach dem one-record-at-time Mechanismus. Hierdurch kann eine sehr geringe Latenz erreicht werden. Die Architektur eines mit

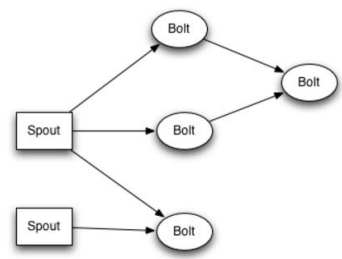

**Abbildung 8: Storm Architektur**

Storm erstellten Systems ist in Abbildung 8 dargestellt und unterteilt die Systemkomponenten in *Spouts* und *Bolts*. Hierbei sind die *Spouts* die Datenquelle, die einen Stream erzeugt und die *Bolts* sind einzelne Streamverarbeitungsschritte, welche nacheinander den Stream verändern. Die Verkettung von einzelnen Spouts und Bolts wird als *topology* bezeichnet und bildet einen wie in Abbildung 8 dargestellten Graphen. Für Storm existiert wiederum ein Unterprojekt namens *Trident,* welches die grundlegenden von Storm angebotenen Verarbeitungsmechanismen abstrahiert und einen Micro-Batch-Ansatz mit *exactly-once* Fehlersemantik anbietet. Storm selbst bietet hingegen nur eine *at-least-once* Semantik.[Apac15a]

## 4.2   Flink

Flink ist eine verteilte Stream- und Batchverarbeitungsplattform und wird seit März 2015 von der Apache Foundation betreut [Apac15b]. Flink steht somit im direkten Wettbewerb zu Spark. Flink baut im Kern auf einer Streaming-Engine auf und abstrahiert diese für die Batchverarbeitung. Für die Definition der jeweiligen Datenverarbeitungsschritte bietet Flink eine mit Spark vergleichbare Programmierschnittstelle an. Auch hier werden die einzelnen Verarbeitungsschritte zunächst *lazy* definiert und erst beim Ausführen einer *execute* Anweisung gestartet. Für die Streamverarbeitung bietet Flink umfangreiche Verarbeitungsmöglichkeiten und unterstützt z.B. Event Timing, wodurch die einzelnen Elemente eines Streams nicht nach der Ankunftszeit im System, sondern nach einem in der Nachricht enthaltenen Zeitstempel sortiert werden. Hierdurch ist es möglich falsche Sortierungen zu beheben, die z.B. auf den Übertragungsweg zurückzuführen sind. Des Weiteren unterstützt Flink iterative Algorithmen und kann diese direkt durch native Schleifen auf den einzelnen Worker abbilden. In Bezug auf die Fehlertoleranz bietet Flink eine *exactly-once* Semantik und realisiert diesen über einen auf verteilten Schnappschüssen[1] basierenden Checkpointing-Mechanismus [KoEM15]. Des Weiteren implementiert Flink ein eigenes Memory-Management, um unter anderem die Arbeit des Java-Garbage-Collectors zu optimieren. [ABEF14, Apac16g, KoEM15]

## 4.3     Vergleich der Streaming-Systeme

Abschließend folgt ein Vergleich der zuvor vorgestellten Systeme mit Spark Streaming:

|  | Spark Streaming | Flink | Storm |
|---|---|---|---|
| Verarbeitungsart | micro-batch | one-record-at-time | one-record-at-time |
| Latenz | Sekunden | Millisekunden | Millisekunden |
| Durchsatz | hoch | hoch | gering |
| Fehlersemantik | *exactly-once* | *exactly-once* | *at-least-once* |
| Contributer im letzten Monat | 109 | 31 | 23 |
| Windowing/Eventtime | ja/nein | ja/ja | nur mittels Trident / nein |

**Tabelle 1: Vergleich zwischen Spark Streaming, Flink und Storm**

Zaharia zeigte, dass mit Spark Streaming, im Vergleich zu Storm, ein 4 mal höherer Durchsatz erreicht werden kann, also 4 mal mehr Datensätze pro Sekunde verarbeitet werden können [ZDLH13]. Hierbei ist von Goetz angemerkt worden, dass die Studie mittlerweile veraltet ist und an Storm und Spark viele Verbesserungen vorgenommen wurden, wodurch der Vergleich eventuell nicht mehr zutreffend ist. Außerdem wäre ein Vergleich zwischen Trident und Spark Streaming realistischer, da beide Projekte die Micro-Batch Verarbeitung nutzen und somit vergleichbarer sind [GoHo14]. Generell lässt sich aber sagen, dass Spark das mittlerweile häufiger genutzte Projekt ist und durch das angebotene Ökosystem für eine Vielzahl an Anwendungsfällen nutzbar ist. Storm hingegen ist auf die Streamverarbeitung beschränkt. Andererseits muss bei der Nutzung von Spark Streaming abgewogen werden, ob die höhere Latenz noch den Anforderungen des Anwendungsfalls entspricht.  Bei dem Vergleich zwischen Spark Streaming und Flink zeigen sich zunächst die vielen Gemeinsamkeiten der beiden Plattformen. So ermöglichen beide Plattformen sowohl Batch-Verarbeitung als auch Stream-

---

[1] Es wird eine abgewandelte Variante des Chandy-Lamport-Algorithmus verwendet.

Verarbeitung und bieten eine in Teilen sehr ähnliche API an. Da Flink aber einen höheren Fokus auf die Stream-Verarbeitung legt, bietet es hier umfangreichere Funktionen als Spark Streaming an. Besonders ist hier die Event-Time-Verarbeitung zu nennen. Des Weiteren erreicht Flink wie Spark Streaming einen hohen Durchsatz, aber kann gleichzeitig eine mit Storm vergleichbare geringe Verarbeitungslatenz bieten [CDEF16]. Des Weiteren optimiert Flink die Performance der Verarbeitungen umfassender als Spark, da es z.B. ein eigenes Memory-Management implementiert. Hierbei ist jedoch zu vermerken, dass Spark mittlerweile auch einige dieser Verbesserungen umsetzt oder plant diese umzusetzen [XiRo15].

# 5 Praxisbeispiel

Moderne Autos sind mittlerweile hochkomplexe verteilte Informationssysteme und bestehen zum Teil aus mehr als 70-100 einzelnen Steuergeräten (Electronic Control Units), welche wiederum aus mehrere Mikroprozessoren bestehen können. Die einzelnen Steuergeräte arbeiten eigenständig und sammeln bzw. generieren Daten, welche von anderen Systemen genutzt werden können. [ALHS12, Char09]
Diese Informationen sind momentan meist nur lokal im Auto vorhanden und werden nur selten, z.B. von Werkstätten für Diagnosezwecke abgerufen.
Doch mittlerweile entstehen immer mehr Systeme, die die Daten verschiedener Autos während der Fahrt auswerten. So unterscheidet man unter anderem zwischen „Car-to-Car-Communication", bei denen die Autos untereinander Informationen austauschen oder „Car-to-Infrastruktur-Communication"- Systemen, bei denen das Auto Informationen mit z.B. Verkehrsleitsystemen austauscht. [Carc15]
Besonders interessant ist die Analyse der Fahrzeugdaten in Echtzeit, denn so können dem Fahrer während der Fahrt für ihn relevante Informationen geliefert werden. Des Weiteren ist oft eine Verzahnung von verschiedenen Datenquellen interessant, denn so lassen sich besonders interessante Anwendungsfälle realisieren.
Das hier zu entwickelnde Car Information System soll für die Realisierung solcher Daten geeignet sein und wird im Folgenden konzipiert und realisiert.
In Punkt 5.1 werden das Car Information System und konkrete Anwendungsfälle für dieses vorgestellt. In Punkt 5.2 wird der Entwurf und die Konzeption des Systems beschrieben
In Punkt 5.3 wird auf die Realisierung des Systems eingegangen.

## 5.1 Hintergrund

Im Folgenden wird das benötigte System definiert und konkret beschrieben, um die genauen Anforderungen an das System ermitteln zu können.
Außerdem werden exemplarisch verschiedene realisierbare User-Stories präsentiert, welche sich mithilfe des Systems umsetzen lassen sollen.

### 5.1.1 Car Information System
Für den Begriff „Car Information System" (kurz CIS) gibt es keine allgemein gültige Definition. So benutzt der japanische Autozulieferer Hitachi Automotive Systems, Ltd. den Begriff als einen Überbegriff für eine Vielzahl von angebotenen Diensten und Services (z.B. Staumeldungen) [Hita15]. Des Weiteren wird der Begriff „Driver Information System" benutzt. Hiermit ist eine Art modernes Armaturenbrett gemeint, welches für den Fahrer relevante Informationen anzeigen kann und diesen auch über Ereignisse benachrichtigen kann. [Audi16]

Im Rahmen dieser Arbeit wird folgende Definition des Begriffes Car Information System verwendet:

> *Ein Car Information System ist ein System, welches alle mit dem Auto zusammenhängenden Daten zentral sammelt und analysiert. Das System wird um weitere Datenquellen erweitert und kann den Fahrer des Autos aktiv über Analyseergebnisse benachrichtigen.*

Ein solches System wird unter anderem von den Startups Dash und Automatic angeboten [Auto16, Dash16]. Hier wird die Dienstleistung oft mit dem Begriff „Smart Driving" vermarktet und dem Autofahrer verschiedene Informationen über eigene Apps angeboten.

### 5.1.2 Rahmenbedingungen

Der Aufbau des hier zu entwickelnden Car Information Systems wird im Folgenden grob skizziert und erläutert:

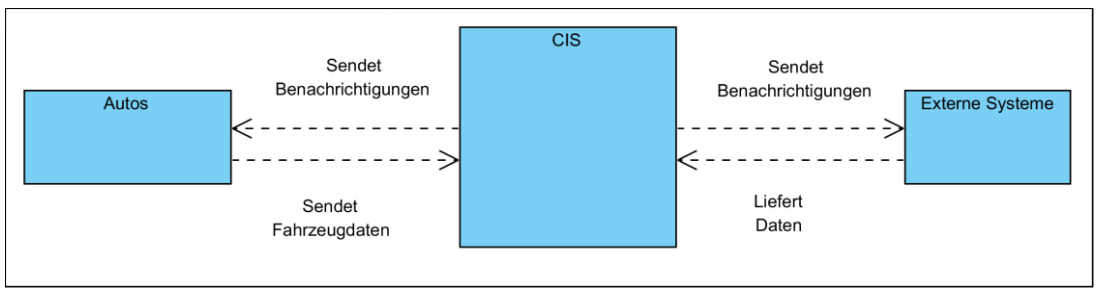

**Abbildung 9: Grober Aufbau des CIS Systems**

Wie in Abbildung 9 dargestellt, ist das CIS System von mehreren anderen Systemen abhängig. So sind die Autos zum einen Empfänger für bestimmte Ereignisse, über welche sie vom CIS benachrichtigt werden und zum anderen sind sie gleichzeitig die wichtigste Datenquelle für das CIS, da sie permanent Fahrzeugdaten zum CIS senden. Die Fahrzeugdaten können unter anderem, die GPS Position des Autos, die Geschwindigkeit, den Tankstand oder die aktuelle Motorumdrehungszahl beinhalten. Dabei ist die Verfügbarkeit der Werten immer davon abhängig, auf welche Art die Daten vom Auto abgerufen werden. So ist z.B. eine direkte Anbindung des CIS seitens der Autohersteller denkbar, sodass hier eine sehr direkte Integration möglich ist und somit auch die Möglichkeit besteht auf fahrzeugspezifische Informationen zuzugreifen. Ein solches System wäre z.B. mit dem von BMW bekannten Dienst ConnectedDrive vergleichbar [Bmwa16]. Da eine solche Integration zunächst das System auf nur einen Autohersteller beschränken würde, wird im Rahmen dieser Arbeit, genau wie bei den Diensten Dash und Automatic, auf die On-Board-Diagnose-2 (OBD2) Schnittstelle gesetzt. Dabei handelt es sich um eine Fahrzeugdiagnoseschnittstelle, die seit 2001 in der EU bei allen Neuwagen verpflichtend ist [Obdn16]. Über diese Schnittstelle kann auf eine Vielzahl standardisierter Werte zugegriffen werden, wobei auch die oben genannten Daten enthalten sind. Mittels eines speziellen OBD2-Bluetooth-Dongles kann jedes Smartphone Daten bei den Fahrzeugen abfragen und diese an das CIS übermitteln. Des Weiteren kann das CIS an weitere externe Systeme angebunden werden. Diese können zum einen Datenquellen sein, sodass das CIS hier entweder permanent Daten empfängt oder selbstständig Daten abruft. Durch die Nutzung dieser weiteren Datenquellen könnten z.B. die Daten der Fahrzeuge angereichert werden, um so neue Erkenntnisse zu gewinnen. Zum anderen ist es vorstellbar, dass bestimmte externe Systeme auch vom CIS über Ereignisse benachrichtigt werden sollen, um die Analyseergebnisse des CIS weiter zu verarbeiten oder auch selbstständig bei dem CIS Analyseergebnisse abfragen.

Aus der vorangegangenen Beschreibung des CIS lässt sich folgender abstrakter Aufbau des CIS ableiten.

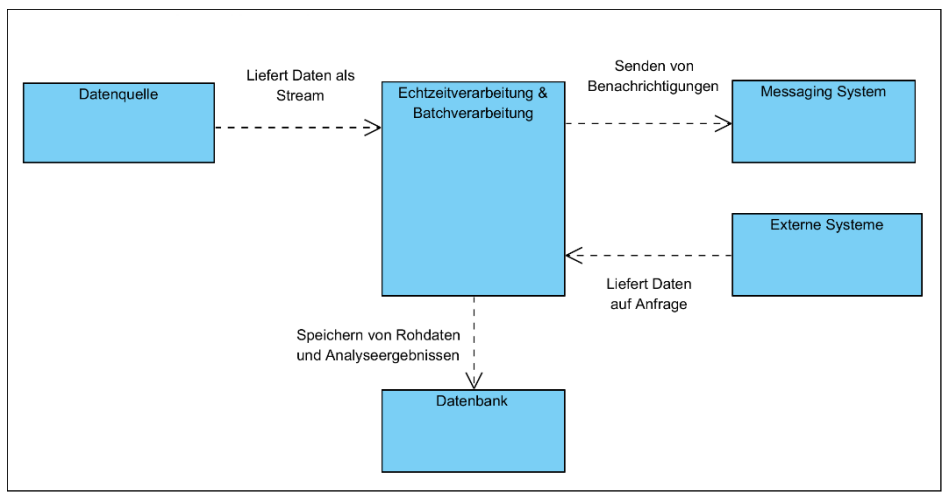

**Abbildung 10: Abstraktion des CIS Verarbeitungssystems**

So soll diese Grafik vor allem die zentrale Rolle der Echtzeitverarbeitung für das CIS verdeutlichen. Hier müssen im Rahmen des CIS alle Fahrzeugdaten unmittelbar nach deren Eintreffen verarbeitet werden. Des Weiteren stellt die Abbildung 10 eine sehr allgemeine Sicht auf das System dar und zeigt, dass das CIS System vor allem permanent Daten aus Datenquellen empfängt, diese mit Daten aus externen Systemen anreichert und Analyseergebnisse entweder über ein System, hier Messaging System, verschickt oder lokal in einer Datenbank speichert. Dies zeigt, dass sich das CIS als ein passendes Beispiel für Streamverarbeitungssysteme eignet. Auch der folgende Systementwurf soll möglichst allgemein gehalten werden, um so eine Plattform bzw. Architektur zu erhalten, die auch für ähnliche Aufgabenstellungen praktikabel ist.

### 5.1.3 User Stories

Im Folgenden sollen einige der für das Car Information System möglichen User-Stories dargelegt werden, um einen Überblick über die Möglichkeiten eines solchen Systems zu erhalten. Teilweise sind die User-Stories zudem an bereits existierende Systeme angelehnt. Dies ist jeweils vermerkt.

**Tankstellensuche:**
> *„Als Autofahrer möchte ich Informationen zu besonders günstigen Tankstellen erhalten. Ich will aber nur benachrichtigt werden wenn ich auch tanken muss, sodass ich keine unnötigen Benachrichtigungen erhalte."*

Für diese User-Story müssen verschiedene Informationsquellen zusammengefasst werden. So müssen zum einen die aktuelle Position des Autofahrers und der Tankstand des Autos vorliegen. Des Weiteren benötigt man Informationen über die aktuellen Tankstellenpreise, sowie Analysen über die historischen Tankstellenpreise, um eine möglichst wertvolle Benachrichtigung aussprechen zu können. Dies ist vergleichbar mit Diensten wie clever-tanken.de, welche ihren Kunden Informationen bezüglich der gerade aktuellen Tankstellenpreise liefern [Info16]. Durch die Fahrzeugdaten des CIS können aber noch mehr Information ausgewertet werden, um die Qualität der Empfehlungen zu verbessern.

**Fahrstilanalyse-System:**
> *„Als Fahrer will ich nach dem Beenden der Fahrt über meinen Fahrstil informiert werden, um diesen eventuell verbessern zu können."*

In dieser User-Story muss das Beenden der Fahrt erkannt werden, um dem Nutzer daraufhin eine Analyse der Fahrt zu übermitteln.
Hierfür müssen die während der Fahrt gesammelten Daten analysiert werden. Dies kann anhand von verschiedenen Kriterien erfolgen und ist für Autofahrer, aber auch für andere Dienstleister, z.B. Versicherungen interessant. [Appd15, Yell15]

**Fahreridentifizierungsanalyse**
> *„Als Versicherung will ich den Fahrer anhand des Fahrstils identifizieren."*

Für eine Versicherung kann es sehr interessant sein zu einem gegebenen Zeitpunkt feststellen zu können, ob ein Versicherter wirklich der Fahrer eines Autos war. Die verschiedenen während der Fahrt gesammelten Daten könnten so genutzt werden, um den Fahrer eindeutig identifizieren. [Axak15]

**Car-to-Car Anwendungsfall**
> *„Als Fahrer will ich vor einer gefährlichen Situation gewarnt werden, damit es nicht zu einem Auffahrunfall kommt."*

Das System muss das Verhalten aller Verkehrsteilnehmer analysieren und so schnell auf Ereignisse reagieren. Wenn z.B. ein Auto auf einer unübersichtlichen Landstraße plötzlich anhält, müssen die nachfolgenden Autos gewarnt werden, um einen Auffahrunfall zu verhindern. [Carc15]

## 5.2 Entwurf

Im Folgenden wird das zu entwickelnde Car Information System entworfen.
Hierbei wird für die prototypische Realisierung des CIS eine spezifische User-Story ausgewählt und anhand dieser ein Durchstich (Vertikale Prototype) erstellt. So wird hierbei das System über alle Schichten der Architektur hinweg konzipiert und realisiert, sodass es die Anforderungen dieser einen User-Story abdeckt und sich komplexe Funktionalitäten des Gesamtsystems demonstrieren und evaluieren lassen [Kuhr09]. Außerdem wäre eine vollständige Umsetzung aller User-Storys für den Rahmen dieser Arbeit zu umfangreich. Für den Durchstich wurde die User-Story „*Tankstellensuche*" ausgewählt, da sie viele Komponenten der anderen User-Stories umfasst und die Integration externer Datenquellen beinhaltet. Zunächst wird in Punkt 5.2.1 die ausgewählte User-Story konkretisiert und der genaue Ablauf dargelegt. In Punkt 5.2.2 bis 5.2.4 wird anschließend auf die Konzeption des Gesamtsystems eingegangen.

### 5.2.1 Konzeption der ausgewählten User-Story
Zur Realisierung der User-Story gelten folgende Anforderungen:
1. Der Fahrer soll über den Tankstellenpreis benachrichtigt werden, wenn er tanken muss → Tank weniger als 50 % gefüllt.
2. Er soll zu der günstigsten Tankstelle in einem Radius von 10 km geleitet werden. (Der Radius kann noch variiert werden)
    a. Hierbei gilt es zu beachten, dass der Weg zur Tankstelle nicht so lang ist, dass sich eine nähere, aber teurere Tankstelle wieder lohnen würde.
3. Es soll ermittelt werden, ob die Tankstelle wirklich günstig ist oder ob es eventuell Sinn macht, noch etwas zu warten bis die Preise wieder fallen.
    a. Hierbei muss der Tankstellenpreis mit den Preisen der letzten Tage verglichen werden.
    b. Es ist jedoch wichtig, dass auch der Tankfüllstand als Faktor einbezogen wird, um bei einem geringen Tankstand trotzdem eine Empfehlung abzugeben.
4. Der Fahrer soll in der Benachrichtigung den zu erwartenden Tankpreis erfahren.
5. Der Fahrer soll nicht mehrmals benachrichtigt werden. Die Benachrichtigung ist für die Fahrt blockiert

Im Folgenden wird der konkrete Ablauf der User-Story in einem Aktivitätsdiagramm dargestellt.

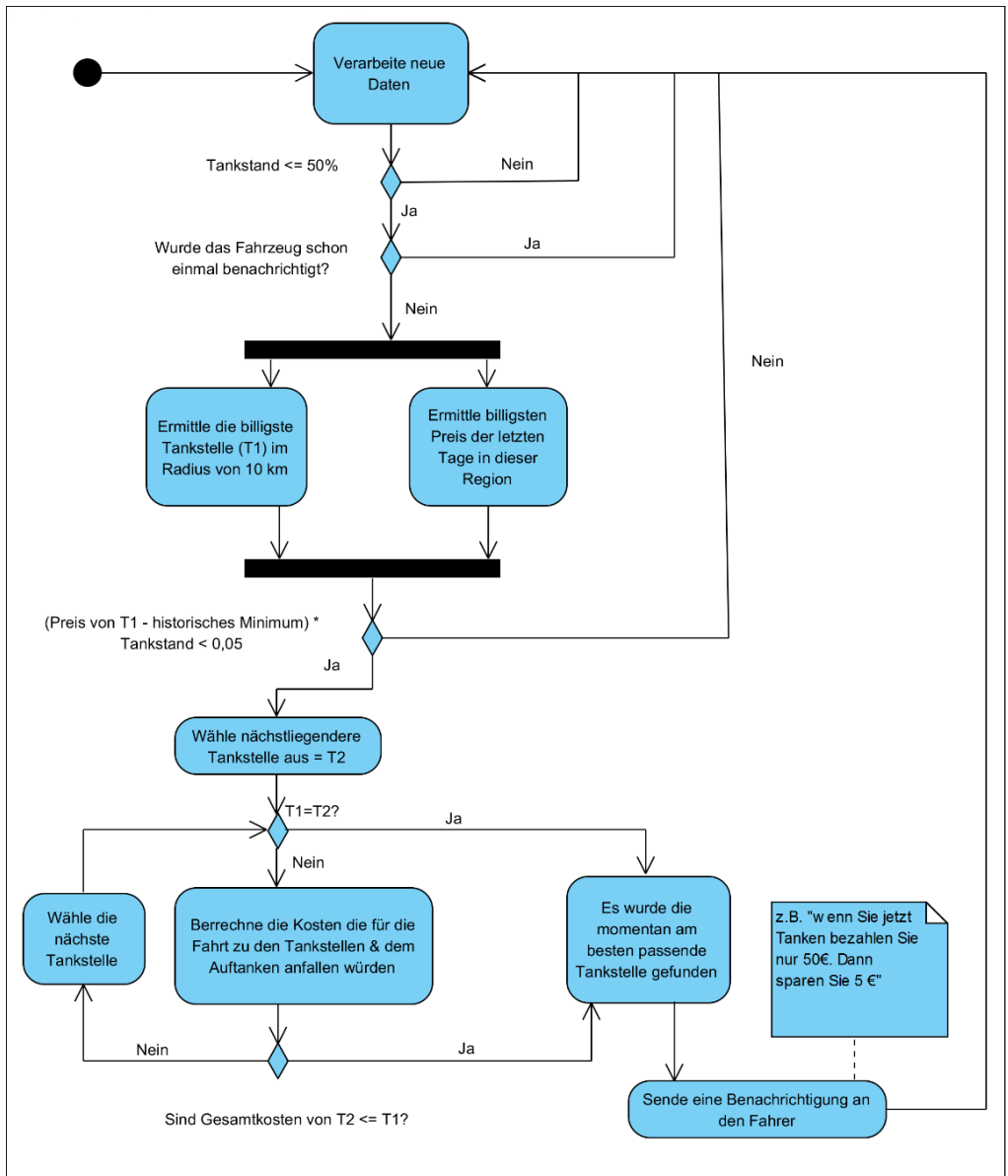

**Abbildung 11: Aktivitätsdiagramm für den Anwendungsfall „Tankpreisbenachrichtigung"**

**Technische Rahmenbedingungen:**

Die zuvor geschilderte User-Story nutzt die folgenden technischen Komponenten.

Um die Daten aus dem Auto an das System zu übermitteln, wird wie in Punkt 5.1.2 beschrieben auf die OBD2-Schnittstelle zurückgegriffen. Um die Daten zu übertragen, braucht es in diesem Konzept einen OBD2-Dongle, der die Daten per Bluetooth an das Smartphone des Fahrers sendet, um diese von dort mittels einer App weiter an das CIS zusenden. Es wäre auch möglich direkt einen OBD2-Dongle zu bauen, der eigenständig über das Mobilfunknetz mit dem CIS kommunizieren kann. Dies ist aber nicht angedacht, da die Konzeption des auf Spark basierenden CIS im Vordergrund der Arbeit steht. Als Quelle der Tankstellenpreise steht die Schnittstelle von tankerkoenig.de zur Verfügung. Hier können sowohl Livedaten als auch historische Daten abgerufen werden [Tank16].

### 5.2.2 Systemkontext

Anhand der zuvor erfolgten Analyse der User-Story ist im Folgenden der Systemkontext dargestellt und beschreibt das CIS mit allen benachbarten Systemen.

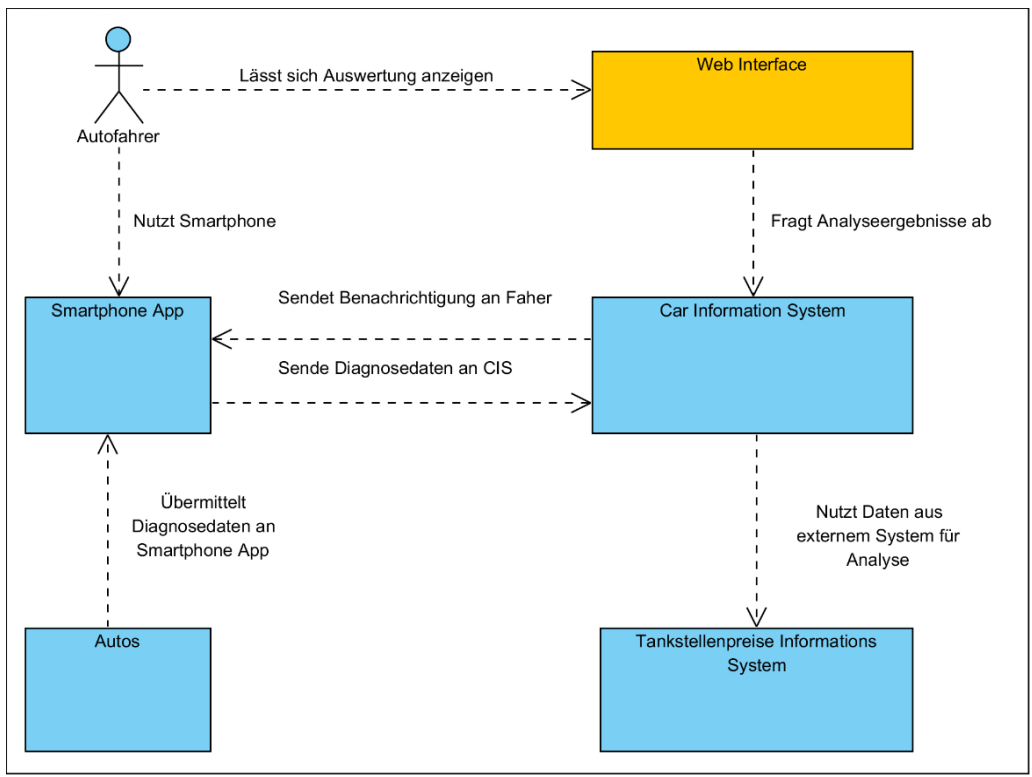

**Abbildung 12: Systemkontext**

Das Übermitteln der Daten an das Smartphone und vom Smartphone an das CIS ist als externes System dargestellt, da es sich in der Realisierung um eine voneinander getrennte Appli-kation handelt. Bei dem Smartphone und dem Tankstellenpreise-Informationssystem handelt es sich um zwei spezifische Datenquellen, wobei das CIS auch noch weitere Systeme anbinden könnte. Auf beide Systeme wird im Rahmen dieser Arbeit nicht konkret eingegangen. Das in Gelb dargestellte Web Interface steht beispielhaft für eine Anwendung, welche Analyseergebnisse aus dem CIS abruft. Es gehört jedoch nicht zum Realisierungsumfang dieser Arbeit und wird daher im Folgenden nicht behandelt.

**Stakeholder:**
Folgende Stakeholder sind für die ausgewählte User-Story interessant.

**Autofahrer / Kunde:**
Übermittelt seine Daten in Echtzeit an das System und will Benachrichtigungen schnell und präzise erhalten.

**Entwickler**:
Das System muss hoch skalierbar sein und außerdem eine hohe Ausfallsicherheit gewährleisten. Die Anwendung soll zudem einfach erweiterbar sein, um andere Anwendungsfälle schnell zu integrieren.

Hieraus werden folgende Anforderungen für das System abgeleitet:
- Die Reaktionszeit auf eintreffende Nachrichten muss in ca. 1-2 Sekunden liegen[2].
- Das System sollte bis zu 70.000 Kunden zeitgleich bedienen können = 70.000 Nachrichten pro Sekunde[3].
- Das System soll trotz Ausfall einzelner Knoten weiter funktionsfähig sein.
- Es soll sowohl Echtzeitanalysen als auch Batchanalysen ermöglichen.
- Andere Systeme sollen benachrichtigt werden können, aber auch Analyseergebnisse abfragen können.
- Das System soll einfach erweiterbar sein und modular aufgebaut sein.

---

[2] Es wird eine maximale Bearbeitungszeit von 1-2 Sekunden festgelegt, um möglichst präzise Benachrichtigungen zu erhalten.

[3] In Deutschland sind mittlerweile ca. 44,5 Millionen PKWs zugelassen. Außerdem haben 16 % der Deutschen angegeben, dass sie sehr interessiert an der Nutzung von Smart-Home-Anwendungen sind. Daraus lässt sich auch schließen, dass wahrscheinlich auch 16 % der Deutschen Autofahrer, also ca. 7 Millionen an einer Smart-Driving-Anwendung interessiert wären. Hiervon soll erstmal 1 % durch dieses System als Kunden gesehen werden. [Kraf15, Stat15]

### 5.2.3 Technische Architektur

Die Kernkomponente des CIS soll mittels Spark bzw. Spark Streaming realisiert werden, da so eine hohe Skalierbarkeit und Ausfallsicherheit erreicht werden kann. Außerdem wird sowohl die Echtzeitfähigkeit als auch die Batchverarbeitung durch Spark unterstützt, wodurch sich Synergieeffekte nutzen lassen. Durch die Verbreitung von Big Data Systemen und durch die neuen Herausforderungen der Echtzeitverarbeitung sind innerhalb der letzten Jahre mehrere Architekturvorlagen entstanden, von denen im Folgenden drei kurz vorgestellt werden.

**Lambda Architektur:**

Die Lambda Architektur wurde 2011 von Nathan Marz als Alternative zu traditionellen datenzentrierten Architekturen vorgestellt. Sie soll vor allem eine besonders schnelle Integration von Echtzeitdaten ermöglichen und zudem eine hohe Fehlertoleranz sowohl gegenüber Hardwarefehlern als auch gegenüber Bugs im System bieten, also menschlichen Fehlern.

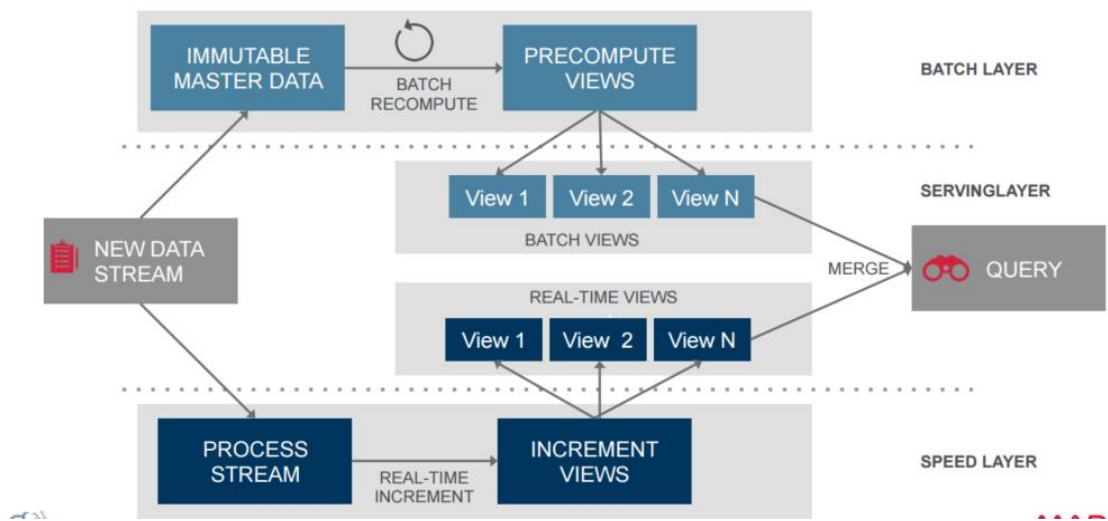

**Abbildung 13: Lambda Architecture (www.mapr.com)**

Eine Lambda Architektur unterteilt sich in drei Komponenten:
Der *Batch Layer* hat die Datenhoheit und speichert alle Daten *immutable,* also unveränderlich ab. Wenn neue Daten im System eintreffen, werden diese nur dem Datenbestand hinzugefügt. Es werden also keine Daten wie in einer klassischen Datenbank verändert. Auf dem Batch Layer werden verschiedene Auswertungen gefahren und deren Ergebnisse in dem Batch-View des Serving Layer abgespeichert. In jeder Batch-Iteration werden alle Auswertungen neu gestartet und alle Daten im Batch-View komplett überschieben. Eine solche Auswertung kann zum Teil sehr langwierig sein.
Der *Serving Layer* steht für Abfragen anderer Systeme bereit und beinhaltet die Analyseergebnisse des Batch sowie des Speed Layers.

Im **Speed Layer** werden die neu eintreffenden Daten direkt verarbeitet. Er hat nur Zugriffe auf alle Daten, welche seit dem letzten Batchzyklus angefallen sind, wodurch die Datenanalyse hier wesentlich schneller ist. Nach einer Analyse werden die Werte im Speed View des Serving Layers aktualisiert und nicht wie im Batch View komplett überschrieben. Durch diesen Ansatz können sehr schnell neue Daten im Serving Layer für Anfragen zur Verfügung stehen. Des Weiteren werden in einem Fehlerfall im Speed Layer, z.B. durch eine fehlerhafte Update-Routine, die Analyseergebnisse im Serving Layer wieder komplett durch den Batch Layer nach der nächsten Iteration überschieben. Da der Batch Layer immer alle Ergebnisse neu berechnet und immer Zugriff auf alle Daten hat, kann es hier auch zu keinem Datenverlust oder Fehlern kommen. [Krep14, Marz11b, Sout14]

**Kappa Architektur:**

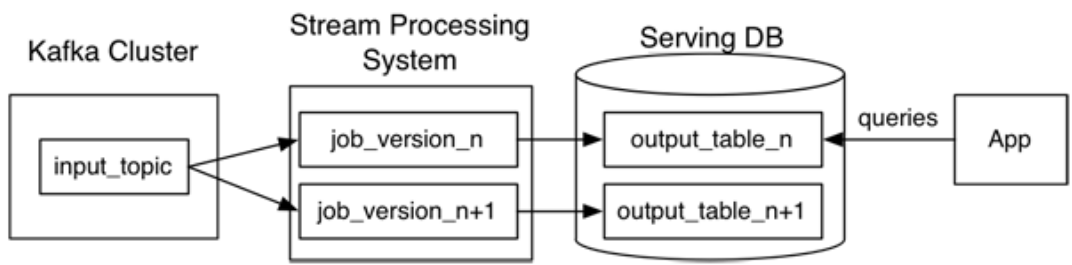

**Abbildung 14: Kappa Architecture (radar.oreilly.com/)**

Die Kappa Architektur ist eine auf Streamverarbeitung konzentrierte Architektur und ist als Gegenentwurf zur Lambda Architektur entstanden. Sie vermindert vor allem die Systemkomplexität, indem sie die komplette Verarbeitung in einem Stream-Processing-System durchführt. Bei der Lambda Architektur sind hingegen immer zwei Systeme für die Stream- und Batchverarbeitung nötig, wobei in beiden Systemen dieselben Analysen implementiert werden müssen.
Die Architektur besteht aus zwei Komponenten.
Zum einem aus dem *Stream Processing System*, welches Daten aus einem Messaging System abruft und analysiert. Zum anderen aus der *Serving DB*, welche die Analyseergebnisse für externe Anwendungen bereithält. Diese ist vergleichbar mit dem Serving Layer der Lambda Architektur. Als Messaging System wird Kafka vorgeschlagen, wodurch sich eine hohe Fehlertoleranz erreichen lässt. So werden alle Inputdaten für einen längeren Zeitraum im Kafka-Cluster gespeichert. Sollte es nun zu einem Fehlerfall oder zu einer Abänderung des Analysealgorithmus kommen, kann das Stream Processing die alten Daten im Kafka System neu verarbeiten. So kann sichergestellt werden, dass keine falschen Daten in der Serving DB abgelegt werden bzw. diese durch eine simple Neuberechnung korrigiert werden können. [Krep14]

**Iot-a (Internet of things architecture)**

Der Begriff „Internet of Things" bezeichnet die Vernetzung von Gegenständen mit dem Internet, damit diese Gegenstände selbstständig über das Internet kommunizieren und so verschiedene Aufgaben für den Besitzer erledigen können. [Lack15]

Die „Internet of Things architecture" ist eine besonders auf die Bedürfnisse dieser Systeme zugeschnittene Architektur und ermöglicht es, deren Daten zu analysieren und die Analyseergebnisse direkt an andere Systeme weiterzureichen, z.B. über einen Message-Broker. Dies ist gerade im IoT-Umfeld interessant, da Systeme hier oft Daten verschiedener Sensoren auswerten und Entscheidungen treffen und wiederum Aktoren über diese benachrichtigt werden.

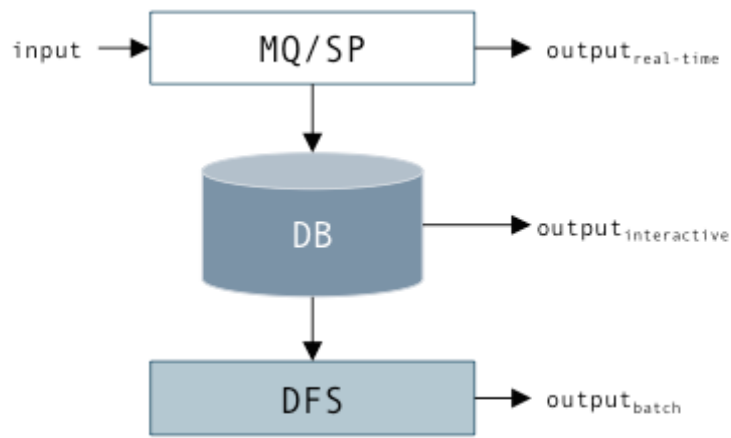

**Abbildung 15: Internet of things architecture (iot-a.info)**

Die Architektur ist in drei Böcke unterteilt:

Im *MQ/SP Block* werden Daten in Echtzeit verarbeitet und an andere Systeme weitergeleitet bzw. an den DB Block übermittelt und dort abgelegt. Hierbei kann es sich auch um Analysen mittels eines Stream-Processing-Frameworks wie Spark Streaming handeln.

Der *DB Block* sammelt vor allem die Analysen des vorherigen Blocks und steht für interaktive Abfragen, z.B. über SQL zur Verfügung.

Im *DFS Block* werden Batchanalysen über die Daten der DB gefahren, so z.B. für Metriken oder Reports. [Haus15]

**Eigene Architektur – Theta Architektur:**

Jede der zuvor vorgestellten Architekturen hat gewisse Vorzüge. Hierbei ist besonders die Fehlertoleranz der Lambda- bzw. Kappa-Architektur interessant. Außerdem passt die IoT-Architektur besonders gut zu dem hier angestrebten Car Information System. Hier wäre das Auto eine Art Sensor, der Daten an das System liefern würde. Das System würde dann zum Teil andere Systeme benachrichtigen.

Im Folgenden wird die technische Architektur des Car Information Systems dargestellt und die einzelnen Komponenten kurz beschrieben. Um dem Namensschema der vorherigen Architekturen zu folgen, wird die hier vorgestellte Architektur *Theta Architektur* genannt.

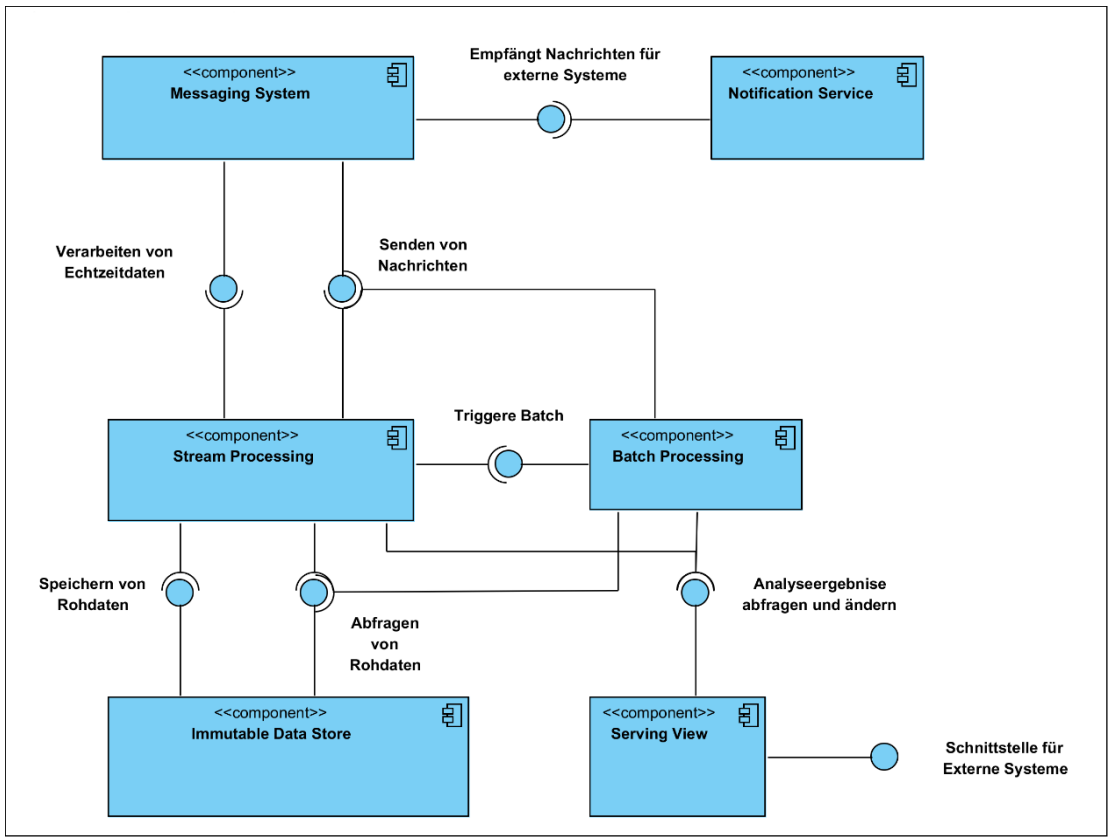

**Abbildung 16: Theta Architektur bzw. technische Architektur des CIS**

Das ***Messaging System*** ist das zentrale Element der Theta Architektur und empfängt die zu verarbeitenden Daten der externen Systeme und reicht diese an das CIS weiter. Hierbei können die empfangenden Daten an das Stream Processing zur Verarbeitung weitergereicht werden und im *Immutable Data Store* abgelegt werden. Außerdem kann das Messaging System für die Kommunikation zwischen den einzelnen Systemkomponenten genutzt werden. Für die Realisierung des Message Broker bieten sich verschiedene Lösungen an. Hier sind vor allem Apache Kafka und RabbitMQ zu nennen. Beide Lösungen sind Messaging Systeme wobei RabbitMQ über ausgeprägte Routingmechanismen verfügt und Apache Kafka auf einen hohen Durchsatz optimiert ist. Beide Lösungen lassen sich einfach in Spark Streaming als DStream verarbeiten.

Im **_Immutable Data Store_** werden verschiedene Rohdaten persistent gehalten. Die Daten im Immutable Data Store sind unveränderbar, sodass neue Daten hier jeweils nur hinzugefügt werden, aber es zu keinen Datenänderungen über Updatefunktionen kommt. Er dient als Datenquelle für die Stream- und die Batch-Analyse. Durch die Speicherung der Inputdaten könnte auch im Nachhinein noch auf die Rohdaten zugegriffen werden, um eventuell neue Auswertungen zu erzeugen. Ob die Eingangsdaten gespeichert werden, ist aber anwendungsfallabhängig.

Im **_Stream Processing_** werden direkt die neuen Daten verarbeitet, um schnell auf neue Ereignisse reagieren zu können. Hierfür wird Spark Streaming genutzt. Die Analyseergebnisse können im Serving View abgespeichert werden oder an den Notification Service weitergereicht werden, um ein externes System oder beispielsweise den Fahrer eines Autos zu benachrichtigen. Des Weiteren kann hier auch eine langwierigere Batchverarbeitung angestoßen werden.

Die **_Batch Processing_**-Komponente beinhaltet Auswertungen über den gesamten Datenbestand und kann durch verschiedene Aktionen bzw. Trigger ausgelöst werden. Sie wird durch Spark realisiert und kann so auch mit der Stream Processing-Komponente eng verknüpft werden und sich so z.B. eine Verarbeitungslogik teilen.

Der **_Serving View_** ist eine Datenbank, in der die Auswertungsergebnisse abgespeichert werden.

Der **_Notification Service_** ist für die direkte Benachrichtigung externer Dienste zuständig. Hierfür wird in diesem Praxisbeispiel auf GoogleCloudMessaging zurückgegriffen. Er könnte aber auch um andere Dienste bzw. Endpunkte erweitert werden.

### 5.2.4   Konzeption der OBD2-Smartphone-App

Da der Fokus dieser Arbeit vor allem auf der Realisierung des CIS und der damit verbundenen Stream-Verarbeitung liegt, wird im Folgenden zwar die Konzeption der Smartphone-App und der Anbindung der OBD2-Schnittstelle beschrieben. Es wird jedoch nicht weiter auf Realisierung der App eingegangen. Das Smartphone des Autofahrers ist aus der Perspektive des CIS die wichtigste Datenquelle, da es für die Übertragung der Fahrzeugdaten zuständig ist und diese jede Sekunde mit jeweils möglichst aktuellen Werten an das CIS gesendet werden müssen. Um diese Daten aber vom Fahrzeug abzurufen, ist hier noch die Kommunikation mit dem OBD2-Bluetooth Dongle nötig, welche in der folgenden Grafik dargestellt ist.

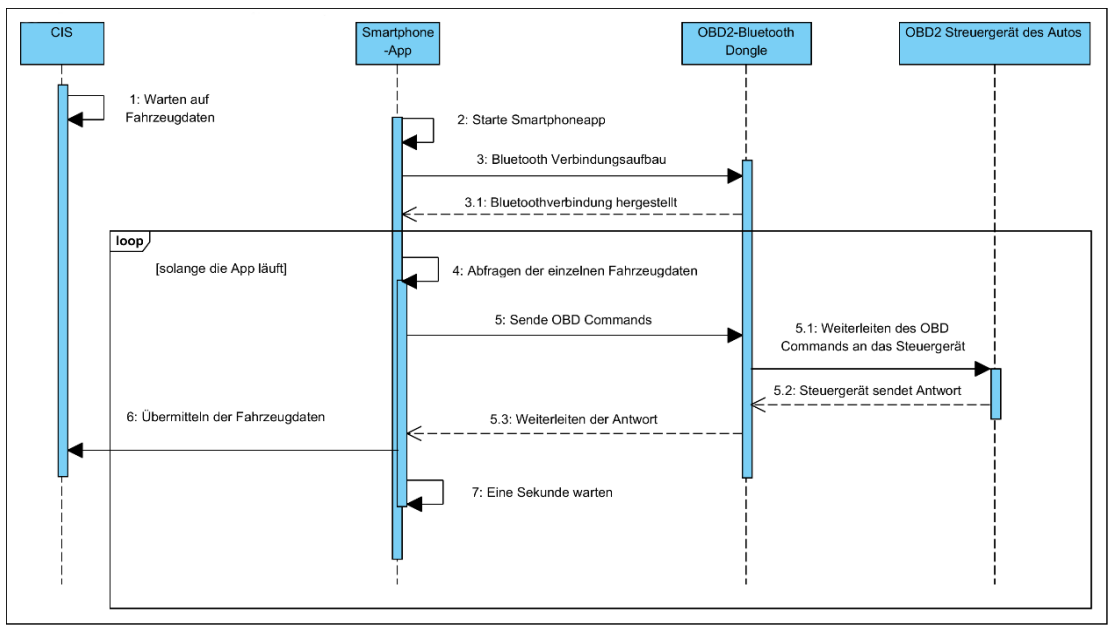

**Abbildung 17: Darstellung der Kommunikation für die Abfrage der OBD2 Daten.**

In der Abbildung 17 ist zunächst zu erkennen, dass es sich bei der Smartphone App und dem OBD2-Bluetooth Dongle um ein eigenständiges System handelt, welches losgelöst vom CIS arbeitet. Nach dem Starten der Smartphone App (2.) versucht diese zunächst eine Verbindung zum OBD2-Bluetooth Dongle herzustellen (3.). Über diese Verbindung wird daraufhin nach dem *SAE J1979*-Protokoll mit dem OBD2-Steuergerät des Fahrzeugs kommuniziert. Dabei kann das Smartphone über bestimmte PIDs die einzelnen Fahrzeugdaten nacheinander abfragen (5.). Die einzelnen PIDs sind in zehn unterschiedliche Modi klassifiziert und jeweils mit einer eindeutigen Bedeutung und der zu erwartenden Antwortlänge im Protokollstandard definiert. Das Abfragen der einzelnen Fahrzeugdaten erfolgt für jeden Wert separat und sequenziell, wobei aufgrund der Kommunikationszeit zwischen App und OBD2-Dongle bzw. OBD2-Steuergerät nicht alle verfügbaren OBD2-Daten jede Sekunde neu abgerufen werden können. So ist es sinnvoll, die Werte nach ihrer Änderungshäufigkeit zu priorisieren und somit z.B. die Geschwindigkeit öfter als den Tankstand abzurufen. Um die Kommunikation mit dem OBD2-Bluetooth Dongle und die Erstellung der App zu vereinfachen, wird hierfür auf das Open-Source Projekt *android-obd-reader*[4] und die Bibliothek *obd-java-api*[5] zurückgegriffen, welche im Rahmen der Arbeit nur leicht erweitert werden. Dies betrifft vor allem das Senden der Fahrzeugdaten an das CIS (6.). Neben dem Abfragen der OBD2-Daten kann das CIS auch Nachrichten an das Smartphone senden. Hierfür wird der von Google betriebene Dienst *GoogleCloudMessaging* genutzt, da so die native Benachrichtigungsinfrastruktur der Android Smartphones genutzt werden kann.

---

[4] https://github.com/pires/android-obd-reader
[5] https://github.com/pires/obd-java-api

## 5.3 Realisierung

Im Folgenden wird die Realisierung des zuvor erstellten Architekturentwurfs thematisiert. Hierbei spielt vor allem die Einbindung der verschiedenen Datenquellen, der konkrete Aufbau des Systems und die Spark-Integration eine wichtige Rolle. Ein wichtiges Ziel ist es hierbei, eine möglichst flexible Software zu erhalten, um so flexibel auf Änderungen der Anforderungen, wie z.B. die Implementierung weiterer Anwendungsfälle, reagieren zu können. Wie zuvor bei der Erstellung des Architekturentwurfs konzentriert sich die Darstellung zunächst nur auf die Userstory „Tankstellensuche".

### 5.3.1 Beschreibung der Datenquellen

Zur Realisierung der gewählten Userstory müssen verschiedene Datenquellen integriert werden. Diese werden im Folgenden konkret beschrieben.

**Fahrzeugdaten:**

Wie in Kapitel 5.2.4 dargestellt, werden die Fahrzeugdaten jeweils einmal pro Sekunde über das Smartphone des Autofahrers an das CIS übermittelt. Wie die Daten über die OBD2-Schnittstelle gewonnen werden, ist für das CIS zunächst unerheblich. Die Fahrzeugdaten werden vom Smartphone als JSON-Datei übermittelt, wobei das Schema wie folgt definiert ist:

```json
{
   "altitude": Double,
   "latitude": Double,
   "longitude": Double,
   "readings": {
     "FUEL_LEVEL": String,
     "ENGINE_RPM": String
   },
   "timestamp": Long,
   "vehicleid": String
}
```

**Listing 5: Fahrzeugdaten JSON Beispiel**

Wie das Schema darstellt, beinhaltet jeder Datensatz neben den über OBD2 abgefragten Fahrzeugdaten (*readings*) auch einige weitere Daten, unter anderem eine eindeutige VehicleID und die aktuellen Geo-Koordinaten des Fahrzeugs, welche über GPS ermittelt wird. Wie zuvor geschildert, hängt die Anzahl der im Feld *readings* übertragenden Werte von den jeweils unterstützten OBD-Funktionen des Autos ab, sodass hier das CIS unvollständige Datensätze verarbeiten können muss.

**Aktuelle Tankstellenpreise:**
Die aktuellen Tanktellenpreise werden über eine REST-Schnittstelle des Dienstes Tankerkönig.de abgerufen. Hierbei verwendet das CIS die eindeutige Tankstellen-ID, um die jeweiligen Tankpreise bei Bedarf abzurufen. Die Antwort wird hier im JSON-Format übermittelt, wobei hierbei die aktuellen Preise mit den Bezeichnungen *e5, e10* und *diesel* übermittelt werden. Da das Abfragen dieser externen Schnittstelle immer Zeit kostet, muss hierbei eine sinnvolle Caching-Strategie gewählt werden, sodass ein unnötig häufiges Abfragen der Schnittstelle vermieden werden kann. Hierbei ist außerdem sicherzustellen, dass nicht mit veralteten Daten im System gearbeitet wird.

**Historische Tankstellenpreise:**
Zur Realisierung der Userstory werden außerdem historische Tankstellenpreise benötigt. Hierfür wird auf einen von Tankkerkönig.de bereitgestellten tagesaktuellen PostgreSQL Datenbank-Dump zurückgegriffen. Die Datensätze beinhalten zum einen Informationen zu den einzelnen Tankstellen wie GPS Position und Adresse, aber auch einen Verlauf über alle Tankpreisänderungen der jeweiligen Tankstellen. Da diese Daten nur als PostgreSQL Datenbank-Dump bereitgestellt werden, müssen sie zunächst in eine PostgreSQL Datenbank importiert werden. Um diese Daten im CIS zu nutzen, ergeben sich verschiedene Möglichkeiten. So wäre die einfachste Lösung mittels Spark SQL und dem angebotenen JDBC-Adapter direkt auf die PostgreSQL Datenbank zuzugreifen. Jedoch hat sich in der Erprobungsphase dieses Ansatzes gezeigt, dass die JDBC-Unterstützung von Spark noch nicht vollständig ist und bestimmte in dem verwendeten Datenbankschema genutzte Datentypen nicht unterstützt werden[6]. In diesem Fall führt ein Importversuch zu einer Fehlermeldung seitens Spark.
Als Alternative zum ersten Ansatz wurde ein CSV-Export der Datenbank erstellt und in dem Immutable Datastore abgelegt. Hierdurch kann nun direkt über Spark mit den Daten gearbeitet werden, da der Immutable Datastore bereits als eine Datenquelle für Analyseschritte vorgesehen ist und so der Aufwand des Anbindens einer weiteren Datenquelle gespart werden kann.

### 5.3.2  Technische Konzeption der Systemkomponenten

Die in 5.2.3 definierte *Theta Architektur* beschreibt verschiedene Systemkomponenten, welche im Folgenden bezüglich deren technischer Realisierung beschrieben werden. Für die selbst zu entwickelnden Komponenten wie Streampocessing und Notification Service, ist es gewünscht, dass die einzelnen Softwarekomponenten möglichst unabhängig voneinander agieren können. Um dies zu realisieren, werden die einzelnen Komponenten als Microservices betrachtet werden, welche über das Messaging-System kommunizieren.
Dadurch lässt sich unter anderem eine separate Skalierung der einzelnen Komponenten erreichen, sodass besonders wichtige und performancekritische Services mehr Ressourcen zu-geteilt bekommen, als nicht so wichtige Services. Außerdem wäre es möglich, dass die verschiedenen Services in unterschiedlichen Programmiersprachen realisiert werden, um z.B. bestimmte Anwendungsfälle effizienter lösen zu können. Im Rahmen des CIS wurde aber für alle Services die Programmiersprache Scala gewählt, da Spark hierfür die besetzte Unterstützung anbietet und die einzelnen Services so auf eine einheitliche Basis aufbauen können. Um die einzelnen Komponenten trotzdem gleichförmig zu gestalten und einen doppelten Implementierungsaufwand zu vermeiden, nutzen alle Softwarekomponenten ein gemeinsames Framework, welches bestimmte Funktionen bereitstellt und abstrahiert. Dieses Paket wird, angelehnt an die Theta Architektur, im Folgenden *Spark Theta Architecture Framework* kurz *STA*-Framework genannt.

---

[6] https://github.com/apache/spark/pull/4549  => *[SPARK-5753] [SQL] add JDBCRDD support for postgres types: uuid, hstore and arrays*

Im Folgenden wird auf die konkrete technische Funktionsweise und Umsetzung der einzelnen Komponenten eingegangen, dabei wurden zum Teil einige Komponenten der Architektur zusammengefasst.

*Messaging System*:

Als Grundlage des Messaging Systems wurde zunächst RabbitMQ als Messaging Plattform gewählt. Die Arbeit mit dieser technischen Plattform wird mithilfe des oben angesprochenen STA-Framework abstrahiert, sodass es nachträglich durch eine andere Plattform, wie z.B. Apache Kafka, ersetzt werden kann.

Über das Messaging System können zum einen Inputdaten von außen an das System übergeben werden. Zum anderen läuft auch die interne Kommunikation der einzelnen Services asynchron über das Messaging System, wobei es auch für das Zwischenspeichern und für die Verteilung von Nachrichten zuständig ist. Durch diese lose Kopplung sind die einzelnen Services nicht mehr direkt voneinander abhängig und können bei einem Ausfall eines Services trotzdem ohne Datenverlust weiterarbeiten. Als Serialisierungs-Format der Nachrichten wird in der momentanen Implementierung des Massaging Systems JSON verwendet. Es hat vor allem den Vorteil, dass dessen Verarbeitung sehr gut durch eine Vielzahl von Bibliotheken unterstützt wird und die einzelnen Nachrichten direkt lesbar sind, wodurch sich die Fehlersuche im System vereinfacht. Als alternatives Nachrichtenformat zu JSON würde sich jedoch auch das von Google hervorgebrachte Protocol Buffers Format eignen, welches ein Binärformat ist und eine besonders schnelle Verarbeitung von Nachrichten ermöglicht [Goog16].

**Immutable Data Store:**

In dem Immuatable Data Store werden verschiedene für die Analysen wichtige Rohdaten persistiert. Hierbei werden nur Daten zu dem existierenden Datenbestand hinzugefügt und im Analyseschritt ausgelesen, wodurch auf Update-Routinen verzichtet werden kann. Hierdurch lässt sich eine sehr einfache Datenbank benutzen, da z.B. die ACID-Fähigkeiten nicht nötig sind. Oft wird für diese Komponente das Hadoop Distributed File System (HDFS) verwendet [Marz11b]. HDFS bietet ein redundantes und verteiltes File-System an und wird auch direkt durch Spark unterstützt. So akzeptiert Spark direkt HDFS Links als Pfadangabe für das Lesen und Schreiben von Dateien [Apac13, Apac16h]. Für die Realisierung des CIS wurde sich nach einer Testphase schließlich gegen eine Nutzung von HDFS entschieden, da sich durch die Nutzung eines Datenbank-Management-Systems (DBMS) die Arbeit und Organisation der Daten erheblich vereinfachen lässt. Außerdem können einfache Vorverarbeitungen auch über das DBMS erledigt werden, wodurch die Verarbeitungsgeschwindigkeit gesteigert werden kann, da Spark nicht mehr auf dem kompletten Datenset arbeiten muss.

Für die Wahl eines DBMS gelten einige Anforderungen. So nutzt HDFS eine Organisation von Daten auf Dateibasis, wodurch es möglich ist, komplett uneingeschränkt verschiedenstrukturierte Daten abzulegen. Dies ist auch für das zu wählende DBMS erforderlich, wodurch die Wahl auf eine dokumentenorientiere Datenbank fällt. Des Weiteren ist eine Integration in das Spark Ökosystem besonders wichtig, um die Arbeit mit Spark direkt auf dem Datenbestand zu ermöglichen. Für eine Auswahl des DBMS wurden ElasticSearch, MongoDB und Couchbase näher betrachtet, da diese Systeme sowohl dokumentenorientiert arbeiten, als auch eine Spark Unterstützung anbieten. Im Folgenden werden die möglichen DBMS Systeme kurz vorgestellt und bewertet.

| **ElasticSearch** | **MongoDB** | **Couchbase** |
|---|---|---|
| + Offizielle Spark Unterstützung<br>+ Rest Schnittstelle & einfache Handhabung | + Einfache Handhabung<br>+ Sehr etabliert | + Offizielle Spark Unterstützung<br>+ Einfache Handhabung |
| - Ist ursprünglich als Suchmaschine konzipiert<br>- Es wird vor Konsistenzproblemen und Datenverlust in gewissen Szenarien gewarnt und von der Nutzung als primäre Datenbank abgeraten. [King15] | - Keine offizielle Spark Unterstützung<br>- Community Bibliothek ist nicht sehr ausgereift und unterstützt nicht Spark > 1.4 [7] | - Nicht so etabliert wie MongoDB |

**Tabelle 2: Vergleich der einzelnen Datenbanksysteme**

Für die Realisierung wurde zunächst Couchbase gewählt, wobei sich bei einer Verbesserung der Spark Unterstützung auch MongoDB eignen würde. Es muss jedoch beachtet werden, dass diese Wahl vom jeweiligen Anwendungsfall abhängig ist. So wäre für sehr datenintensive Anwendungsfälle eventuell eine Lösung mittels HDFS besser geeignet. Für die Nutzung des Immutable Data Stores wurden einige Funktionalitäten im STA-Framework abstrahiert, sodass die gewählte Datenbank einfacher ausgetauscht werden kann.

**Serving View:**
Im Serving **View** werden verschiedene Daten, wie z.B. Analyseergebnisse gesammelt und externen Systemen bereitgestellt. Hierbei handelt es sich nicht, wie beim Immutable Datastore, um die gesamte Menge aller Daten sondern nur um eine klar strukturierte Menge an Informationen. Hier wird in der ersten Version des CIS eine MySQL Datenbank über die JDBC-Schnittstelle von Spark angebunden. Das klassische relationale Schema passt sehr gut zu der Speicherung der Analyseergebnisse, da deren Struktur klar definiert ist. Außerdem wird es so externen Anwendungen erleichtert, die Analyseergebnisse zu verarbeiten. Für die technische Realisierung des Serving View wäre aber auch die Nutzung eines anderen DBMS Systems möglich. Hier ist eine Revision des Systems bei steigenden Anforderungen sinnvoll.

---
[7] https://github.com/Stratio/spark-mongodb/blob/master/doc/src/site/sphinx/about.rst

**Processing Service:**
Die in der Architektur beschriebenen *Stream Processing* und *Batch Processing* Komponenten werden innerhalb der Anwendung als Processing Service implementiert, sodass in diesem Service die jeweiligen Verarbeitungsschritte definiert und abgearbeitet werden. Die beiden Verarbeitungsverfahren sind hier zusammengelegt, da z.B. während einer Stream-Verarbeitung eine Batch-Verarbeitung angestoßen werden kann. Hierbei ist anzumerken, dass diese Verarbeitungsschritte auch auf mehrere separate Processing Services aufgeteilt werden könnten, welche z.B. auf verschiedene Use-Cases spezialisiert sind und somit komplett eigenständig laufen.
Momentan ist nur der *CarDataProcessingService* implementiert, welcher die Verarbeitung der Fahrzeugdaten durchführt. Hier werden die über das Messaging System bereitgestellten Daten als Spark DStream verarbeitet und auf diesem wird die Tankstellenpreisanalyse durchgeführt. Hierfür werden momentan vor allem die Stream-Processing-Komponenten genutzt. Des Weiteren kann der Processing Service über die im STA Framework angebotenen Funktionen mit dem Immutable Datastore, dem Serving View und dem Messaging System kommunizieren. So wird im *CarDataProcessingService* nach einer erfolgreichen Tankstellensuche der Notification Service über das Messaging System angesprochen.

**Notification Service:**
Bei dem Notification Service handelt es sich um einen weiteren separaten Service, welcher auf bestimmte Nachrichten, z.B. Analyseergebnisse wartet und dann für die Benachrichtigung von externen Systemen zuständig ist. Es ist so auch möglich, dass verschiedene Notification Services auf dieselben Nachrichten reagieren und so mehrere externe Dienste benachrichtigen.
Eine Implementierung des Notification Services ist der *GCMNotificationService*, welcher auf die Analyseergebnisse der Tankstellenpreisanalyse wartet und die jeweiligen Tankbenachrichtigungen als Pushnachricht auf die Smartphones der Nutzer sendet. Hierfür reicht ein einfacher HTTP-POST-request an den von Google betriebenen Dienst GCM aus.

### 5.3.3   Spark Theta Architecture Framework
In diesem Punkt wird das zuvor erwähnte *STA Framework* beschrieben. So bietet es vor allem eine Struktur bzw. ein Grundgerüst für die einzelnen Anwendungsservices. Außerdem unterstützt es die Umsetzung der gewählten Architektur und definiert einige Prinzipien, die von den jeweiligen Komponenten eingehalten werden sollen. Das Framework ist sehr allgemein gehalten und nicht auf das Car Information System spezialisiert, wodurch es auch zur Nutzung in komplett anderen Anwendungsfällen denkbar wäre. Folgendes Klassendiagramm veranschaulicht den Aufbau des STA Frameworks, wobei es nur die wichtigsten Klassen darstellt und Beziehungen teilweise vereinfacht.

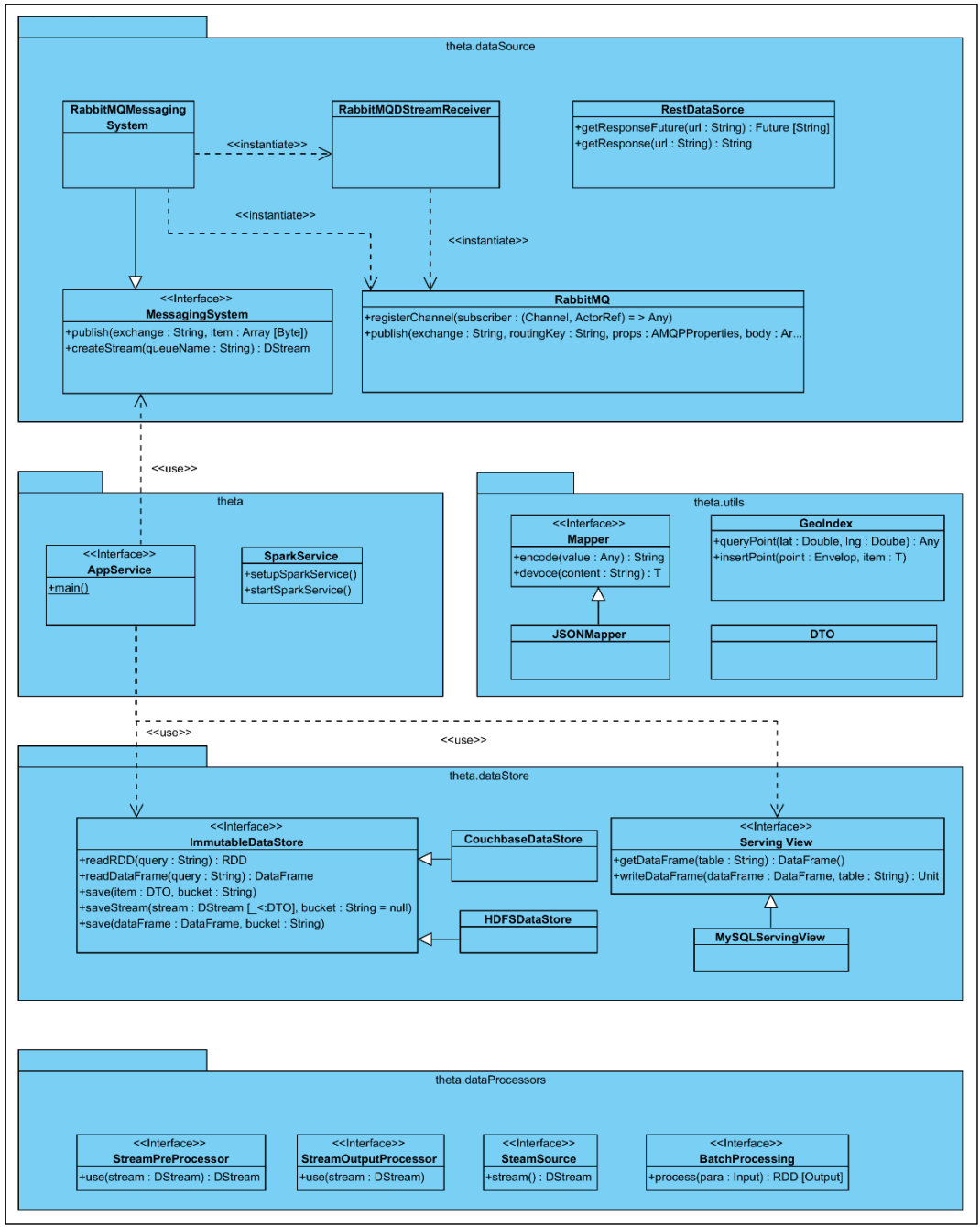

**Abbildung 18: Klassendiagramm des STA Frameworks**

Das abgebildete Klassendiagramm veranschaulicht die fünf Hauptpakete des STA Frameworks, welche im Folgenden erläutert werden.

**Theta Package:**
Dies ist das Hauptpaket des STA Frameworks und bietet vor allem den Trait[8] *AppService* und *SparkService* an. Die Hauptroutine eines mit dem STA Framework gestalteten Prozesses bildet immer ein Scala Objekt, welches von dem Scala *APPService* erbt. Dies wird so z.B. von den zuvor vorgestellten *Processing Services* und *Notification Services* genutzt. Der *APPService* bietet eine generische Main-Methode, sodass das erbende Objekt direkt als Scala Programm gestartet werden kann. Des Weiteren initialisiert der *APPService* verschiedene für das Framework wichtige Elemente und erzwingt die Konfiguration des Messaging Services und des Immutable Data Stores in dem Anwendungsservice. Neben dem *APPService* ist es auch möglich, dass der jeweilige Service den *SparkService* Trait implementiert, welcher Spark bzw. Spark Streaming für die komplette Anwendung initialisiert. Dies ist allerdings nur für die *Processing Services* sinnvoll.

**Theta.dataStore Package:**
Die Klassen in diesem Paket erleichtern die Arbeit mit dem *Immutable Data Store* und dem *Serving Layer*, wobei hier für beide Arten ein Interface angeboten wird, welches von der Anwendung verwendet wird. So lassen sich rein technische Details, wie z.B. die jeweilige verwendete Datenbank für den *Immutable Data Store* leichter ändern. Hier sind momentan die Klassen *HDFSDatastore* und *CouchbaseDatastore* zu nennen, welche das *ImmutableDatastore* Interface implementieren.

**Theta.dataSource Package:**
Dieses Paket bietet Hilfe bei der Nutzung von Datenquellen. So wird mittels dem Trait *MessagingSystem* eine Schnittstelle angeboten, welche die Arbeit mit dem Messaging System vereinfacht. Genau wie beim Immutable Datastore und beim Serving Layer ist die technische Realisierung des Messaging Systems für die nutzenden Klassen unerheblich, wodurch sich Änderungen an dieser vereinfachen lassen. In der aktuellen Ausführung ist die Klasse *RabbitMQMessagingSystem* jedoch die einzige Implementierung dieser Schnittstelle. Des Weiteren bietet dieses Paket Grundlagen für die Nutzung von anderen Datenquellen an. Hier ist z.B. die *RestDataSource* zu nennen, welche einige Funktionalitäten zur Kommunikation mit einem REST-Webservice anbietet.

**Theta.util Package:**
In diesem Paket werden mehrere allgemeine Funktionalitäten angeboten, welche die Arbeit mit dem Framework und die Implementierung von Verarbeitungsschritten erleichtern. Hier ist beispielsweise der Mapper *Trait* zu nennen, welcher mit Hilfe seiner Implementierung Methoden für die Serialisierung und Deserialisierung von Objekten anbietet. Des Weiteren wird hier das DTO-Interface definiert, welches für alle Datenobjekte verwendet werden sollte, um so eine einheitliche Basis darzustellen.

---

[8] Trait = In Scala vergleichbar mit Java Interfaces

**Theta.dataProcessor Package:**
Dieses Paket beinhaltet verschiedene vom STA Framework angebotene Abstraktionen für die Umsetzung der Stream- bzw. Batch-Verarbeitungsschritte. Dies ermöglicht eine leichtere Trennung der einzelnen Verarbeitungsschritte und dadurch eine höhere Modularisierung des Gesamtsystems. So besteht beispielsweise eine Streamverarbeitung hier immer aus der Nutzung einer StreamSource als Datenquelle, welche durch mehrere StreamPreProcessor verarbeitet wird und schlussendlich über einen StreamOutputProcessor ausgegeben oder an eine andere Anwendung weitergereicht wird.

### 5.3.4 Definition der Streamverarbeitung

Nach der ausführlichen Beschreibung der jeweiligen Komponenten und deren technischer Konzeption wird nun die Spark Integration genau beschrieben. Dies bezieht sich vor allem auf den Processing Service.
Als eine der Kernanforderungen des Systems wurde die gute Erweiterbarkeit des Zielsystems gewünscht. Hierdurch wird eine hohe Modularisierung und Unabhängigkeit der einzelnen Komponenten nötig, welches auch bei der Nutzung von Spark berücksichtigt werden muss. So ist es durch die vom STA Framework angebotenen Schnittstellen möglich, die eigentlichen Verarbeitungsschritte in separate Module zu kapseln und somit einzelne Verarbeitungsschritte in verschiedenen Kontexten zu verwenden. Ohne diesen Aufbau könnte es leicht zu sehr umfangreichen sequenziellen Definitionen der einzelnen Analyseschritte kommen. Um die Modularisierung zu ermöglichen, wird für die Verkettung der Verarbeitungsschritte eine Konstruktion verwendet welche sich an das Decorator Pattern[9] orientiert.
Im Folgenden wird zunächst die Integration von verschiedenen Abarbeitungsschritten anhand des *CarDataProcessingService* exemplarisch gezeigt, um darauf aufbauend die Modularisierung der einzelnen Verarbeitungsschritte zu erklären.

```
1. val carDataSource = new CarDataStreamSource()
2. carDataSource.stream.saveToImmutableDatastore()
3. gasStationNotificationOutputProcessor.use(
     gasStationSearchStreamProcessing.use(
       carDataFirstAppearanceFilter.use(
         carDataFuelLevelFilter.use(
           car.stream
         )
       )
     )
   )
```

Listing 6: Beispiel für die Verkettung von einzelnen Verarbeitungsschritten im CIS

---

[9] Das Decorator pattern ist eines durch die Gang of Four definiertes Entwurfsmuster.

Dieses Beispiel ist der Hauptroutine des *CarDataProcessingService* entnommen und beschreibt, wie die einzelnen Verarbeitungsschritte miteinander in Beziehung gesetzt werden können.

In Zeile 1 wird die *CarDataStreamSource* erzeugt. Hierbei handelt es sich um eine Datenquelle, welche die OBD-Daten über das Messaging System empfängt und als Spark Streaming DStream intern verwaltet.

In Zeile 2 wird der Stream verwendet und in den Immutable Data Store gespeichert.

In Zeile 3 wird der Stream in einem anderen Kontext verwendet. So wird der Stream zunächst vom *carDataFuelLevelFilter* und *carDataFirstAppearanceFilter* bearbeitet und dessen Ergebnisstream im *gasStationSearchStreamProcessor* als Eingabestream verwendet.

Der Zusammenhand der einzelnen Module bzw. deren Datenfluss ist auch noch einmal in folgender Grafik Dargestellt.

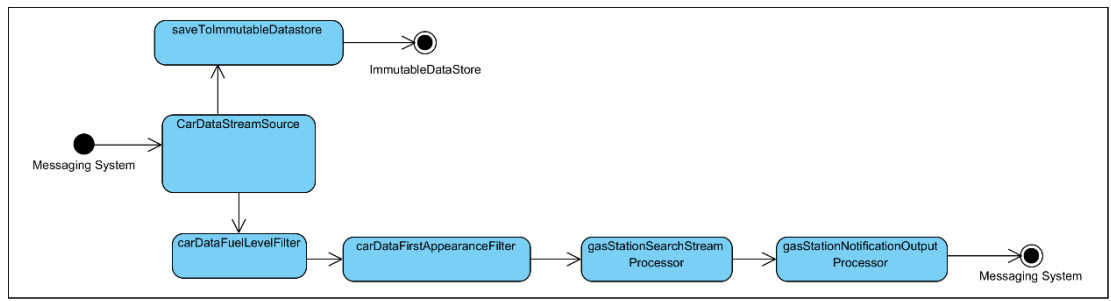

**Abbildung 19: Darstellung der Verarbeitungshierarchie**

Durch diese Struktur wird es nun möglich, verschiedene Verarbeitungsschritte auszutauschen bzw. die Abfolge durch weitere Schritte zu ergänzen. Außerdem wäre es möglich, Zwischenverarbeitungsschritte in verschieden Kontexten zu nutzen. So wird in diesem Beispiel, wie in Abbildung 19 dargestellt, der *CarDataStreamSource* nicht zweimal erzeugt, aber zweimal in verschiedenen Kontexten genutzt. Da es sich bei den Spark DStreams um unveränderliche Datenstrukturen handelt, ist es möglich auf einem DStream verschiedene Bearbeitungsschritte durchzuführen, ohne dass diese sich beeinflussen. Um diese Art der Verarbeitung zu ermöglichen, werden innerhalb des STA Frameworks zwei Arten der Stream Verarbeitungsschritte unterschieden. Dies sind zum einen sogenannte *StreamPreProcessors*. Hiermit sind Vorverarbeitungsschritte, wie z.B. m*ap*, *filter* und *reduce* gemeint, welche jeweils einen neuen Stream erzeugen. Zum anderen werden *StreamOutputProcessors* genutzt, welche als Eingabe einen DStream erhalten und auf diesem Spark Output Operationen durchführen und keinen direkten Rückgabewert haben.

Beide Arten sind über Interfaces definiert und werden in den jeweiligen Verarbeitungsschritten implementiert. Als Ein- bzw. Rückgabewerte werden hier immer klar typisierte DStreams verwendet. Hierdurch kann sichergestellt werden, dass bestimmte Verarbeitungsschritte nur auf Streams vom richtigen Typ möglich sind. Um klare Datentypen zu erzeugen, werden für die jeweiligen Daten eigene DTOs definiert. So ist es in dem oben genannten Beispiel nur möglich, den *gasStationSearchStreamProcessor* mit einem Stream zu nutzen, welcher auch Elemente vom Typ CarDataDTO enthält. Hierdurch sind die zu erwartenden Daten klar definiert sind und die weiteren Verarbeitungsschritte können sich so auf ein konkretes Schema verlassen. Es wäre aber auch möglich, nicht so strikt typisierte Daten von einem Verarbeitungsschritt in den Nächsten zu übertragen. Dies könnte z.B. über die Nutzung der von Spark SQL eingeführten DataFrames geschehen. Die Nutzung von nicht strikt typisierten Daten hat den Vorteil, dass sich sehr heterogen strukturierte Daten in einem Stream leichter verarbeiten lassen.

### 5.3.5 Datenverarbeitung und –analyse

Im Folgenden werden die konkreten Datenverarbeitungs- bzw. Analyseschritte beschrieben, die für die Realisierung der Tankstellensuche-Userstory nötig sind. Hierfür wird auf die konkrete Implementierung der einzelnen in 5.3.4 dargestellten Verarbeitungsschritte eingegangen. Zunächst eine Darstellung der aufeinander folgenden Verarbeitungsschritte und deren Beschreibung:

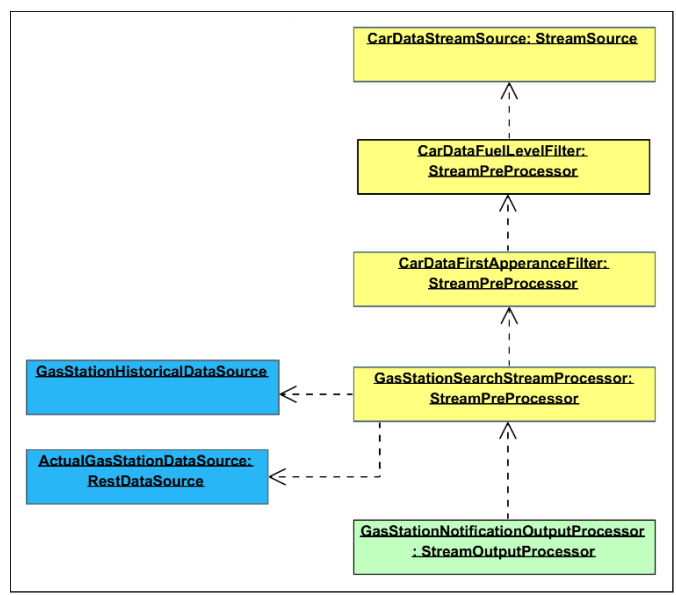

**Abbildung 20: Darstellung der einzelnen Verarbeitungsschritte für die Implementierung der Tankstellensuche**

Hierbei stellen die gelben und grünen Blöcke die in Punkt 5.3.4 geschilderte Verzahnung der einzelnen Verarbeitungsschritte dar. Die blau hinterlegten Blöcke sind weitere Abhängigkeiten. Wie beschrieben, nutzt die *CarDataSource* das Messaging System, um die Fahrzeugdaten zu empfangen. Die CarDataSource deserialisiert außerdem die eingehenden JSON-Daten in CarDataDTO Objekte. Dabei wurde die eigentliche Deserialisierung nicht als Spark Streaming Map Operation realisiert, sondern geschieht direkt im RabbitMQ-Receiver des Massaging Systems, bevor die eigentlichen Daten an Spark weitergereicht werden. Hierdurch müssen die Daten nicht erst zu Beginn des zeitkritischen Spark Streaming Micro-Batch-Verarbeitungsschritts konvertiert werden. Zunächst werden im *CarDataFuelLevelFilter* alle Fahrzeugdaten aus dem Stream gefiltert welche noch einen mehr als halbvollen Tank haben, da hier eine Benachrichtigung nicht sinnvoll wäre. Daraufhin werden die restlichen Daten vom

*CarDataFirstAppearanceFilter* weiterverarbeitet. Dieser Verarbeitungsschritt sorgt dafür, dass nur die jeweils erste Nachricht eines Autos pro Fahrt vom System weiterverarbeitet wird. Alle weiteren Nachrichten werden hier aus dem Stream gefiltert. Dieses Vorgehen hat zwei Gründe: So wird zum einen der Nutzer vor erneuten und damit unnötigen bzw. störenden Tankbenachrichtigungen geschützt, zum anderen werden auch weniger Daten an den nächsten Verarbeitungsschritt weitergereicht, sodass hier weniger Verarbeitungsaufwand entsteht. Bei der Implementierung dieser Anforderung hat sich hier die Spark *updateStateByKey*-Funktion schlussendlich am effizientesten herausgestellt. Hierbei wird für jede FahrzeugID in einem Zustand festgehalten, ob für diese FahrzeugID schon einmal Daten in einem bestimmten Zeitintervall vom System empfangen wurden. Falls dies der Fall ist, werden die betreffenden Nachrichten aus dem Stream gefiltert. Ein Problem dieser Lösung ist, dass über die Zeit immer mehr Zustände gesammelt werden, wodurch der Aufwand für die Zustandsberechnung steigt. Hier wäre eine Evaluation der mit Spark 1.6 eingeführten Funktion *mapWithState* eventuell sinnvoll, da sie die Performance für die Zustandsaktualisierung verbessern soll [DaZh16].

Der nun gefilterte Stream wird an das *GasStationSearchStreamProcessor* Modul weitergereicht. In diesem Modul wird die eigentliche Vorhersage der zu empfehlenden Tankstelle durchgeführt. Um dies zu realisieren, nutzt dieses Modul folgende Abhängigkeiten:

Das *ActualGasStationDataSource* Modul ermöglicht die Abfrage des aktuellen Tankpreises anhand einer bestimmten Tankstelle. Hierfür nutzt es die vom Dienst Tankerkönig.de bereitgestellte API und erbt von der *RestDataSource,* um die schon vorgefertigten Funktionalitäten zur Kommunikation mit einer REST-API, zu nutzen. Die Antworten der Tankerkönig.de API werden zwischengespeichert, sodass zum Teil der teure Aufruf der externen Schnittstelle vermieden werden kann.

Außerdem wird das *GasStationHistoricalDataBatch* Modul genutzt. In diesem Modul werden verschiedene Informationen über die eigentlichen Tankstellen bereitgestellt. Diese werden zunächst aus dem *Immutable Data Store* geladen und während der Laufzeit des *Processing Services* im Speicher gehalten, um die Abfragegeschwindigkeit auf diesen Daten zu erhöhen. Zum einen kann das Modul Informationen über die historischen Preise der Tankstellen liefern und zum anderen kann es anhand einer GEO Position in der Nähe befindliche Tankstellen finden. Hierfür könnte zwar auch die API des Dienstes Tankerkönig.de genutzt werden, aber durch die Selbstimplementierung des Suchalgorithmus lässt sich dieser besser konfigurieren. Außerdem ist es so möglich, diesen Algorithmus durch ein anderes Verfahren zu ersetzen, wie z.B. die Vorhersage der zu fahrenden Route.

Um eine passende Tankstelle anhand der aktuellen Position eines Fahrzeugs zu finden, wurden verschiedene Möglichkeiten getestet. So wurde zunächst eine Lösung mittels Spark SQL umgesetzt. Hierbei stellte sich heraus, dass die von Spark intern verwendete join Operation eine Komplexität von $O(anzahlTankstellen * anzahlElementImBatchIntervall)$ und damit einen sehr großen Aufwand zur Folge hatte. In einer verbesserten Lösung wird nun zunächst der Datensatz der Tankstellen (z.B. 30.000 Einträge) über Spark SQL in den Speicher geladen, um dann aus diesen Daten eine RTree-Indexstruktur zu erstellen. Dadurch kann mit dem Aufwand von $O(\log(n))$ anhand einer Geoposition nach Tankstellen gesucht werden.

Für die Erzeugung des Indexes wird auf die Java Bibliothek JTS Topology Suite[10] zurückgegriffen. Nach der Erzeugung des Indexes wird dieser mittels der Spark-broadcast-Funktion auf alle Executer verteilt. Hierdurch kann jeder Executer lokal auf dem Index arbeiten, sodass während der Tankstellensuche kein Datenaustausch zwischen den einzelnen Rechnern im Cluster nötig ist. Im nächsten Schritt werden nun anhand der gefundenen Tankstellen die aktuellen Tankstellenpreise über das *ActualGasStationDataSource* Modul abgefragt. Nachdem alle Daten bereitstehen, kann die beste Tankstelle, wie unter 5.2.1 dargestellt, ermittelt werden und das Ergebnis über das Messaging System bereitgestellt werden. Eine Ablage der Analyseergebnisse in den Serving Layer ist hier momentan umgesetzt.

### 5.3.6 Deployment der Anwendung

Im Folgenden wird das Deployment der CIS Anwendung dargestellt. Hierbei wird die AWS-Cloud-Infrastruktur von Amazons Dienst genutzt. Die Anzahl und Art der einzelnen Systeme ist aber von der Infrastruktur unabhängig.

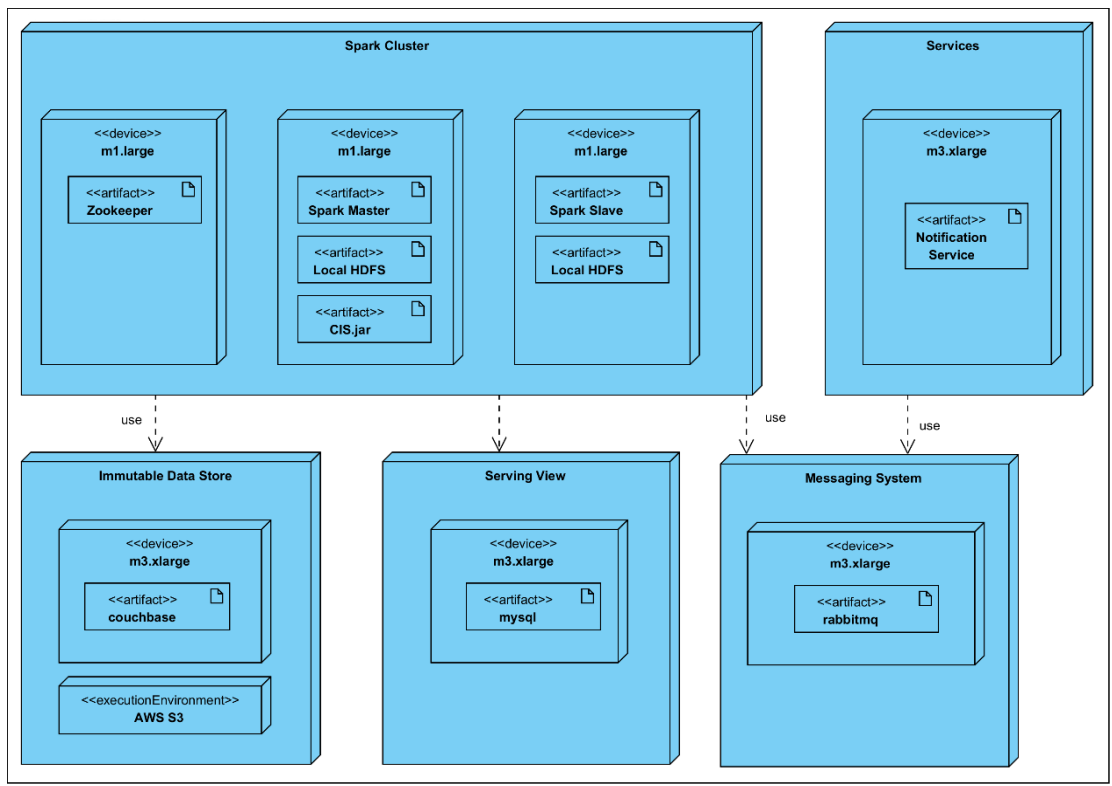

**Abbildung 21: CIS Anwendungs Deployment**

---

[10] http://www.vividsolutions.com/jts/JTSHome.htm

In der oben dargestellten Grafik sind die einzelnen physikalischen Systeme und die darauf laufenden Software Artefakte aufgezeigt. Bei den fünf Hauptkomponenten Spark Cluster, Immutable Data Store, Serving Layer, Services und Messaging System handelt es sich um komplett getrennte Systeme. Um eine hohe Ausfallsicherheit des gesamten Systems zu erreichen, müssen somit auch alle Einzelsysteme ausfallsicher ausgelegt werden. Um das Management der Infrastruktur möglichst einfach zu gestalten, wurde eine hohe Automatisierbarkeit angestrebt. So wurde zunächst in Betracht gezogen, Apache Mesos als grundlegende Plattform zu verwenden, da sich so alle Softwarekomponenten, einschließlich Spark, auf dieser einheitlichen Plattform hätten betreiben lassen können. Jedoch zeigte sich, dass Mesos zum Zeitpunkt der Erstellung dieser Arbeit noch nicht gut für den Betrieb von zustandsbehafteten Anwendungen, wie z.B. Couchbase geeignet war. Daher wird in der jetzigen Umsetzung zwischen dem Spark-Cluster und allen restlichen Anwendungen unterschieden. Für Couchbase, MySQL, RabbitMQ, und den Notificationservice werden Docker Container genutzt, welche auf angemieteten Instanzen von AWS ausgeführt werden. Für Spark wurde der Standalone-Cluster-Manager gewählt. Hier kann mittels dem Tool spark-ec2, sehr einfach eine AWS Infrastruktur erstellt werden. Ein solches Setup besteht in der Regel aus einem Spark-Master und einer beliebigen Anzahl von Spark-Slaves, wodurch Anwendungen auf diesem Cluster aber nur resistent gegen den Ausfall der Spark-Slaves sind. Um den Ausfall des Masters zu kompensieren, muss auch dieser redundant ausgelegt sein. Hierfür ist es nöglich, dass sich mehrere Master über einen Zookeeper-Server abstimmen und einen primären Master wählen. Um eine hohe Ausfallsicherheit zu garantieren, wurde sich dafür entschieden, die einzelnen Systeme mindestens dreimal zu replizieren, um so den Ausfall von bis zu 2 Komponenten kompensieren zu können. Die Anzahl an Spark Slaves richtet sich vor allem nach der zu erwartenden Last, wobei hier genug Reserven eingeplant werden sollten. Des Weiteren läuft auf jeder Spark Instanz ein lokaler HDFS-Node, welcher vor allem für das für Spark Streaming nötige Checkpointing genutzt wird. Als Alternative könnte hier aber auch AWS S3 genutzt werden, welches API-kompatibel zu HDFS ist.

# 6 Diskussion

Im folgenden Kapitel wird das entwickelte System bewertet und evaluiert.
Hierfür wird zunächst in 6.1 das Vorgehen vorgestellt und in 6.2 und 6.3 dieses auf das Car Information System bzw. auf Spark Streaming angewendet. Hierbei liegt der Fokus vor allem auf der Nutzbarkeit des Systems und darauf, wie sich das System bezüglich der in 5.3 aufgestellten Anforderungen verhält.

## 6.1 Beschreibung des Vorgehens

Für eine fundierte Qualitätsermittlung wird die Definition der Softwarequalität nach ISO/IEC 9126 genutzt.

> *„Softwarequalität ist die Gesamtheit der Merkmale eines Softwareprodukts, die sich auf dessen Eignung beziehen, festgelegte und vorausgesetzte Erfordernisse zu erfüllen."* [Inte01]

Die Softwarequalität ermittelt sich demnach aus der Qualität verschiedener Qualitätsmerkmale einer Software, welche auch in der ISO/IEC 9126 Norm definiert sind und im Folgenden veranschaulicht werden.

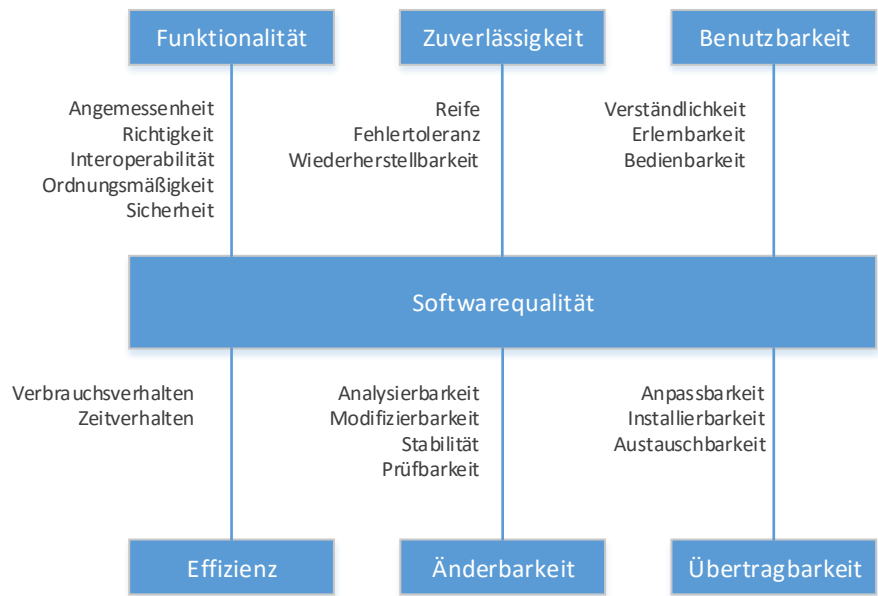

**Abbildung 22: Softwarequalität nach Norm ISO/IEC 9126**

Die jeweiligen Softwarequalitätsmerkmale und deren Bedeutungen ergeben sich aus der zuvor dargestellten Grafik. Auf bestimmte Merkmale muss jedoch gesondert eingegangen werden, um sie in den Kontext des hier zu bewertenden Systems zu stellen.

So wird unter dem Punkt Benutzbarkeit eines Systems oft die Bedienbarkeit der Benutzeroberfläche bzw. der Begriff Software-Ergonomie verstanden [Kelt07], da aber beide der hier zu bewertenden Komponenten über keine Oberfläche verfügen, wird hier die Benutzbarkeit aus Sicht eines Entwicklers bewertet. Hiermit ist z.B. der Aufwand gemeint, den ein neuer Entwickler hat, um sich in das System einzuarbeiten. Außerdem ist gefragt, wie intuitiv das System bzw. die Programmierschnittstellen verwendet werden können. Dies ist nicht mit dem Punkt Änderbarkeit zu verwechseln, welcher die Erweiterbarkeit des Systems bzw. den Aufwand für eine Erweiterung des Systems beschreibt.

Die einzelnen Qualitätsmerkmale lassen sich sowohl qualitativ als auch quantitativ bewerten. Mittels Metriken können z.B. quantitative Daten gesammelt werden, welche direkt einen konkreten Zahlenwert für die Bewertung der Software angeben. Hierdurch können sie direkt in die Bewertung des Gesamtsystems einfließen. Es ist jedoch anzumerken, dass die jeweiligen quantitativen Daten immer in einen spezifischen, fachlichen und technischen Kontext gesetzt werden müssen, um eine Vergleichbarkeit der Daten zu gewährleisten [Drst14]. Außerdem muss genau überprüft werden, wie die jeweiligen Daten zustande kommen, um eventuelle Fehlinterpretationen zu vermeiden [Drst14]. Des Weiteren lassen sich durch kontinuierliche Messungen Änderungen am System mit Änderungen in den Testergebnissen vergleichen, um so langfristiger die Auswirkungen von bestimmten Systemanpassungen zu beobachten[Drst14].

Bei der Bewertung einer Software ist es essenziell, die eigentliche Bewertung und nicht das reine Erheben von Metriken im Blick zu halten. So wird von Basili und Rombach der Goal-Question-Metric (GQM) Ansatz vorgestellt, welcher sich auf spezifische Ziele für das Erheben einer Metrik fokussiert [BaRo94]. Hierdurch wird hervorgehoben, dass es sich bei dem Erheben von Metriken nicht um einen Selbstzweck handelt, sondern dieses immer einem übergeordneten Ziel dienen muss, z.B. der Evaluation eines Produktes [AsKP02]. Das GQM Paradigma beschreibt 6 Phasen in einem systematischen top-down Ansatz, welche im Folgenden beschrieben werden:

**1. Phase**
Es werden anhand eines Verbesserungsziels mehrere Messziele (Goals) abgeleitet. Hierbei sind Verbesserungsziele übergeordnete Ziele, wie beispielsweise Geschäftsziele. Die Messziele werden jeweils durch die 5 Aspekte *Objekt*, *Zweck*, *Qualitätsfokus*, *Perspektive* und *Kontext* ausgedrückt und mittels eines sogenannten *Abstraction Sheet* veranschaulicht.

Der *Abstraction Sheet* ist wie folgt aufgebaut:

| Objekt | Zweck | Qualitätsfokus | Perspektive | Kontext |
|--------|-------|----------------|-------------|---------|
|        |       |                |             |         |

| QUALITÄTSFAKTOREN:<br>Welche Faktoren/Metriken müssen betrachtet werden? | EINFLUSSFAKTOREN:<br>Welche Variablen haben einen Einfluss auf die Qualitätsfaktoren? |
|---|---|
| HYPOTHESE:<br>Was sind die momentanen Erwartungen bezüglich der Qualitätsfaktoren? | EINFLUSS AUF HYPOTHESE:<br>Wie beeinflussen die Einflussfaktoren die Qualitätsfaktoren? |

**Abbildung 23: Abstraction Sheet nach** [AsKP02]

Die 5 Aspekte des Messziels:

**Objekt**: Das Objekt, welches analysiert werden soll, z.B. Produkt oder Prozess
**Zweck**: Der Zweck der Analyse, z.B. Evaluation oder Verbesserung
**Qualitätsfokus**: Der Qualitätsfokus beschreibt, welche Attribute analysiert werden sollen
**Perspektive**: Die Perspektive, aus der die Analyse durchgeführt wird, z.B. Entwickler oder Kunde
**Kontext**:  Einen bestimmten Kontext, in dem die Messergebnisse gelten.

Des Weiteren werden festgelegte Fragen bezüglich des Messziels beantwortet, um dieses weiter zu spezifizieren. Da die einzelnen Fragen aufeinander aufbauen, wird hierbei in einer festgelegten Reihenfolge vorgegangen. Es werden zunächst die *Qualitätsfaktoren* definiert, daraufhin wird für die einzelnen *Qualitätsfaktoren* eine *Hypothese* festgelegt, um die eigene Erwartungshaltung bezüglich des Ausgangs der Messung im Nachhinein überprüfen zu können. Nach dem Erstellen der *Hypothese* muss geprüft werden, welche Faktoren einen Einfluss auf diese haben könnten. Hieraus können dann konkrete Einflussfaktoren abgeleitet werden.

**2. Phase**
Nun werden anhand der festgelegten Messziele und des *Abstraction Sheets* mehrere konkrete Fragen (Questions) abgeleitet. Hierbei werden die Qualitäts- und Einflussfaktoren genauer betrachtet [AsKP02]. Jede Frage sollte präzise klären, was über das Objekt bzw. die Eigenschaft im Qualitätsfokus des Ziels gelernt werden soll.
Für die einzelnen Fragen empfehlen Assmann und Kalmar folgende Eigenschaften:
- Die Frage muss mit dem Zweck und dem Blickwinkel übereinstimmen.
- Die Frage muss quantifizierbar sein.
- Die Bedeutung der möglichen Antwort muss klar sein.
- Alle Begriffe müssen klar definiert sein
- Einflussfaktoren sollten beachtet werden, aber nicht im Vordergrund stehen.

**3. Phase**
Um nun die zuvor festgelegten Fragen beantworten zu können, werden für jede Frage passende Metriken ausgewählt. Hierbei ist das Beachten der zuvor gefunden Einflussfaktoren wichtig, da es sonst zu einer Fehlinterpretation der Metrik-Ergebnisse kommen könnte.

**4. Phase**
Anhand der Fragen und der jeweiligen Metriken wird ein Messplan erstellt, welcher den konkreten Ablauf der Messung darstellt und alle Bedingungen für die Messung definiert.

**5. Phase**
Es werden die eigentlichen Daten erfasst und nachfolgend analysiert und interpretiert. Hierbei ist eine Validierung der Daten vor der Analyse nötig, sodass Ausreißer die Messergebnisse nicht verfälschen.

**6. Phase**
Zum Abschluss können mittels eines *Lessons Learned* Erfahrungen, Entwicklungen, Hinweise, Fehler und Risiken zusammengetragen werden.

Das hier zu bewertende Projekt beinhaltet zwei voneinander getrennte Teilgebiete. Zum einen die Bewertung der CIS Anwendung bzw. der hier gewählten Theta Architektur. Zum anderen wird die Arbeit mit Spark bewertet. Für beide Aspekte wird das GQM-Modell verwendet.

## 6.2 Evaluation des CIS

In diesem Kapitel wird das CIS und dessen Theta Architektur mithilfe des GQM Paradigmas evaluiert und bewertet. Hierfür wurden die 6 Phasen des GQM-Modells durchgeführt, diese werden aus Gründen der Übersichtlichkeit jedoch nicht separat voneinander dargestellt. Für die Wahl der Qualitätsfaktoren wird auf die zuvor dargestellten Softwarequalitätsmerkmale nach ISO 9200 zurückgegriffen. Für die Nutzung des GQM-Modells müssen zunächst die jeweiligen Messziele formuliert werden. Hierbei gelten für alle Messziele folgende Punkte:
**Zweck**: Evaluation
**Objekt**: CIS System
**Perspektive**: Entwickler des Systems
**Kontext**: Das CIS als Einzelsystem in einer AWS-Umgebung
Bezüglich des Qualitätsfokus wurde sich für *Fehlertoleranz*, *Änderbarkeit*, *Übertragbarkeit* und *Skalierbarkeit* entschieden, da diese Qualitätskriterien zum einen die architektonischen Entscheidungen bewerten und zum anderen auch das zugrundeliegende Spark System betreffen und so der Zielsetzung dieser Arbeit entsprechen. Im Folgenden sind die separaten Messziele als *Abstraction Sheets* und darauf aufbauend die jeweiligen Fragen und Metriken dargestellt.

### 6.2.1 Skalierbarkeit

| CIS System | Evaluation | Skalierbarkeit | Entwickler | AWS Umgebung |
|---|---|---|---|---|

| | |
|---|---|
| **Qualitätsfaktoren:**<br>QF1: Maximaler Durchsatz pro Sekunde | **Einflussfaktoren:**<br>EF1: Art der Nachrichten<br>EF2: Betrachtete Komponente |
| **Ausgangshypothesen:**<br>Der maximale Durchsatz steigt linear mit der Anzahl an beteiligten Rechnerknoten an | **Einflusshypothesen:**<br>Der Durchsatz hängt von der jeweils betrachteten Verarbeitungskomponente und von der Art an Nachrichten ab. |

| Frage | Metrik |
|---|---|
| F1: Wie verhält sich der maximale Durchsatz in Bezug auf die Anzahl an Rechnerknoten? | Lasttest |
| F2: Wie verhält sich der maximale Durchsatz in Bezug auf die Art der Last? | Lasttest |
| F3: Wie verhält sich der maximale Durchsatz in Bezug auf die Komponente (Vorverarbeitung / Tankstellensuche)? | Lasttest |

**Messumgebung:**
Die zuvor festgelegten Fragen bezüglich der Skalierung des Systems sollen mithilfe eines Lasttests beantwortet werden. Hierfür wird auf das in 5.3.6 vorgestellte Setup aufgebaut, wobei auf die Ausfallsicherheit verzichtet wurde, da diese hier nicht im Fokus steht und somit ein erhöhter Konfigurierungsaufwand vermieden werden kann. Für die einzelnen Knoten wurden folgende AWS-Instanzen verwendet:

- Messaging System (RabbitMQ): eine m3.xlarge Instanz [11]
- Immutable Data Store (Couchbase): eine m3.xlarge Instanz
- Spark Standalone Cluster: eine Master m1.large Instanz & N Slave m1.large Instanz[12]

---

[11] Beschreibung der AWS-Instanz-Typen unter b
[12] Die m1.large Instanz ist die Standardinstanz für die AWS Spark Konfiguration und wird hier daher als Ausgangspunkt verwendet.

Bei dem hier beschriebenen Lasttest wird eine horizontale Skalierung betrachtet, es wird also über die Anzahl an separaten Spark Slave Instanzen skaliert. Durch bessere AWS-Instanz wäre auch eine vertikale Skalierung prinzipiell möglich, sodass eine höhere CPU-Kernanzahl bzw. mehr Arbeitsspeicher pro einzelner Instanz zur Verfügung stehen würde. Dies wird aber nicht näher betrachtet, da nur die horizontale Skalierung im Vergleich theoretisch nicht begrenzt ist und Spark speziell für diese ausgelegt ist. Außerdem ist die Vergrößerung des pro Instanz zur Verfügung stehenden Arbeitsspeichers für den hier gewählten Anwendungsfall nicht relevant.

**Ablauf der Messung**
Das System wird in zwei separaten Konfigurationen getestet. So werden die beiden in 5.3.5 beschriebenen Komponenten Vorverarbeitung (*carDataFirstAppearanceFilter*) und Tankstellensuche (*GasStationSearchStreamProcessor*) unabhängig voneinander untersucht. Dieses Vorgehen liefert zudem direkt die nötigen Informationen für F3. Außerdem entsteht unter realen Bedingungen eine sehr unterschiedliche Lastverteilung auf den einzelnen Komponenten, wodurch eine Betrachtung der jeweiligen Skalierungseigenschaften interessant ist.

**Lastszenarien:**
Für die Beantwortung von F2 müssen verschiedene Lastszenarien getestet werden. Hierfür besteht die Hypothese, dass die Performance der Vorverarbeitungskomponente mit der Anzahl an unterschiedlichen Nachrichtenquellen, sprich Fahrzeugen, zusammenhängt. Hiermit ist nicht die Anzahl der empfangenen Nachrichten pro Sekunde gemeint, sondern vielmehr von wie vielen verschiedenen Fahrzeugen Nachrichten in einem Zeitintervall empfangen wurden. Dies könnte einen großen Effekt auf die Verarbeitungsgeschwindigkeit dieser Komponente haben, da hier rückwirkend über einen längeren Zeitraum Daten anhand der FahrzeugID verglichen werden müssen.

Daraus ergeben sich für das Testen der Vorverarbeitung drei Szenarien:
LS1:    Verarbeiten von gleichen Nachrichten -> Alle Nachrichten an das System stammen von einem Fahrzeug
LS2:    Verarbeiten von simuliertem Nutzerverhalten -> Es wird das realistische Nachrichtenaufkommen von N gleichzeitig fahrenden Fahrzeugen simuliert
LS3:    Verarbeiten von unterschiedlichen Nachrichten -> Alle Nachrichten stammen von unterschiedlichen Fahrzeugen

Für die Tankstellensuche wird kein gesondertes Testszenario benötigt, da hier die einzelnen Nachrichten komplett losgelöst voneinander verarbeitet werden. Nachdem alle Lastszenarien durchgespielt wurden, wird die Anzahl an Spark Slaves verdoppelt und das System neu gestartet. Für jeden Test wird das Nachrichtenaufkommen in 500er Schritten bis zu einer Obergrenze von 10.000 Nachrichten/Sekunde erhöht.

**Datenerhebung:**
Der zuvor definierte Lasttest wird anhand der beschriebenen Lastszenarien automatisiert ausgeführt und aufgezeichnet. Die jeweiligen Kennzahlen werden als CSV-Datei gespeichert. Die Ergebnisse der einzelnen Lasttests sind im Anhang dieser Arbeit dargestellt.

**Interpretation:**
Im Folgenden werden die Ergebnisse des Lasttests interpretiert und die drei ursprünglichen Fragen beantwortet.

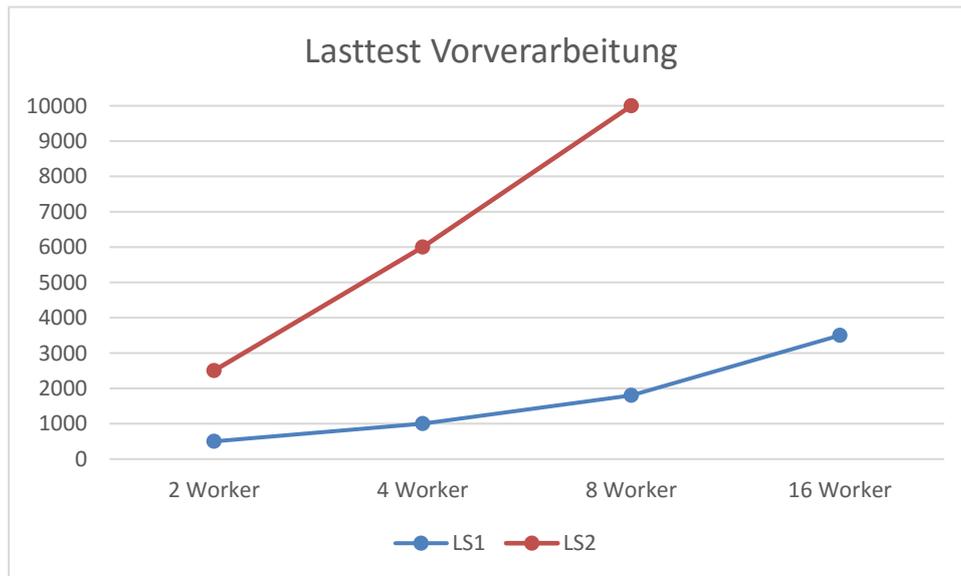

**Abbildung 24: Auswertung Lasttest der Vorverarbeitungskomponente**

Abbildung 24 veranschaulicht die Ergebnisse des Lasttests der Vorverarbeitungskomponente, wobei jeweils der maximal zu erreichende Durchsatz für den jeweiligen Test dargestellt ist. Falls der Durchsatz über diesen Punkt ansteigt, wird der Aufwand für ein Spark Bache-Intervall zu groß und die nachfolgenden Intervalle werden verzögert. In der Grafik sind jeweils nur LS2 und LS3 bis zu einem maximalen Durchsatz von 10.000 Nachrichten pro Sekunde dargestellt. LS1 erreicht bereits auf einem System mit zwei Workern dieses Maximum von 10.000 Nachrichten und wird daher hier nicht näher betrachtet. Hierbei zeigt sich, dass unabhängig vom Lastszenario eine Steigerung des maximalen Durchsatzes bei Erhöhung der Knotenzahl ermöglicht wird. Außerdem ist zu erkennen, dass bei dieser Komponente der Unterschied des maximalen Nachrichtendurchsatzes zwischen den einzelnen getesteten Lastszenarien sehr hoch ist. So zeigt sich, dass sich bei LS2 ein Durchsatz von mehr als 10.000 Nachrichten pro Sekunde auf einem System mit 8 Spark Workern erreichen lässt. Hingegen wird bei LS3 auf einem System mit derselben Anzahl an Workern nur ein Durchsatz von ca.

1.800 Nachrichten pro Sekunde erreicht. Dies resultiert daraus, dass das System in diesem Fall wesentlich mehr separate Zustände erfassen muss.[13]

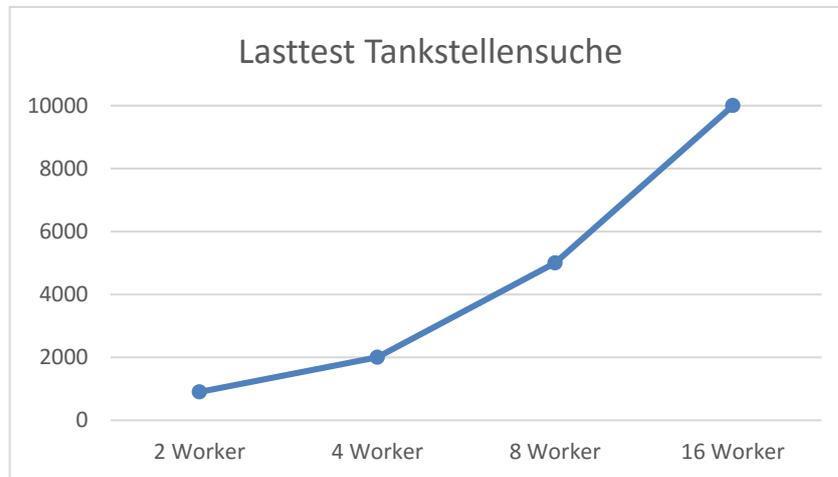

**Abbildung 25: Auswertung Lasttest der Tankstellensuche**

Abbildung 25 veranschaulicht die Tankstellensuche und zeigt die Skalierbarkeit dieser Komponente. So konnte der maximale Durchsatz von ca. 900 Nachrichten pro Sekunde bei 2 Workern auf mehr als 10.000 Nachrichten pro Sekunde mit 16 Workern gesteigert werden. Dies resultiert vor allem aus der Beschaffenheit der Aufgabe. So wird bei der Tankstellensuche jede Nachricht separat voneinander verarbeitet, wodurch keine aufwendigen und teuren Join- bzw. Gruppierungsoperationen nötig werden. Der größte Aufwand entsteht durch die Suche der jeweiligen Tankstellen anhand der GeoPosition der Fahrzeuge.

**Gesamte Testergebnisse:**

| Messung | 2 Worker | 4 Worker | Δ | 8 Worker | Δ | 16 Worker | Δ |
|---|---|---|---|---|---|---|---|
| LS1 | 500 | 1000 | 100% | 1800 | 80% | 3500 | 94% |
| LS2 | 2500 | 6000 | 140% | 10000 | 67% | | |
| Tankstellensuche | 900 | 2000 | 122% | 5000 | 150% | 10000 | 100% |

**Tabelle 3: Darstellung der Messergebnisse des Lasttests**

Anhand der Daten bzw. der Grafiken ist die deutliche Skalierbarkeit der Anwendung zu erkennen. Im Folgenden werden die einzelnen Fragen anhand der Messergebnisse beantwortet:

---

[13] Für LS2 resultieren 1.000 Nachrichten pro Sekunde nach 30 Sekunden in 1.000 Zuständen.
Für LS3 resultieren 1.000 Nachrichten pro Sekunde nach 30 Sekunden in 1.000*30 Zuständen.

F1: *Wie verhält sich der maximale Durchsatz in Bezug auf die Anzahl an Rechnerknoten?*
Es ist festzustellen, dass in jedem Fall der maximale Durchsatz durch das Hinzufügen weiterer Rechnerknoten gesteigert werden konnte. Im Durchschnitt wurde eine Steigerung von ca. 114 % erreicht[14].

F2: *Wie verhält sich der maximale Durchsatz in Bezug auf die Art der Last?*
Anhand der Messergebnisse ist für die Vorverarbeitung eine deutliche Beziehung zwischen Durchsatz und Lastszenario erkennbar. So ist der maximale Durchsatz bei LS2 im Schnitt ca. 6mal höher als in LS3. Des Weiteren liefert der Extremfall LS1 ein Verhalten, bei dem auf jeder Systemkonfiguration mehr als 10.000 Nachrichten pro Sekunde verarbeitet werden können. Das wiederum zeigt, wie groß die Beziehung ist.

F3: *Wie verhält sich der maximale Durchsatz in Bezug auf die Komponente (Vorverarbeitung / Tankstellensuche)?*
In Bezug auf die beiden einzelnen Komponenten Vorverarbeitung und Tankstellensuche lassen sich sehr unterschiedliche Lastkurven erkennen. So erreicht die Tankstellensuche zwar pro Systemkonfiguration einen geringeren maximalen Durchsatz als das LS2 bei der Vorverarbeitung, sie erreicht jedoch in allen Tests einen höheren Durchsatz als in LS3 der Vorverarbeitung.

---

[14] Es wird die Steigung aller Einzelmessungen betrachtet, welche einen Durchsatz kleiner als 10.000 Nachrichten haben und anschließend der Durchschnitt gebildet.

### 6.2.2 Fehlertoleranz

| CIS System | Evaluation | Fehlertoleranz | Entwickler | AWS Umgebung |
|---|---|---|---|---|
| **Qualitätsfaktoren:** QF1: Funktionalität des Systems während eines teilweisen Systemausfalls ||| **Einflussfaktoren:** EF1: Funktion der Komponenten EF2: Laststeigerung nach einem Ausfall ||
| **Ausgangshypothesen:** Das System wird den Ausfall einzelner Komponenten kompensieren ||| **Einflusshypothesen:** Die Funktion der ausgefallenen Komponenten könnte einen Einfluss auf die Funktionstüchtigkeit des Restsystems haben. Die Last könnte für das Restsystem zu hoch sein, sodass dieses zusammenbricht ||

| Frage | Metrik |
|---|---|
| F2: Wie viele Testfälle schlagen durch den Ausfall einzelner Knoten fehl? | M1: Anzahl der positiven Tests pro jeweiligem Ausfallszenario |
| F1: Wie verhält sich der Durchsatz bei dem Ausfall einzelner Knoten? | M2: Lasttest während des Ausfalls eines Knotens. |
| F3: Hängen die Auswirkungen eines Ausfalls von der Aufgabe des jeweiligen Rechners ab? | M3: Vergleich des Ausfalls eines Master-, Driver-, Slave-Knotens. |

**Messumgebung:**
Die Messung wird genau wie die Evaluation der Skalierbarkeit auf einem AWS System durchgeführt. Hierfür wird auf das unter 5.3.6 beschriebene ausfallsichere Spark Setup zurückgegriffen.

**Messablauf:**
Für M1 und M2 wird das System zunächst einem Failover-Test unterzogen. Dabei werden einzelne Knoten des Systems deaktiviert. Während der Deaktivierung werden verschiedene Tests auf dem System durchgeführt, um so die Auswirkungen eines Ausfalls zu beobachten. Es wird hier zwischen dem Ausfall eines Spark Workers, des Spark Drivers und des Spark Master Knoten unterschieden.

**Messergebnisse:**

Hierfür zunächst eine Darstellung der drei unterschiedlichen Ausfallarten und ihre Konsequenzen:

**Ausfall eines Spark Workers:**

In diesem Fall werden die Auswirkungen des Ausfalls eines Spark Workers untersucht. Die Testergebnisse sind in folgenden Grafiken veranschaulicht:

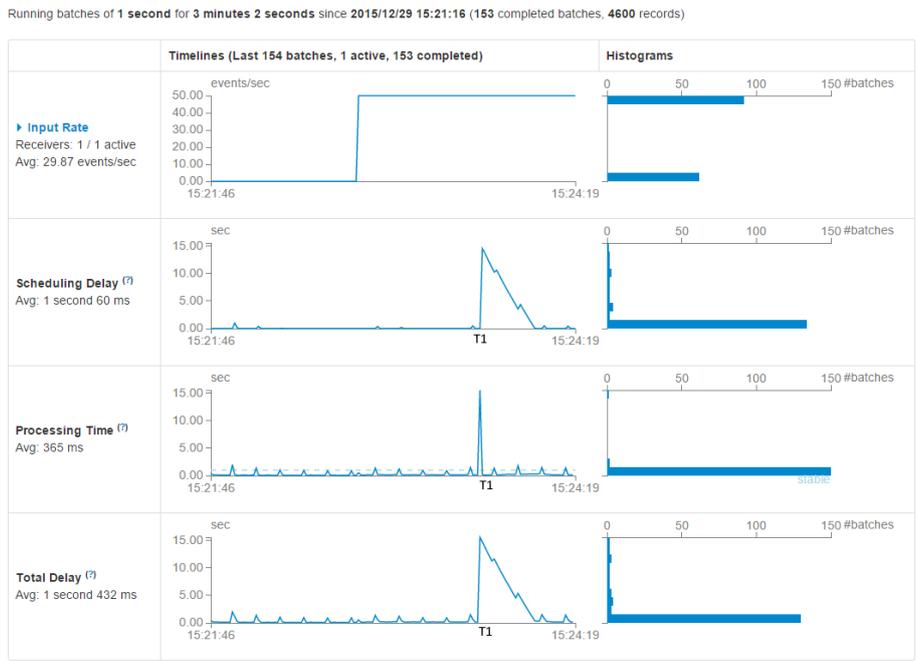

**Abbildung 26: Ausfall eines Workers Spark UI**

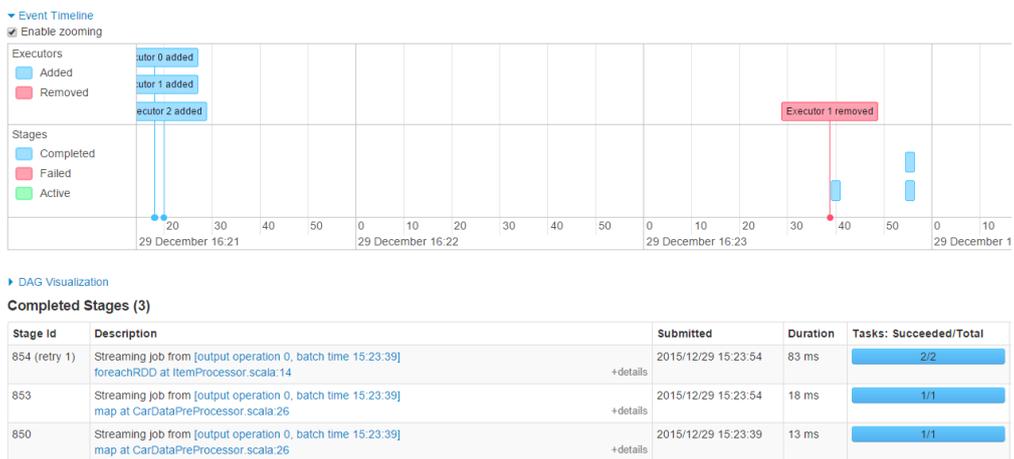

**Abbildung 27: Ausfall eines Workers Event UI**

In Abbildung 26 wird der Zustand des gesamten Systems veranschaulicht und es ist zu erkennen, dass das System nach dem Ausfall des Worker Knotens bei T1 weiterhin funktionstüchtig bleibt und 50 Nachrichten pro Sekunde verarbeitet. Des Weiteren ist festzustellen, dass im Moment des Ausfalls die Verarbeitungsdauer des momentan aktiven Batch Intervalls sprunghaft auf ca. 15s ansteigt. Hierdurch müssen die nachfolgenden Batch Intervalle zurückgestellt werden, sodass ein Delay entsteht. In der Abbildung 27 ist außerdem der Verlauf des aktiven Batches im Moment des Ausfalls dargestellt. Hier ist zu erkennen, dass der Executer 1 entfernt wurde und daraufhin die Stage mit der ID 854 auf einem anderen Executer erneut ausgeführt werden musste. Außerdem ist zu erkennen, dass zwischen der ersten *Stage* und den beiden weiteren ca. 15 Sekunden vergehen. Dies lässt sich vor allem durch die Neuverteilung der Stages auf die noch funktionstüchtigen Executer und das erneute Berechnen von Zwischenergebnissen zurückführen. So erstellt die Vorverarbeitung alle 10 Sekunden einen Checkpoint und muss somit in diesem Fall die Zwischenergebnisse erneut bis zu dem letzten Checkpoint berechnen. Für alle nachfolgenden Batch Intervalle erhöht sich die Verarbeitungsdauer hingegen nur geringfügig, was durch den Wegfall der Rechenkapazität des Workers erklärbar ist.

**Ausfall des Spark Masters:**
In diesem Fall werden die Auswirkungen des Ausfalls des Spark Master Knotens untersucht. Die Testergebnisse sind in folgenden Grafiken veranschaulicht.

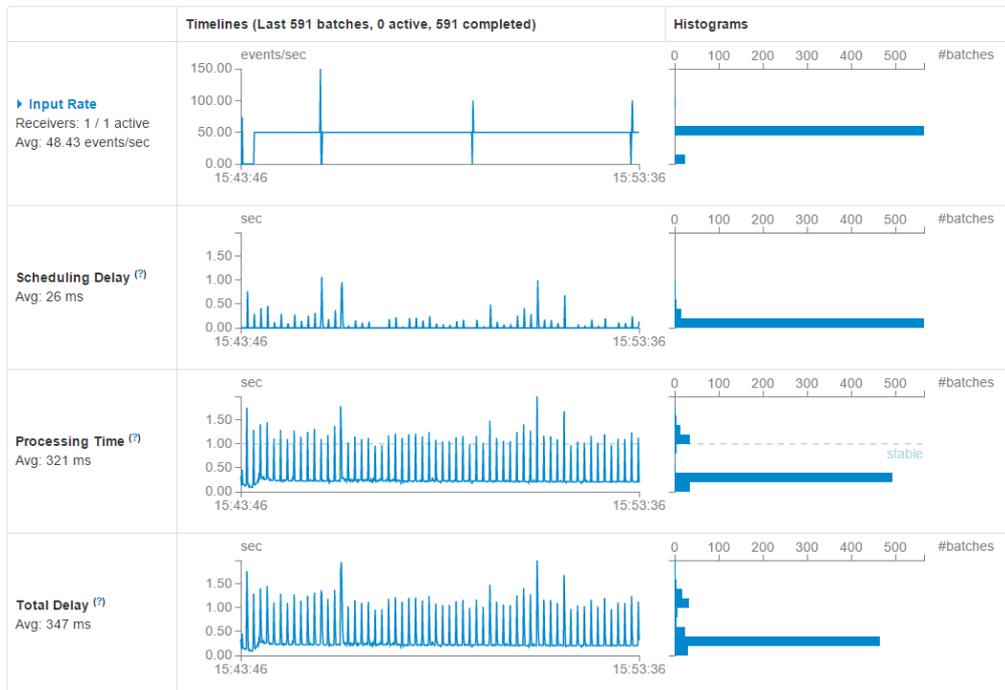

**Abbildung 28: Ausfall Spark Master Knotens**

Die Abbildung 28 veranschaulicht, dass der Ausfall des Master Knotens keine direkten Auswirkungen auf die eigentliche, schon im Spark-Cluster laufende Anwendung hat. So ist die Aufgabe des Masters vor allem auf das initiale Starten und das Überwachen einer Anwendung beschränkt. Die Ausfallsicherheit des Masters ist aber essenziell, da nur er auf den Ausfall des Drivers reagieren kann.

**Ausfall des Spark Drivers:**
In diesem Fall werden die Auswirkungen des Ausfalls des Spark Driver Knotens untersucht. Die Testergebnisse sind in folgenden Grafiken veranschaulicht. Hierbei ist es nicht möglich, auf die normale Spark UI zurückzugreifen, da diese über den Spark Driver ausgeführt wird und somit auch ausfällt.

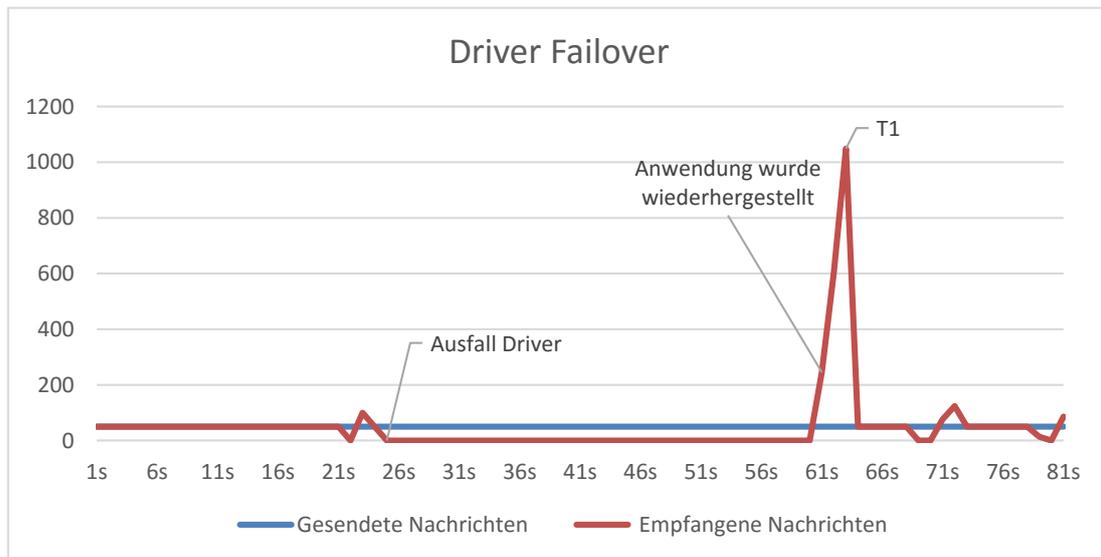

**Abbildung 29: Spark Driver Failover**

Die Abbildung 29 veranschaulicht, wie viele Nachrichten an das System gesendet bzw. wie viele Antworten empfangen wurden. Hier wurden die gesendeten Nachrichten so gewählt, dass bei einem funktionierenden System für jede gesendete Nachricht eine Nachricht empfangen werden sollte.
Die Abbildung zeigt, dass durch den Ausfall des Drivers keine Nachrichten mehr vom System verarbeitet werden konnten, da die gesamte Anwendung beendet wurde. Daraufhin startete der Spark Master das Driver Programm neu. Währenddessen wurden weiter Daten an das System geschickt, welche vom Messaging System bis zur Wiederherstellung des Drivers zwischengespeichert wurden. Diese erklärt auch die hohe Anzahl an empfangenen Nachrichten bei T1.

**Gesamtergebnis Analyse:**
Insgesamt lässt sich feststellen, dass das CIS System auf den Ausfall einzelner Knoten reagieren kann. Abhängig von der Funktionalität des jeweiligen Knotens sind jedoch verschiedene Auswirkungen festzustellen. So hat der Ausfall eines Workers bzw. des Drivers immer eine Auswirkung auf die CIS Anwendung, sodass für einen gewissen Moment der Durchsatz an verarbeiteten Nachrichten einbricht.
Der Ausfall eines Master Knotens hat für das CIS keine direkten Konsequenzen. Hierbei ist es aber wichtig zu beachten, dass eine Ausfallsicherheit des Masters trotzdem unablässig ist, da sonst andere Ausfälle nicht mehr kompensiert werden können.

F1: *Wie viele Testfälle schlagen durch den Ausfall einzelner Knoten fehl?*
Durch den Ausfall eines Knotens schlagen keine Testfälle fehl. Das Gesamtsystem ist trotz einer Phase der Kompensierung immer noch in einem funktionstüchtigen Zustand. Während des Ausfalls eintreffende Nachrichten werden vom Messagingsystem zwischengepuffert oder sind redundant im Spark System gespeichert.

F2: *Wie verhält sich der Durchsatz bei dem Ausfall einzelner Knoten?*
Der Durchsatz hängt im Moment des Ausfalls von der Funktion der einzelnen Knoten ab. Bei dem Ausfall eines Worker bzw. Driver Knotens braucht das momentan aktive Batch-Intervall länger. Für die nachfolgenden Batch Intervalle wird der maximale Durchsatz nur bei dem Ausfall eines Worker Knotens eingeschränkt, da nun weniger Ressourcen für das Gesamtsystem zur Verfügung stehen.

F3: *Hängen die Auswirkungen eines Ausfalls von der Aufgabe des jeweiligen Rechners ab?*
Die Auswirkungen hängen stark von der Funktion des Knoten ab:
- Worker-Knoten -> neu Berechnung von Zwischenergebnissen auf anderen Workern, dadurch Verlangsamung des Systems
- Master-Knoten -> keine direkten Auswirkungen auf die Anwendungen
- Driver-Knoten -> Anwendung wird komplett neu gestartet.

### 6.2.3 Änderbarkeit

| CIS System | Evaluation | Änderbarkeit | Entwickler | AWS Umgebung |
|---|---|---|---|---|
| **Qualitätsfaktoren:**<br>QF1: Aufwand für das Umsetzen einer neuen Funktion<br>QF2: zyklomatische Komplexität<br>QF3: Anzahl der Abhängigkeiten zwischen Modulen | | | **Einflussfaktoren:**<br>Art der Änderung | |
| **Ausgangshypothesen:**<br>QF1: Für das Umsetzen einer neuen Funktion müssen keine Änderungen an der Theta Bibliothek durchgeführt werden und nur ein neuer Verarbeitungsalgorithmus angepasst werden. Hierdurch entsteht nur geringer Aufwand.<br><br>QF2: Es wird keine hohe zyklomatische Komplexität für die einzelnen Module gemessen<br><br>QF3: Geringe Abhängigkeit zwischen den Verarbeitungsmodulen untereinander, hohe Abhängigkeit im Verarbeitungsmodul zum Theta Framework, keine Abhängigkeit zwischen Theta Framework zu den Verarbeitungsmodulen | | | **Einflusshypothesen:**<br>Bei grundlegenden Änderungen muss möglicherweise das Theta Framework erweitert werden, wodurch ein höherer Aufwand entsteht | |

| Frage | Metrik |
|---|---|
| F1: Wie viele Änderungen müssen an der bisherigen Architektur durchgeführt werden, um eine neue Funktion zu implementieren? | M1: Durchführung der Konzeption einer neuen Funktion und Aufsummierung der nötigen Änderungen. |
| F2: Wie viele unterschiedliche Arten von Änderungen lassen sich erkennen? | M1: Durchführung der Konzeption einer neuen Funktion mit anschließender Bewertung der nötigen Änderungen. |
| F3: Wie hoch ist die Komplexität der einzelnen Module der Anwendung? | M2: Messung der Komplexität mittels einer statischen Code Analyse |
| F4: Wie hoch ist die Abhängigkeit der einzelnen Module untereinander? | M3: Zählen der Abhängigkeiten zwischen den Kern Modulen des CIS. |

**Messablauf:**

Für die Beantwortung dieser Fragen wurden mehrere Metriken festgelegt. Zum einen wird die Komplexität des CIS mittels der McCabe-Metrik bzw. des Zählens von Abhängigkeiten ermittelt. Des Weiteren wird eine konzeptionelle Umsetzung einer weiteren Funktionalität für das CIS durchgeführt. Hierbei werden die nötigen Änderungen am CIS analysiert.

**Messergebnisse:**

**M1: Konzeption einer weiteren User-Story:**

Folgende User-Story soll für das System umgesetzt werden:
> *Eine Versicherung will anhand des Fahrverhaltens des Nutzers dem Kunden eine halbjährliche Prämie auszahlen.*

Hieraus ergeben sich folgenden Anforderungen:
- Das System muss alle für die Analyse des Fahrverhaltens nötigen Informationen sammeln.
- Die Versicherung muss mindestens einmal im halben Jahr auf die Daten zugreifen können.

Dieses Anfordern ermöglichen zwei verschiedene Umsetzungsmöglichkeiten
1. Halbjährliche Batch-Analyse der im Immutable Datastore gespeicherten Fahrevents.
2. Permanente Analyse des Fahrverhaltens und Speicherung der reinen Analyseergebnisse am Ende einer Fahrt.

Beide Lösungsansätze haben einige Vor- bzw. Nachteile. So müssen die für die erste Lösungsmöglichkeit benötigten Daten im Immutable Datastore abgelegt werden, was über eine *saveToImmutableDatastore()* Funktion einfach realisierbar ist. Außerdem muss hier nur eine gesonderte Batchroutine erstellt werden, welche einmal im halben Jahr angestoßen wird. Hierbei ist aber zu beachten, dass die Batchroutine dann auf einem sehr großen Datenbestand arbeiten würde und somit ein extra System nötig ist, um diesen einmaligen Aufwand zu bewerkstelligen und den Ablauf des eigentlichen Systems nicht zu beeinflussen. Der zweite Ansatz hat den Vorteil, dass das Fahrverhalten permanent analysiert wird, so erhöht sich zwar der kontinuierlichere Aufwand zum Betrieb der Anwendung. Andererseits wird eine höhere Flexibilität erreicht, da so jederzeit ad hoc auf die Auswertung des Fahrverhaltens zugegriffen werden kann. Für die Umsetzung dieses Ansatzes müsste eine neue Verarbeitungskomponente realisiert werden, welche sich an der Tankstellenpreissuche orientiert.

Für die Umsetzung der neuen Anforderungen wird im Folgenden der zweite Lösungsansatz gewählt, da er zum einen eine flexiblere Nutzung der Analyseergebnisse erlaubt und zum anderen keine Speicherung aller Daten benötigt.

Nachfolgend sind die nötigen Anpassungen an dem Systemaufbau dargestellt.

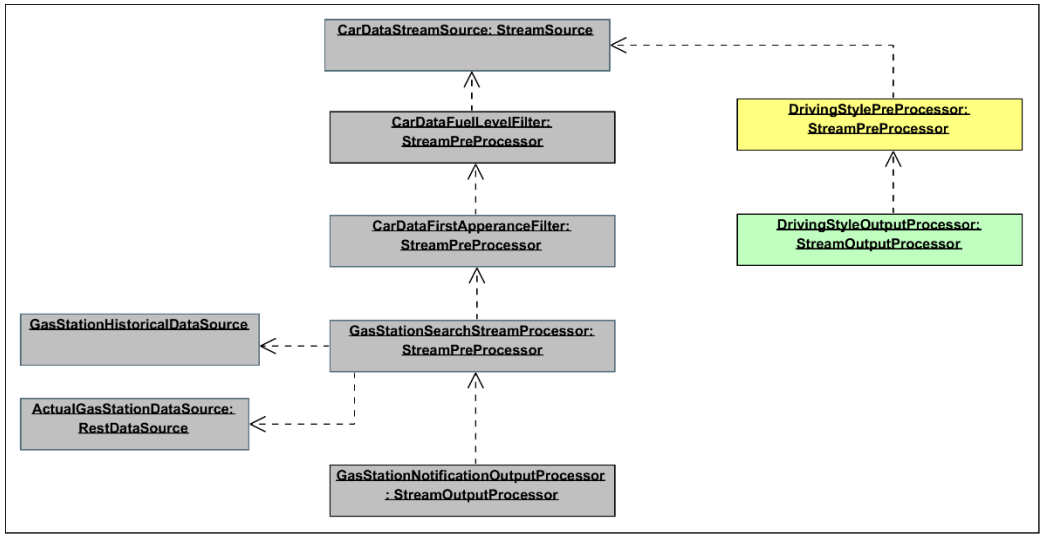

**Abbildung 30: Angepasste Systemarchitektur**

Um die Anforderung umzusetzen, sind keine Änderungen an der grundlegenden Theta Architektur bzw. des STA Frameworks nötig. Es entstehen nur zwei zusätzliche Verarbeitungsschritte. Im *DrivingStylePreProcessor* werden die Fahrzeugdaten einer Fahrt kontinuierlich analysiert und zu einem Wert zusammengefasst, der das Fahrverhalten auf dieser Fahrt beschreibt. Im *DrivingStyleOutputProcessor* werden nach einer Fahrt die Daten aufbereitet und im ServingView abgespeichert.

**M2: Messung der Komplexität**

Für die Messung der zyklomatischen Komplexität des CIS wird auf das Tool scalastyle zurückgegriffen. Hierbei handelt es sich um die Scala Variante des Java Tools StyleChecker. Folgende Werte ergaben sich bei der Auswertung:

| Mittelwert über alle Module | 2,416666667 |
|---|---|
| Mittelwert über alle Verarbeitungsmodule | 2,636363636 |
| Maxwert über alle Module | 4 |
| Maxwert über alle Verarbeitungsmodule | 4 |

**Tabelle 4: Statische Code-Analyse des CIS**

**M3: Abhängigkeiten zwischen Modulen**

Die einzelnen Verarbeitungsmodule haben zum Teil Abhängigkeiten untereinander. So benötigt das *CarDataGasStationSearchStreamProcessing* Modul zwei weitere Verarbeitungsmodule. Des Weiteren benötigen alle Verarbeitungsmodule mindestens ein Modul des Theta Frameworks. Diese Abhängigkeiten werden aber in allen Fällen über Dependency Injection realisiert, wodurch keine direkte Koppelung entsteht.

**Messergebnisse:**

In M1 wurden für die Umsetzung der neuen Anforderung zwei weitere Module benötigt. Beide Module orientieren sich stark an der schon bestehenden Tankstellensuche, wodurch der Aufwand für die Umsetzung einer solchen Anforderung als gering eingeschätzt werden kann. Hier bildet vor allem die Abstraktion der einzelnen Verarbeitungsschritte und die Aufteilung zwischen Theta Framework und der separaten Definition der Verarbeitungsschritte einen Vorteil.

M2 zeigt, dass die Komplexität der einzelnen Module sehr gering ist. So erreicht nur der in Kapitel 5.2.1 vorgestellte Algorithmus einen Wert von 4. Laut Thomas J. McCabe müsste jedoch erst ab einem Wert von 10 über die Überarbeitung eines Moduls nachgedacht werden [Mcca76]. Es gilt zu beachten, dass die McCabe Metrik vor allem für prozedurale Programmiersprachen entwickelt wurde und somit nicht exakt für funktionale Programmiersprachen übernommen werden kann. So wird zum Teil von der Scala Community empfohlen *if-then-else* Konstrukte in *higher-order functions* zu überführen, um so einen McCabe Wert von 1 zu erhalten [SpIr11].

M3 zeigt, dass momentan nur geringe Abhängigkeiten zwischen den Modulen bestehen. Außerdem wird durch die Verwendung von *Dependency Injection* keine Instanziierung von Abhängigkeiten in den einzelnen Verarbeitungsmodulen realisiert, wodurch hier die Komplexität sinkt. Es gilt auch hier zu beachten, dass es sich in der momentanen Realisierung um eine einfache Konstruktor *Dependency Injection* handelt, sodass alle Abhängigkeiten zentral in der Hauptroutine der Anwendung verwaltet werden. Hier wäre bei einer größeren Anwendung eventuell die Nutzung eines *Dependency Injection* Frameworks, wie Google *Guice*, hilfreich.

F1: *Wie viele Änderungen müssen an der bisherigen Architektur durchgeführt werden, um eine neue Funktion zu implementieren?*
Es müssen keine Änderungen an der Architektur durchgeführt werden. Es werden lediglich zwei zusätzliche Verarbeitungsschritte integriert, wodurch zudem keine Anpassungen an die schon bestehenden Verarbeitungsschritte nötig werden.

F2: *Wie viele unterschiedliche Arten von Änderungen lassen sich erkennen?*
Im Fall dieses Anwendungsfalls ist nur die Erstellung neuer Verarbeitungsschritte nötig. Eine Anpassung des STA Frameworks wird vor allem durch die Integration neuer Techniken nötig. Hierbei ist jedoch zu erwarten, dass nur das jeweils betroffene Modul angepasst werden muss.

F3: *Wie hoch ist die Komplexität der einzelnen Module der Anwendung?*
Die Komplexität der einzelnen Module ist zunächst als gering zu betrachten. Hierbei sollte eine genauere Untersuchung der einzelnen Module erfolgen, um eventuelle Optimierungsmöglichkeiten zu nutzen.

F4: *Wie hoch ist die Abhängigkeit der einzelnen Module untereinander?*
Die Abhängigkeit der einzelnen Module untereinander ist gering.

## 6.2.4 Übertragbarkeit

| CIS System | Evaluation | Übertragbarkeit | Entwickler | AWS Umgebung |
|---|---|---|---|---|
| **Qualitätsfaktoren:** Nötige Änderungen am System, um von Spark auf eine andere Verarbeitungsplattform zu wechseln | | | **Einflussfaktoren:** Art der nötigen Änderungen | |
| **Ausgangshypothesen:** Hoher Aufwand, da sich die Verarbeitungsschritte zum Teil direkt auf Spark beziehen und so eine hohe Kopplung besteht. | | | **Einflusshypothesen:** Abstraktionen des Theta Frameworks könnten jedoch auch viele nötige Änderungen erleichtern. | |

| Frage | Metrik |
|---|---|
| F1: Wie viele Änderungen an der Systemarchitektur sind nötig um die Verarbeitungsplattform Spark gegen eine Alternative zu ersetzen? | Konzeptionelle Planung der Umsetzung auf einer anderen Verarbeitungsplattform und Zählen der nötigen Änderungen. |
| F2: Welche Art von Änderung muss am System durchgeführt werden? | Konzeptionelle Planung der Umsetzung auf einer anderen Verarbeitungsplattform und Bewertung der nötigen Änderungen. |

**Messablauf:**
Um die zuvor definierten Fragen zu beantworten, wird die Systemarchitektur konzeptionell so angepasst, dass sie mit der Verarbeitungsplattform Flink kompatibel ist. Flink wurde gewählt, da es - wie im Kapitel 4.2 aufgeführt - eine direkte Konkurrenz zu Spark darstellt. Hierbei sind die jeweiligen Änderungen an der Architektur und die daraus resultierenden technischen Änderungen aufzuführen. Anschließend werden die Art und die Anzahl an nötigen Änderungen bewertet. Da es sich bei Flink um eine komplett andere Verarbeitungsplattform handelt, müssen alle Spark relevanten Operationen durch das jeweilige Flink äquivalent ersetzt werden.

**Migration des STA Frameworks:**
Im Folgenden werden zunächst die einzelnen Komponenten des aktuellen Systems auf die Kompatibilität mit Apache Flink untersucht. Hierbei ist gewünscht, die generelle Theta Architektur und die einzelnen Systeme beizubehalten. So muss das Spark spezifische STA Framework für die Nutzung von Flink angepasst werden.

*Messaging System*:
Für die Nutzung des Messaging System kann direkt auf Flink Funktionalitäten zugegriffen werden. So kann Flink RabbitMQ, aber auch Kafka als Datenquellen nutzen und daraus sogenannte DataStreams[15] erstellen. Hierdurch muss das Messaging Modul im STA Framework angepasst werden.

*Immutable Data Store:*
Der Immutable Data Store wird momentan mittels einer Couchbase Datenbank umgesetzt. Für den Zugriff auf diese Datenbank wird momentan ein natives Spark Modul genutzt, um direkt auf Spark RDDs arbeiten zu können. Flink bietet hier zum Zeitpunkt der Erstellung dieser Arbeit noch keinen nativen Connector an, sodass nur auf die JDBC Funktionalität von Flink zurückgegriffen werden könnte.

*Serving Layer:*
Die Anbindung des Serving Layers lässt sich genauso wie die Anbindung des Immutable Data Store realisieren.

*Processing Service:*
Der Processing Service bildet einen Rahmen für die Batch- bzw. Stream-Verarbeitung, welcher durch die einzelnen Verarbeitungsschritte implementiert wird. Diese Abstraktion muss nicht durch die Nutzung von Flink verändert werden.

**Notification Service:**
Dieses Modul ist vollkommen losgelöst von der Verarbeitungsplattform und nutzt nur das Messaging System, sodass keine Änderungen nötig werden.

**Migration der Tankstellenpreissuche:**
Des Weiteren müssen die auf dem STA Framework aufbauenden Verarbeitungsschritte auf Flink migriert werden. Hierfür wird dies exemplarisch anhand des Tankstellenpreis-UseCases konzeptionell durchgeführt. Für viele der genutzten Spark Streaming Funktionalitäten existieren unter Flink gleichbedeutende Gegenstücke. Dies gilt jedoch beispielsweise nicht für die genutzte *updateStateByKey()* Funktion. In Flink kann der State jedoch über direkte Zustandsbehaftete Operationen realisiert werden. Der *CarDataFirstAppearanceFilter* könnte unter Flink z.B. wie folgt umgesetzt werden:

```
sourceStream.map {(_, 1)}    // Jedes Element wird zu einem Tupel umgewandelt
    .keyBy(0)                // Das Tupel wird nach dem 1. Wert gruppiert.
    .sum(1)                  // Der 2. Wert wird pro Gruppe aufsummiert.
    .filter(x=>x._2==1)      // Es werden alle Elemente gefiltert, die
                             // öfter als einmal im Stream aufgetaucht sind.
```

**Listing 7: Beispiel für die Umsetzung der Vorverarbeitung mittels Flink**

---

[15] DataStreams sind in Flink die interne Stream Abstraktion, vergleichbar zu den von Spark bekannten DStreams.

Dieser Ansatz kann bei Flink genutzt werden, da Flink eine korrekte Record-by-Record Semantik bietet und so jede Nachricht einzeln verarbeitet wird. Dies ist aufgrund des Micro-Batch Ansatzes in Spark Streaming nicht möglich, da hier in einem Bachintervall mehrere Nachrichten desselben Autos auftreten könnten. So würde das Fahrzeug schon im ersten Intervall aus dem Stream gefiltert werden.

**Messergebnisse:**
Generell lässt sich schlussfolgern, dass die Migration des CIS auf die Flink Plattform einen sehr großen Aufwand zur Folge hätte. So müssen viele Module im STA Framework angepasst und alle Verarbeitungsroutinen überarbeitet werden. Hierbei ist es von Vorteil, dass Flink und Spark für die Stream Verarbeitung und für die Batchverarbeitung sehr ähnliche Operationen anbieten, sodass die einzelnen Verarbeitungsschritte meist nur leicht angepasst werden müssen. Leider ist es bei Flink nicht möglich, die Batchverarbeitung (DataSet API) mit der Streamverarbeitung (DataStream API) so leicht wie in Spark zu vereinigen, sodass zum Teil eine Umstrukturierung des Verarbeitungsalgorithmus nötig wird.

F1: *Wie viele Änderungen an der System Architektur sind nötig um die Verarbeitungsplattform Spark gegen eine Alternative zu ersetzen?*
Es müssen in jedem Fall die vier Module, *Messaging System*, *Immutable Data Store, Serving Layer und Processing Service,* des STA Frameworks angepasst werden. Des Weiteren muss jede Verarbeitungsroutine überarbeitet werden.

F2: *Welche Art von Änderung muss am System durchgeführt werden?*
- Änderungen an dem STA Framework – mittlerer Aufwand
- Ersetzen der Spark Verarbeitung Funktionalitäten durch äquivalente Operationen in Flink – geringer Aufwand
- Überarbeiten von Verarbeitungsfunktionalitäten, da Flink bestimmte Arten nicht anbietet – hoher Aufwand.

## 6.3 Evaluation von Spark

Um die Arbeit mit Spark bzw. Spark Streaming selbst zu beurteilen, wird erneut auf das GQM Modell zurückgegriffen, wobei hier vor allem bei den zu bewertenden Qualitätsfaktoren ein anderer Fokus gesetzt wird.

Die folgenden Punkte gelten für alle Messziele:
**Zweck**: Evaluation
**Objekt**: Spark Streaming
**Perspektive**: Nutzer von Spark Streaming
**Kontext**: Spark als Grundlage eines anderen Systems

Bezüglich des Qualitätsfokus wurde hier die Erlernbarkeit, die Bedienbarkeit und die Stabilität von Spark untersucht. Dies entspricht einer Bewertung des Gebrauchs von Spark und dem Gesichtspunkt, wie sich dieses System bei der Integration in ein anderes System verhält. Die für Spark besonders entscheiden Punkte Skalierbarkeit und Fehlertoleranz werden in dieser Evaluation nicht bewertet, da diese Punkte schon für das CIS bewertet wurden und dieses auf Spark aufbaut.

### 6.3.1 Erlernbarkeit

| Spark Streaming | Evaluation | Erlernbarkeit | Nutzer | Spark System |
|---|---|---|---|---|
| **Qualitätsfaktoren:**<br>QF1: Aufwand zum Erlernen der nötigen Konzepte<br>QF2: Anzahl der Probleme, die beim Erlernen des Systems entstehen<br>QF2: Qualität der Dokumentation | | | **Einflussfaktoren:**<br>Wissensstand des Nutzers | |
| **Ausgangshypothesen:**<br>Das System erfordert zwar das Erlernen einiger neuer Konzepte, bietet aber insgesamt eine gute Erlernbarkeit. | | | **Einflusshypothesen:**<br>Der Wissenstand des Nutzers hat einen großen Einfluss auf den Aufwand zum Erlernen des Systems. | |

| Frage | Metrik |
|---|---|
| F1: Wie viel Aufwand ist zum Erlernen des Systems nötig? | M1: Anzahl der nötigen Konzepte |

| F2: Wie hoch ist die Qualität der Dokumentation? | M2: Anzahl an Funktionen, die nicht in der Hauptdokumentation erläutert werden |
|---|---|
| F2: Wie viele Probleme entstehen bei dem Erlernen des Systems? | M3: Stack Overflow Fragen bezüglich Spark |

**Messablauf:**

M1:
Analyse des Sparks bzw. Spark Streaming Entwickler Guides nach vorgestellten Konzepten.

M2:
Von Spark wird ein Programming Guide angeboten, der grundlegende Spark Konzepte vorstellt, aber auch viele API Funktionen direkt erklärt. Es wird untersucht, wie viele API Funktionen in dem Programming Guide vorgestellt werden, wie viele durch Alias-Funktionen erwähnt werden und wie viele gar nicht genannt werden. Hierbei wird zwischen dem Spark Core Guide (RDD Einführung und API) und dem Spark Streaming Guide (DStream Einführung und API) unterschieden.

M3:
Analyse der Suchtreffer bezüglich Spark in der Suchmaschine von Stack Overflow.

**Messergebnisse:**

M1:
Für die Nutzung von Spark Core bzw. Spark Streaming werden die folgenden acht Konzepte benötigt: Resilient Distributed Datasets, Shuffle Operations, RDD Persistence, Shared Variables, Local vs. cluster mode, Discretized Streams, Receivers, Caching/Persistence.

M2:
Die einzelnen API Funktionen mit der jeweiligen Zuordnung sind im Anhang aufgeführt.

|  | In Guide erwähnt | Nicht erwähnt | Alias wurde erwähnt |
|---|---|---|---|
| RDD Funktionen | 35 | 50 | 3 |
| DStream Funktionen | 27 | 9 | 1 |

**Tabelle 5: Analyseergebnisse des Spark Programming Guide**

M3:
Stack Overflow verfügt über 23.040 Fragen bezüglich Spark und 75 Fragen, die jünger als einen Tag sind.

**Interpretation:**
Wie mit M1 dargestellt, müssen für die Nutzung von Spark acht Konzepte erlernt werden, wovon einige wie z.B. RDDs und Local vs. Cluster Mode sehr Spark spezifisch sind. Des Weiteren bauen aber auch Konzepte wie DStreams auf RDDs auf, wodurch der Aufwand des Lernens hier nicht mehr so groß ist.
In M2 wurde festgestellt, dass der Spark Programming Guide für RDDs ca. 40 % und für DStreams ca. 72 % aller Funktionen abdeckt. Bei genauer Betrachtung der Daten zeigt sich, dass einerseits sehr spezielle Funktionen wie z.B. *glom()* nicht angesprochen werden. Dies ist auch sehr plausibel, da diese für das Erlernen von Spark nicht nötig sind. Andererseits werden aber auch ganze Funktionskategorien nicht erwähnt. Wie z.B. *Approx* (für Berechnungen von Näherungswerten), *DoubleRDDFunctions* (bietet verschiedene statische Funktionen an) und *AsyncRDDActions* (ermöglicht asynchrone RDD Output-Operationen). Hier wäre eine Erwähnung im Entwicklerguide durchaus sinnvoll, da so auf gewisse Besonderheiten dieser Funktionen eingegangen werden könnte. Momentan kann ein Entwickler nur über das Java Doc oder den Quellcode auf diese Module aufmerksam werden. Des Weiteren werden als experimentell gekennzeichnete Funktionen, wie z.B *mapWithState()* in der DStream API, nicht im Programming Guide vorgestellt. Dies ist wiederrum nachvollziehbar, da diese Funktionen eventuell noch nicht stabil genug sind. Andererseits wird an anderer Stelle auf sie aufmerksam gemacht, z.B. in Form der Beschreibung der neuen Spark Features in einem Changelog[16]. Dies sorgt so für eine unnötige Diskrepanz.
Die Analyse der Stack Overflow-Fragen in M3 ergibt, dass Spark eine große Anzahl an Stack Overflow-Fragen generiert hat. Dies zeigt vor allem die momentane Marktpräsenz von Spark, aber auch, dass viele Nutzer Probleme bei der Nutzung von Spark haben. Vor allem zeigt die Anzahl der tagesaktuellen Fragen, dass immer wieder neue Probleme bei der Spark Nutzung auftreten, welche nicht durch schon bestehende Antworten gelöst werden konnten.

F1: *Wie viel Aufwand ist zum Erlernen des Systems nötig?*
Bei Spark müssen zwar einige spezifische Konzepte zunächst erlernt werden, trotzdem ist der Einstieg relativ leicht.

F2: *Wie hoch ist die Qualität der Dokumentation?*
Die Qualität des Entwicklerguides ist hoch, wobei bestimmte Themen hier nicht behandelt werden.

F3: *Wie viele Probleme entstehen bei dem Erlernen des Systems?*
Bei dem Erlernen von Spark können Probleme an sehr unterschiedlichen Stellen auftreten. Hierbei verfügt aber Spark über eine sehr große Community auf Plattformen wie Stack Overflow, sodass viele typische Probleme hier schon erklärt wurden.

---

[16] Die Spark Release Notes für Version 1.6.0 stellen das neue Feature mapWithState() vor. Es ist aber nicht im Programming Guide enthalten.
http://spark.apache.org/releases/spark-release-1-6-0.html#spark-streaming Abgerufen am 11.1.2016

## 6.3.2 Bedienbarkeit

| Spark Streaming | Evaluation | Bedienbarkeit | Nutzer | Spark System |
|---|---|---|---|---|

| **Qualitätsfaktoren:** Art der Probleme, die beim Entwickeln des CIS auftraten | **Einflussfaktoren:** Wissensstand des Nutzers |
|---|---|
| **Ausgangshypothesen:** Die Bedienbarkeit von Spark ist zunächst sehr intuitiv, birgt aber einige Problemquellen. | **Einflusshypothesen:** Die Art der Probleme hängt vom Wissensstand des Spark Nutzers und der betrachteten Anwendung ab. |

| Frage | Metrik |
|---|---|
| F1: Wie viele verschiedene Arten von Problemen sind auf bei der Entwicklung des CIS in Bezug auf Spark aufgetreten? | Analyse der aufgetretenen Probleme |

**Messablauf:**
Auflistung und Kategorisierung der bei der Entwicklung des CIS aufwertenden Probleme mit Spark.

**Messergebnisse:**
Im Folgenden werden exemplarisch einige Probleme beschrieben, welche bei der Realisierung des CIS aufgetreten sind.

**Local- vs. Clustermode**
Spark lässt sich zum einen auf einem einzelnen Rechner lokal, aber auch auf einem Rechen-Cluster verteilt ausführen und will hierfür eine gleichförmige Umgebung schaffen, sodass keine Anpassungen am Code nötig sind, um auf ein Cluster zu wechseln. Teilweise entstehen jedoch Probleme, da sich im Loca- Modus, Driver und Executer eine JVM teilen und so auch gemeinsam auf Variablen arbeiten können. Im Cluster-Modus hingegen verfügt jede Komponente über eine eigene JVM, sodass hier nicht so einfach auf Variablen gearbeitet werden kann. Bei der Umsetzung des CIS wurde zunächst eine Lösung gewählt, bei der der Spark Context auch auf den Executern verfügbar gemacht werden musste. Dies sollte über die Serialisierung des Spark Contextes geschehen. Bei einer Erprobung des CIS auf einem Cluster stellte sich heraus, dass dieses nicht möglich war, wodurch hier einige Änderungen nötig wurden, sodass der Spark Context nur noch auf dem Driver verfügbar sein musste.

**Performanceoptimierung**
Durch die Vielzahl bei von Spark angebotenen Funktionen lassen sich dieselben Ziele meist über eine Vielzahl an Wegen realisieren. Hierdurch kam es bei der Realisierung des CIS zum Teil zu Problemen, so ist hier z.B. das in Kapitel 5.3.5 beschriebene Beispiel zu nennen. Hier wurde in einer ersten Realisierung eine Lösung gewählt, welche schon bei einem geringen Anstieg des Nachrichtenaufkommens zu einer Überlastung des Clusters führte und somit nicht skalierbar war. Erst der Verzicht dir aufwendige *join()* Funktion konnte hier eine bessere Lösung bieten.

**Spark Ökosystem**
Spark verfügt mittlerweile über ein umfangreiches Ökosystem, zu dem mehr als 173 Third-Party Projekte[17] gehören. Hierbei kann es zu Problemen kommen, da diese Projekte nicht dieselbe Stabilität wie Spark bieten und die Weiterentwicklung von Spark mit einem sehr hohen Tempo vorangetrieben wird, sodass die Community nicht mit den nötigen Anpassungen an neue Spark Versionen nachkommt und so Inkompatibilitäten entstehen. Während der Realisierung des CIS trat dieses Problem zweimal auf. So war der von der Community angebotene RabbitMQ Connector nur mit Spark 1.3 kompatibel und somit nicht unter Spark 1.5 nutzbar. Hierdurch musst ein eigener RabbitMQ Connector entwickelt werden. Des Weiteren zeigt sich dieses Problem anhand des Couchbase Connectors, der zum einen Kompatibilitätsprobleme mit dem Spark ec-2 Skript hatte und weitere kleine Fehler enthielt, welche vor einer Nutzung gelöst werden mussten. Außerdem kam es, wie unter 5.3.1 beschrieben, zu Problemen mit dem Spark SQL JDBC Adapter und der Nutzung von PostgreSQL. Dies zeigt, dass Spark zum Teil viele Funktionen unterstützt, welche aber zum Teil noch nicht ganz ausgereift sind.

**Interpretation:**
Die gesammelten Projekterfahrungen zeigen, dass bei der Nutzung von Spark verschiede Probleme auftreten können.

F1: *Wie viele verschiedene Arten von Problemen sind auf bei der Entwicklung des CIS in Bezug auf Spark aufgetreten?*
Bei der Umsetzung des CIS sind vor allem drei Arten von Problemen aufgetreten: Probleme bezüglich des Local- bzw. Clustermodes und der Skalierbarkeit von Verarbeitungsalgorithmen, diese wurden mit der Zeit und dem Zugewinn an Erfahrung im Umgang mit Spark aber immer seltener. Außerdem würde hier ein *Continuous Deployment* Vorgehen helfen, da so immer geprüft werden kann, ob die momentan aktuelle Version auch auf einem Produktivsystem funktionstüchtig ist. Das Problem der Third-Party Projekte ist durchaus größer, da man persönlich keinen direkten Einfluss auf deren Entwicklung hat. Aufgrund des Open Source Charakters der meisten Spark Projekte ist aber eine Anpassung dieser immerhin möglich. Trotzdem sollte dies immer vor der Integration eines Third-Party Projektes bedacht werden.

---
[17] 173 Spark Projekte sind über die Seite spark-pakages.org gemeldet. Abgerufen 11.01.2016.

### 6.3.3 Stabilität

| Spark Streaming | Evaluation | Stabilität | Nutzer | Spark System |
|---|---|---|---|---|
| **Qualitätsfaktoren:**<br>QF1: Menge an bekannten Fehlern<br>QF2: Zeit zwischen dem Melden und Beheben von Fehlern<br>QF3: Anzahl von Fehlern zwischen einzelnen Versionen | | | **Einflussfaktoren:**<br>Unbekannte Fehler | |
| **Ausgangshypothesen:**<br>Die Entwicklung von Spark schreitet schnell voran und es sollten nur wenige Fehler im System vorhanden sein. | | | **Einflusshypothesen:**<br>Neben den bekannten Fehlern kann es noch mehrere unbekannte Fehler geben | |

| Frage | Metrik |
|---|---|
| F1: Wie hoch ist die Anzahl an noch ungelösten Fehlern? | M1: Analyse des Spark Issue Trackers und der Changelogs |
| F2: Wie viele Fehler hat im Durchschnitt eine Spark Major-Versionen? | M2: Zählen der Fehler, die im Durchschnitt durch eine Spark Minor-Version behoben wurden. |
| F3: Wie viel Zeit vergeht zwischen dem Release von Versionen? | M3: Analyse der Release Daten von Spark minor-Version und Major-Version |

**Messablauf:**
Alle Daten wurden am 11.01.2016 anhand des öffentlichen Jira-Portals von Spark erfasst.
Für M3 werden die Fehler einer Spark Major-Version gezählt, welche durch eine darauf aufbauende Minor-Version gefixt wurden. Hierbei handelt es sich meist um kritischere Fehler, da sie noch vor dem nächsten Major-Release behoben werden mussten. Daher lässt sich hierdurch abschätzen, wie viele Fehler durch die neue Spark Hauptversion entstanden sind.

**Messergebnisse:**

| M1: Anzahl ungelöster Fehler | 707 |
|---|---|
| M2: Durchschnittliche Fehlerzahl einer Spark Major-Version | 57,5 |
| M3: Durchschnittliche Zeit zwischen Spark Versionen | ca. 1,5 Monate |
| M3: Durchschnittliche Zeit zwischen Major Spark Versionen | ca. 3 Monate |

**Tabelle 6: Analyseergebnisse der Spark release Daten**

**Interpretation:**

F1: *Wie hoch ist die Anzahl an noch ungelösten Fehlern?*
Die Anzahl momentan ungelöster Fehler betrug zum Zeitpunkt der Datenerhebung 707. Hierbei ist vor allem zu beachten, dass die Daten kurz nach dem Release von Spark 1.6.0 erhoben wurden, wodurch eventuell hier besonders viele neue Fehler gemeldet wurden. Des Weiteren zeigen die Daten aber auch, dass Spark noch ein großes Verbesserungspotenzial beinhaltet.

F2: *Wie viele Fehler hat im Durchschnitt eine Spark Major-Versionen?*
Im Durchschnitt lässt sich feststellen, dass eine Spark Major-Version im Durchschnitt 57 Fehler beinhaltet, welche durch eine Minor-Version behoben werden müssen. Dies zeigt auch, dass durch die Nutzung der ersten Version eines neuen Hauptversionszweiges also z.B. Version 1.5.0 Fehler entstehen können. Hier ist ein ausführliches Testen ratsam. Außerdem könnte vor einer Migration auf eine neue Hauptversion erst auf die erste Minor-Version gewartet werden, sodass die kritischsten Fehler bereits beseitigt wurden.

F3: *Wie viel Zeit vergeht zwischen dem Release von Versionen?*
Im Schnitt vergehen zwischen dem Release von Spark Major-Versionen ca. 3 Monate, wobei ca. jeden Monat eine Minor-Version erscheint. Dies verdeutlicht noch einmal das hohe Tempo, mit dem die Spark Entwicklung vorangetrieben wird. Außerdem zeigt es, dass ein möglichst guter Migrationsprozess erstellt werden sollte, sodass ohne großen Aufwand auf neue Spark Versionen gewechselt werden kann. Sollten nämlich zu viele Versionen ausgelassen werden, kann es eventuell zu größeren Migrationsproblemen kommen, sodass hier der Aufwand weiter steigt.

## 6.4 Auswertung

Im Folgenden wird die abschließende Auswertung der vorherigen Evaluationen durchgeführt. Dafür werden zunächst die Kernergebnisse der Evaluation in folgender Tabelle dargestellt:

| Qualitätsfokus | Ergebnis |
| --- | --- |
| **Skalierbarkeit** | - Maximaler Durchsatz steigt mit Knotenanzahl an<br>- Durchsatz in Abhängigkeit zur Lastart<br>- Durchsatz in Abhängigkeit zur Softwarekomponente |
| **Fehlertoleranz** | - Master: Keine Auswirkung auf die laufende Anwendung<br>- Worker: Verlangsamung des Gesamtsystems<br>- Driver: Komplettes Neustarten der Anwendung |
| **Änderbarkeit** | - Der Aufwand für die Umsetzung einer neuen User-Story ist aufgrund der Modularisierung des CIS gering<br>- Die CIS & STA Framework Module sind in ihrer Komplexität gering |
| **Übertragbarkeit** | - Hoher Aufwand für die Migration der Streamverarbeitungsplattform auf Flink, da alle Komponenten des STA Frameworks und die Verarbeitungsschritte angepasst werden müssen |
| **Erlernbarkeit** | - Für Spark müssen mehrere spezifische Konzepte erlernt werden<br>- Die Dokumentation ist umfangreich, wobei einige Themen nicht angesprochen werden |
| **Bedienbarkeit** | - Probleme bei der Nutzung von Spark können mit dem Clusterbetrieb, der Performanceoptimierung und des Spark Ökosystems entstehen |
| **Stabilität** | - Spark entwickelt sich schnell weiter, jede Version beinhaltet neue Fehler, welche aber auch schnell behoben werden. |

**Tabelle 7: Ergebnisse der Evaluation**

Die Evaluation des CIS zeigte vor allem, dass hier die im Kapitel 5.2.2 festgelegten Anforderungen an das System erfolgreich umgesetzt wurden, so ist das System skalierbar und fehlertolerant. Hierbei ist jedoch anzumerken, dass diese Kriterien nicht uneingeschränkt gelten. So zeigt der Test der Skalierbarkeit im Kapitel 6.2.1, dass diese unter anderem stark von der Umsetzung mittels Spark und dem jeweiligen zu lösenden Problem abhängt. Außerdem wurde in Punkt 6.3.2 gezeigt, dass eine vermeintlich einfache Problemlösung mitunter zu großen Skalierungsproblemen führen kann. Bezüglich der Fehlertoleranz zeigt die Evaluation in Kapitel 6.2.2, dass diese durchaus erreichbar ist, aber ein aufwendigeres System-Setup fordert. Des Weiteren zeigt dieser Test, dass der Ausfall eines Knotens zwar kompensiert werden kann, es dadurch aber für einen gewissen Zeitraum zu einer Verlangsamung des Gesamtsystems kommt. Bezüglich der in 6.2.3 evaluierten Änderbarkeit wurde gezeigt, dass das STA Framework und das darauf aufbauende CIS eine gute Änderbarkeit bieten. So zeigte sich anhand der Konzeption einer weiteren User Story, dass diese ohne Anpassungen am schon bestehenden System realisiert werden kann. Hierbei ist darauf hinzuweisen, dass dies von der jeweiligen zu realisierenden User Story abhängt. Generell erleichtern aber die vom STA Framework angebotenen Abstraktionen solche Anpassungen. Des Weiteren wurde anhand zweier Code Metriken festgestellt, dass die einzelnen Module eine geringe Komplexität und Koppelung zu anderen Modulen aufweisen. Bezüglich der Übertragbarkeit wurde in Punkt 6.2.4 untersucht, mit welchem Aufwand sich das CIS von Spark auf die Verarbeitungsplattform Flink übertragen ließe. So wurde hierbei festgestellt, dass zwar viele der benötigten Funktionen durch Flink angeboten werden und in vielen Teilen nur die jeweiligeren STA Framework Module angepasst werden müssten, aber trotzdem ein hoher Aufwand durch den Plattformwechsel entsteht. Es ist festzustellen, dass die Wahl der Verarbeitungsplattform sehr wichtig ist und eventuell nur schwer revidierbar ist. Es wäre also wünschenswert, über eine Möglichkeit zu verfügen, die Verarbeitungslogik abstrakter zu formulieren, sodass die eigentliche Verarbeitungsplattform einfacher ausgetauscht werden kann. Hier wäre eine zukünftige Erprobung des von Google vorgestellten Beam Projektes denkbar. Dieses Projekt hat sich zum Ziel gesetzt, mittels einer *Domain Specific Languages* Verarbeitungsschritte auf verschiedenen Plattformen zu realisieren [Apac16i]. Des Weiteren wäre es interessant, das STA Framework für verschiedene Verarbeitungsplattformen anzupassen, sodass hier pro Verarbeitungsservice eine andere Plattform genutzt werden kann. Dies würde es ermöglichen, die einzelnen Verarbeitungsplattformen flexibel anhand ihrer Eignung für den jeweiligen Use Case auszuwählen.

Die Evaluation von Spark in Punkt 6.3 hat einige interessante Punkte näher betrachtet. So wurde unter anderem die Erlernbarkeit von Spark untersucht, wobei festgestellt wurde, dass Spark zum einen einige spezielle Konzepte nutzt, die für eine Nutzung verinnerlicht werden müssen. Zum anderen bietet Spark auch eine umfangreiche Dokumentation, die die wichtigsten Konzepte darstellt. Hierbei wurde angemerkt, dass diese noch über tiefergehende Themen erweitert werden könnte, damit auch alle Möglichkeiten von Spark zumindest angerissen werden. In Punkt 6.3.2 wurde auf die Bedienbarkeit von Spark eingegangen, wobei hier der Fokus auf die bei der Nutzung von Spark entstandenen Probleme gelegt wurde. Nach einer Beschreibung einiger exemplarischer Probleme und deren Lösung wurde hier geschlussfolgert, dass ein großer Teil dieser durch den fortschreitenden Erfahrungsgewinn im Lauf eines Projektes vermieden werden können. Außerdem wird ein schneller Erprobungszyklus z.B. mittels *Continuous Deployment* empfohlen, um schnell auf mögliche Probleme reagieren zu können. So kann durch Continuous Deployment und automatischen Lasttest z.B. die Skalierbarkeit für neue Verarbeitungsalgorithmen überprüft werden. Des Weiteren wurde auch ausgeführt, dass Spark Third-Party Bibliotheken zu Problemen werden können. So wird hier zum einem nicht dasselbe Maß an Stabilität wie bei Spark geboten und es kann zu Inkompatibilitäten bei neueren Spark Versionen kommen. Dieser Punkt wurde auch bei der Evaluation der Stabilität in Punkt 6.3.3 betrachtet. So wurde hier das extrem schnelle Fortschreiten der Spark Entwicklung hier betrachtet, wodurch es vorkommen kann, dass

Third-Party Entwickler, aber auch man selbst nicht mit den Anpassungen an neue Spark Versionen nachkommen. Beim Auslassen mehrerer Major-Versionen steigt der Migrationsaufwand wahrscheinlich weiter. Des Weiteren wurde hier aber auch gezeigt, dass durch jede neue Spark Major-Version neue Fehler entstehen, welche erst in der darauffolgenden Minor-Version behoben werden. So kann es Sinn machen, vor einer Migration auf eine neue Hauptversion immer erst das erste Minor-Update dieser abzuwarten. Durch die hohe Taktung an Spark Releases ist eine möglichst automatisierbare Migration und der damit verbundenen Tests empfehlenswert, da nur so der Migrationsaufwand begrenzt werden kann.

# 7 Fazit und Ausblick

Ziel der vorliegenden Arbeit war es, Spark Streaming mittels eines praxisrelevanten Anwendungsfalls zu erproben und systematisch zu bewerten. Hierfür wurde das Car Information System (CIS) anhand der „Tankstellensuche"-User-Story konzipiert, realisiert und anschließend evaluiert. Das CIS wertet Fahrzeugdaten fahrender Autos in nahezu Echtzeit aus und eignet sich daher als exemplarischer Vertreter der breiten Anwendungsklasse der Stream Verarbeitungssysteme, welche mittlerweile eines der relevantesten Bereiche im Rahmen der Big Data Analyse bilden. Mittels diesen Stream Verarbeitungssystemen wird es z.B. erst möglich kontinuierlich wachsende Datenmengen innerhalb einer kurzen Latenzzeit zu verarbeiten. Mittlerweile existieren verschiedenste Stream Verarbeitungssysteme, welche jeweils verschiedene Vor- und Nachteile haben. Für die Umsetzung des CIS wurde hierbei Spark Streaming verwendet, da es eines der momentan am häufigsten genutzten Open-Source Stream Verarbeitungssysteme ist.

Als Systemarchitektur wurde die Theta-Architektur anhand mehrerer im Big Data Umfeld bekannter Architekturen hergeleitet, diese kombiniert Architektureigenschaften der Lambda-, Kappa- und IoT-Architektur. Um eine gleichförmige Basis für die einzelnen Softwarekomponenten zu erhalten und somit eine gute Änderbarkeit des Systems zu erreichen, wurde das Spark Theta Architektur (STA) Framework erstellt, welches die einzelnen in der Theta Architektur definierten Systemkomponenten abstrahiert und feste Schnittstellen für die Kommunikation mit diesen anbietet. Zum Abschluss der Arbeit wurde in einer Diskussion das CIS und Spark mit Hilfe des *Goal Question Metric*-Vorgehensmodells evaluiert. Hierbei wurde sich auf verschiedene Softwarequalitätskriterien konzentriert. Für das Testen der Fehlertoleranz und der Skalierbarkeit wurde das CIS auf einer AWS Infrastruktur aufgesetzt. Die Ergebnisse der Evaluation zeigten, dass mithilfe der Abstraktionen des STA Framework eine gute Änderbarkeit und mittels Spark Streaming eine hohe Fehlertoleranz und Skalierbarkeit erreicht werden konnte. Anderseits zeigte die Auswertung auch auf, dass das schnelle Voranschreiten der Spark Entwicklung einige Probleme bereiten kann und somit eine hoch automatisierte Infrastruktur für die Entwicklung eines größeren Projekts nötig wird. Des Weiteren zeigt sich, dass z.B. durch Flink eine starke Konkurrenz zu Spark Streaming besteht, sodass vor der Wahl der jeweiligen Verarbeitungsplattform eine genaue Prüfung der eigenen Anforderungen durchgeführt werden sollte, da ein nachträglicher Wechsel der Verarbeitungsplattform viel Aufwand kosten kann. So fordern bestimmte Anwendungsfälle unter anderem sehr geringe Latenz oder die Verarbeitung eines ungeordneten Streams, sodass hier nicht Spark Streaming verwendet werden kann. Andererseits bietet Spark Streaming im Vergleich zu anderen Systemen eine große Verbreitung, wodurch bei Nutzung von Spark Streaming auf eine größere Community zurückgegriffen werden kann.

Während dieser Arbeit haben sich einige interessante Punkte ergeben, welche in der Zukunft weiter untersucht werden sollten. So wurde im Rahmen dieser Arbeit eine grundlegende Architektur für die Umsetzung von Stream Verarbeitungsanwendungsfällen und eine darauf aufbauende Referenzimplementierung umgesetzt. Diese könnte beispielsweise erweitert werden, um so andere Verarbeitungsplattformen zu unterstützen. Hierbei wäre aber auch eine genaue Untersuchung von Spark Streaming mit vergleichbaren Verarbeitungsplattformen interessant, um so Vor- und Nachteile der jeweiligen Produkte genauer darzustellen und somit Nutzern eine Entscheidungshilfe anzubieten. Es wäre außerdem interessant, eine abstraktere Verarbeitungsplattform zu schaffen, durch welche es möglich wäre die jeweilige grundlegende Plattform austauschbar zu machen. Ein weiterer interessanter Punkt ist die Realisierung von größeren Softwareprojekten auf der Basis von Spark. Hier wäre zum einen eine Verbesserung der Automatisierbarkeit der nötigen Infrastruktur, z.B. mittels *Continuous Deployment* Tools denkbar. Des Weiteren fehlt Spark bzw. Spark Streaming, nach Meinung des Autors, eine umfangreiche Testunterstützung. So ist dies besonders bei der Streamverarbeitung interessant, da es hier möglich sein müsste, bestimmte Testszenarien definieren zu können, um so die jeweiligen Verarbeitungsalgorithmen zu überprüfen. Abschließend lässt sich feststellen, dass die Nutzung von Streamverarbeitungsplattformen im Big Data Bereich weiter zunehmen wird und somit eine weitere Untersuchung dieser Plattformen oder damit verbundener Tools eine hohe Praxisrelevanz aufweist.

# Glossar

Cluster - Der Begriff Cluster bezeichnet im Rahmen dieser Arbeit den Zusammenschluss von mehreren Rechnern zu einem System, welche zusammen einen gewissen Dienst erbringen.

Deployment - Dies ist die Verteilung bzw. Inbetriebnahme einer Software auf einer gegebenen Infrastruktur.

Dependency Injection - Dies ist ein spezielles Entwurfsmuster in der objektorientierten Programmierung, wobei Abhängigkeiten an einer zentralen Stelle instanziiert werden und dann an die nutzenden Klassen zur Laufzeit weitergereicht werden.

Docker – Ist ein Produkt, welches es ermöglicht Anwendungen in Container zu isolieren und diese auf verschiedenen Plattformen lauffähig zu machen. Es ist vergleichbar mit der Virtualisierung von Anwendungen.

DTO – Sind Datentransferobjekte und kapseln Datenstrukturen über Modulgrenzen hinweg.

HDFS - Hadoop Distributed File System: Ein über mehrere Rechenknoten verteiltes hochverfügbares Dateisystem, welches das Speichern großen Datenmengen erlaubt.

In-Memory – Hierbei wird versucht alle Daten im Arbeitsspeicher des Rechners vorzuhalten um einen zeitintensiven Festplattenzugriff zu vermeiden. Eine Herausforderung ist es hierbei eine hohe Fehlertoleranz zu gewährleisten, da der Arbeitsspeicher nicht persistent ist.

Jira – Jira ist eine Webanwendung die zum Verwalten von Software-Projekten eingesetzt wird. Sie ermöglicht unter anderem eine Fehlerverwaltung und Aufgabenplanung.

Microservices – Microservices bilden einen Teil eines speziellen Architekturmusters bei denen das Gesamtsystem in verschiedene unabhängige und entkoppelte Prozesse unterteilt wird, welche jeweils eine Teilaufgabe erledigen.

Write-Ahead-Log - Bevor eine Nachricht vom System weiterverarbeitet wird bzw. ihr empfang an die Datenquelle quittiert wird, wird diese zunächst persistent gespeichert, um in einem Ausfall diese hier wiederherstellen zu können.

# Abbildungsverzeichnis

# Tabellenverzeichnis

# Listings

# Anhang

## a     Lasttest CIS

In diesem Kapitel werden die einzelnen Messergebnisse des Lasttests dargestellt.

**Testgrafiken**
Folgend eine Vorstellung der verwendeten Grafiken mit kurzer Erläuterung:

**Verarbeitete Nachrichten / Zurückgestellte Batches**
Anhand dieser Grafik ist die Anzahl der zu verarbeitenden Nachrichten im Vergleich zu den wartenden Batches zu erkennen. In Spark Streaming findet die Verarbeitung der jeweiligen Batches immer sequenziell statt. Sollte nun die Verarbeitungszeit eines Batches länger als das konfigurierte Batch Intervall sein, dann muss mit der Verarbeitung von neuen Batches gewartet werden. Es kann auch vorkommen, dass einzelne Batches mal länger dauern und dass hierdurch auch wartende Batches entstehen. Dies ist jedoch erst kritisch, sobald die Anzahl an wartenden Batches so stark wie im nebenliegenden Schaubild ansteigen. Die Verarbeitungsdauer einzelner Batches sollte also im Mittel unter der Batch Intervalldauer liegen, um eine zuverlässige Verarbeitung zu ermöglichen.

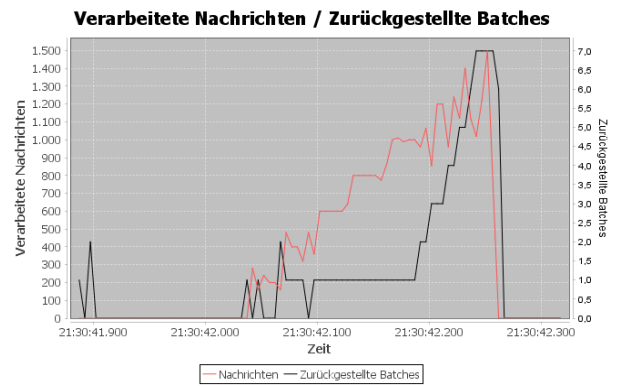

**Abbildung Anhang 1: Beispiel Grafik 1**

**Verarbeitungsdauer / Nachrichten**
In der rechts abgebildeten Grafik wird die Verarbeitungsgeschwindigkeit der jeweiligen Batches im Vergleich zu den verarbeiteten Nachrichten dargestellt. Hierbei ist besonders der Anstieg der Verarbeitungszeit im Vergleich zum Anstieg der verarbeiteten Nachrichten interessant. Außerdem gilt für alle Tests eine Batch Intervalldauer von einer Sekunde. Zudem lässt sich hier erkennen, wann die Verarbeitungsdauer die kritische Schwelle der Intervalldauer überschreitet.

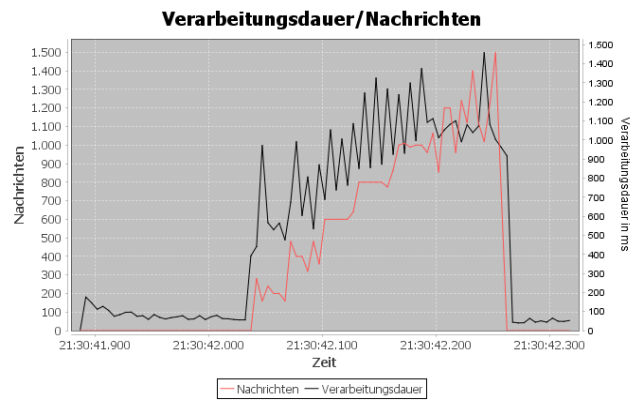

**Abbildung Anhang 2 Beispiel Grafik 2**

## a.a 2 Worker

Im Folgenden wird ein Spark System mit zwei Worker Knoten und einem Master Knoten betrachtet.

**Vorverarbeitungsschritt**:

- **LS1**

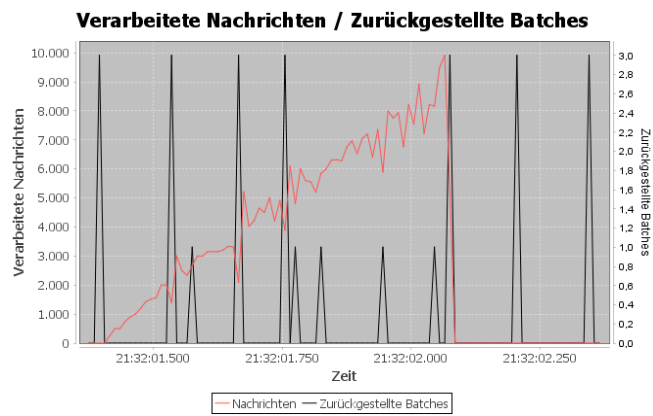

Abbildung Anhang 3: 2 Worker - LS1 - Grafik 1

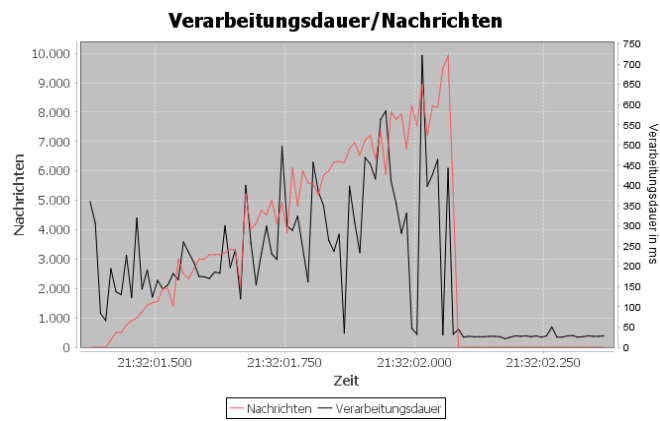

Abbildung Anhang 4: 2 Worker - LS1 - Grafik 2

Bei diesem ersten Test mit gleichförmigen Daten lässt sich feststellen, dass
diese Art von Daten sehr wenig Last für die Anwendung erzeugt. So lassen sich hier 10.000 Nachrichten pro Sekunde verarbeiten und es werden nie mehrere Batches hintereinander zurückgestellt. Auch in der zweiten Grafik ist zu sehen, dass die Verarbeitungsdauer im Schnitt bei ca. 350ms bleibt.

- **LS2:**

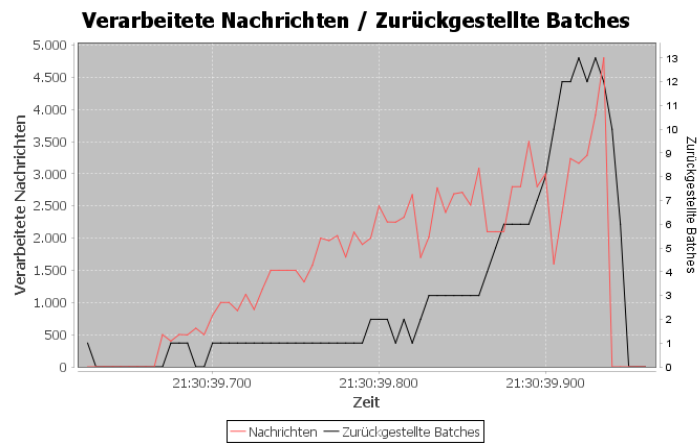

**Abbildung Anhang 6: 2 Worker - LS2 - Grafik 1**

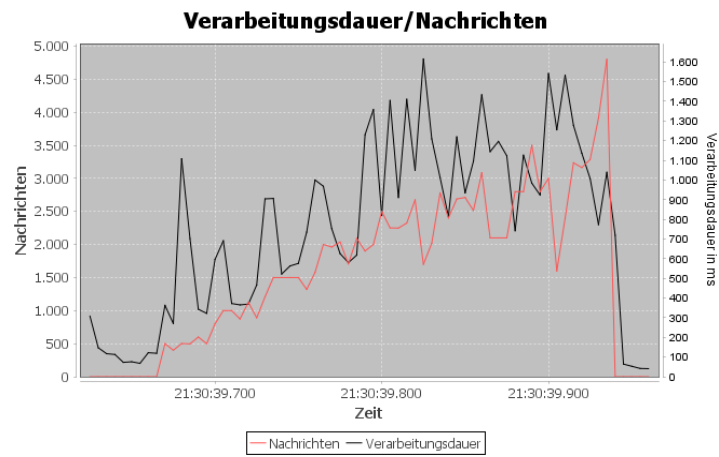

**Abbildung Anhang 5: 2 Worker - LS2 - Grafik 2**

In der ersten Grafik ist ein deutlicher Anstieg der zurückgestellten Batches ab ca. 4000 zu verarbeitenden Nachrichten zu erkennen. Dies spiegelt sich in der zweiten Grafik wider. Hier ist zu sehen, dass die Verarbeitungsdauer stark mit dem Wachsen der Nachrichten Anzahl ansteigt und gegen Ende im Durchschnitt über 1000 ms liegt.

- **Einzigartige Daten**

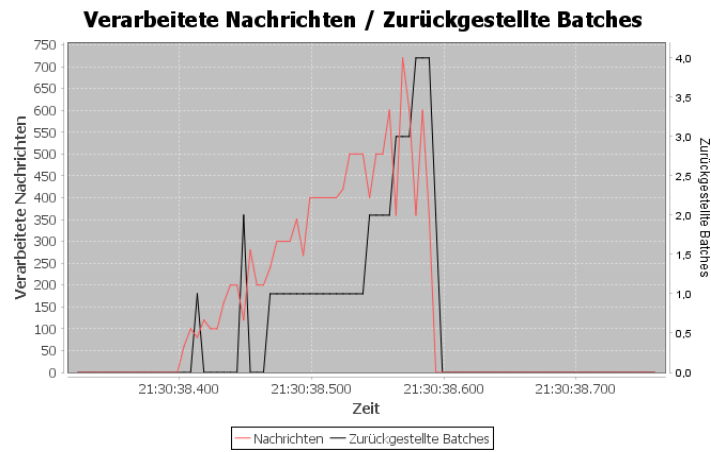

**Abbildung Anhang 7: 2 Worker - LS3 - Grafik 1**

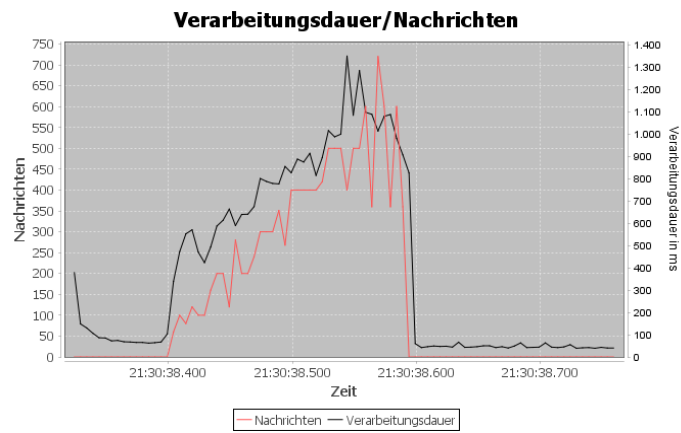

**Abbildung Anhang 8: 2 Worker - LS3 - Grafik 2**

In diesen Grafiken ist zu erkennen, dass dieses Lastszenario den Arbeitsaufwand extrem ansteigen lässt. So wird schon in der zweiten Lasterhöhungsstufe bei ca. 400 Nachrichten der kritische Punkt der Anwendung erreicht und es werden einzelne Batches zurückgestellt.

Schlussendlich lässt sich besonders durch diesen Test zeigen, dass die Art der Lastszenarien einen erheblichen Einfluss auf die Performance des Systems hat. So ist festzustellen, dass bei gleichförmigen Daten ein sehr hoher Durchsatz erreicht wird, da selbst in einer Konfiguration mit 2 Workern das System mehr als 10.000 Nachrichten verarbeiten konnte. Daher wird im Folgenden auf dieses Lastszenario verzichtet.

**Tankstellensuche**:

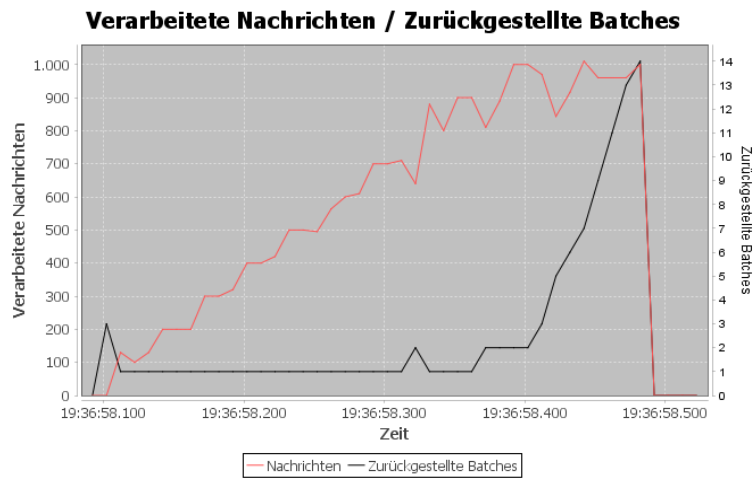

**Abbildung Anhang 9: 2 Worker - Tankstellensuche - Grafik 1**

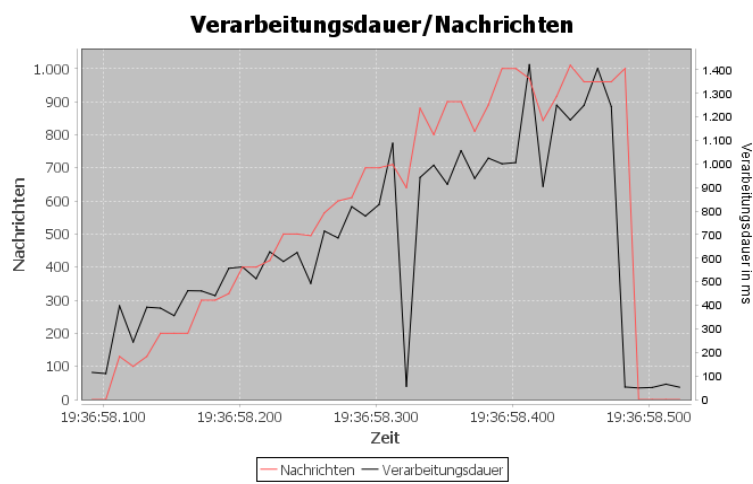

**Abbildung Anhang 10: 2 Worker - Tankstellensuche - Grafik 2**

Bei den Daten der Tankstellensuche lässt sich zunächst feststellen, dass die Verarbeitungsdauer eines Batches direkt mit der Anzahl an zu verarbeitenden Nachrichten wächst. So wird ab ca. 800 zu verarbeitenden Nachrichten bzw. zu suchenden Tankstellen die Verarbeitungszeit länger als die Batch Intervalldauer, sodass Batches zurückgestellt werden.

## a.b 4 Worker

Im Folgenden wird ein Spark System mit 4 Worker Nodes betrachtet.
**Vorverarbeitungsschritt**:

- **LS2:**

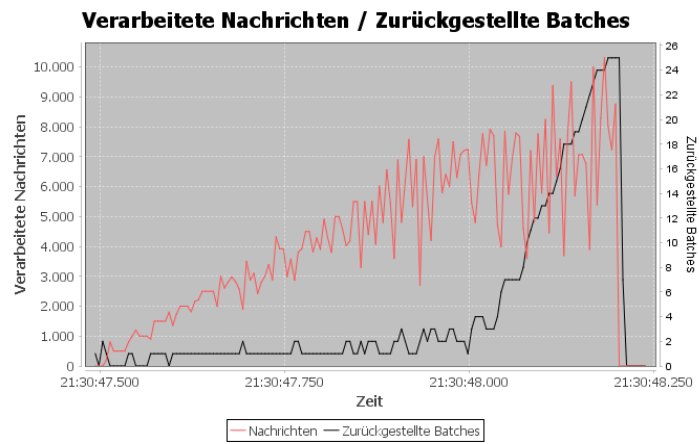

**Abbildung Anhang 12: 4 Worker – LS2 - Grafik 1**

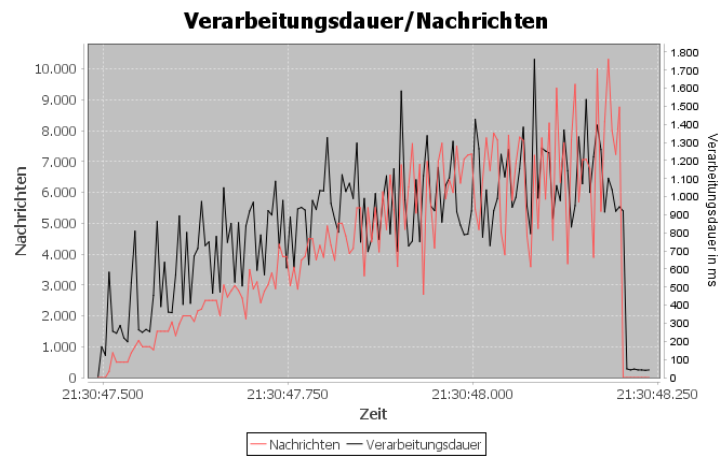

**Abbildung Anhang 11: 4 Worker – LS2 - Grafik 2**

In der ersten Grafik ist ein deutlicher Anstieg der Waiting Batches ab ca. 5500 zu verarbeiteten Nachrichten zu erkennen. Dies spiegelt sich auch in der zweiten Grafik wider. Hier ist zu sehen, dass Verarbeitungsdauer stark mit dem Wachsen der Nachrichtenanzahl ansteigt.

- **LS3:**

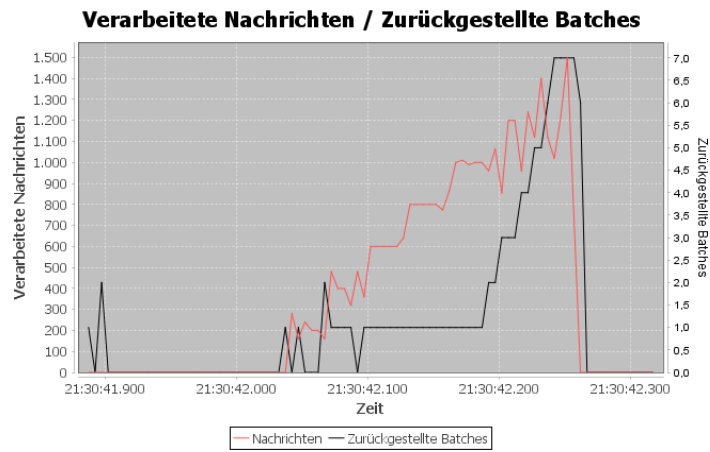

**Abbildung Anhang 13: 4 Worker - LS3 - Grafik 1**

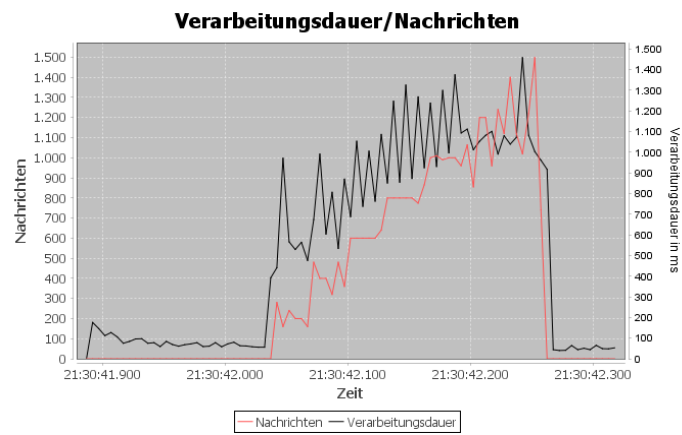

**Abbildung Anhang 14: 4 Worker - LS3 - Grafik 2**

Hierbei ist zu erkennen, dass bei dieser Art von Workload der Arbeitsaufwand extrem ansteigt. So wird schon in der zweiten Lasterhöhungsstufe bei ca. 1000 Nachrichten der kritische Punkt der Anwendung erreicht und es werden einzelne Batches zurück-gestellt.

In beiden Tests konnte durch die Skalierung des Systems eine Erhöhung des Durch-satzes erreicht werden.

**Tankstellensuche**:

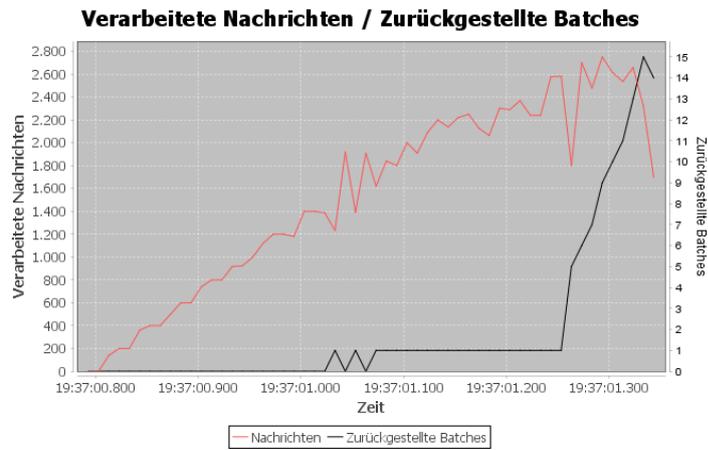

**Abbildung Anhang 15: 4 Worker - Tankstellensuche - Grafik 1**

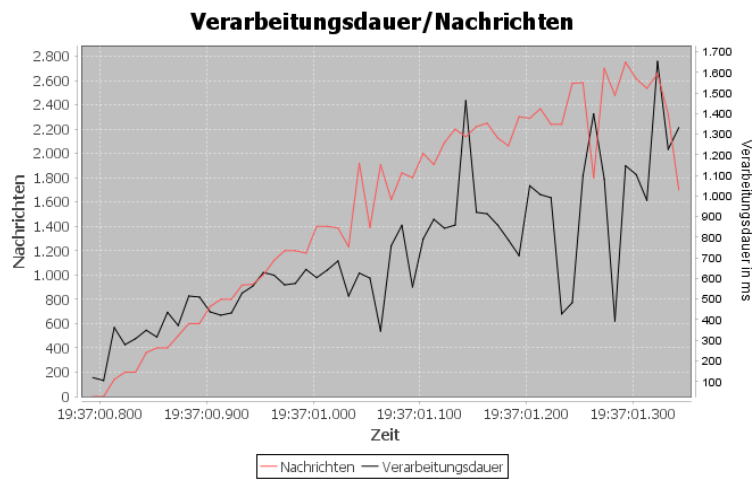

**Abbildung Anhang 16: 4 Worker - Tankstellensuche - Grafik 2**

In dieser Konfiguration können ca. 2600 Nachrichten verarbeitet werden.

## a.c    8 Worker

Im Folgenden wird ein Spark System mit 8 Worker Nodes betrachtet.

**Vorverarbeitungsschritt**:

- **LS2:**

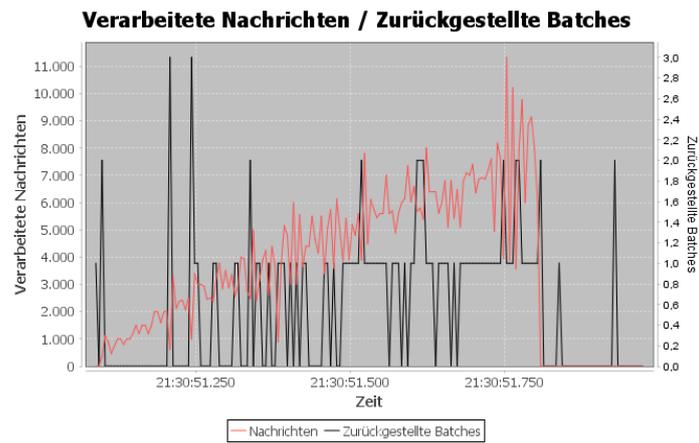

Abbildung Anhang 17: 8 Worker - LS2 - Grafik 1

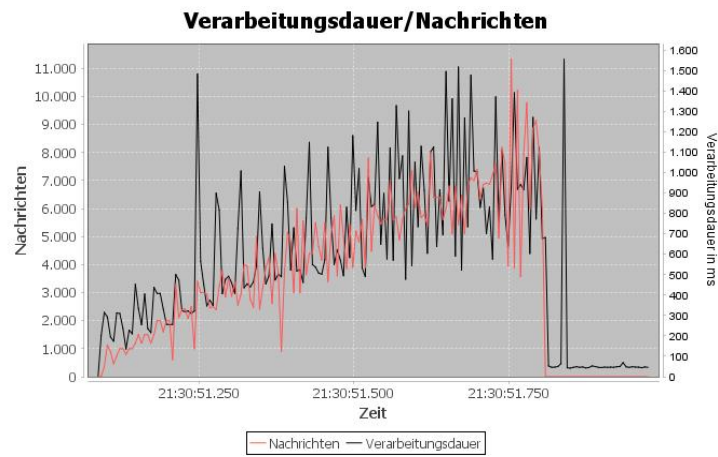

Abbildung Anhang 18: 8 Worker - LS2 - Grafik 2

Die Grafiken zeigen, dass bis zu der hier getesteten Last von 10.000 Nachrichten keine Batches verzögert werden.

- **LS3:**

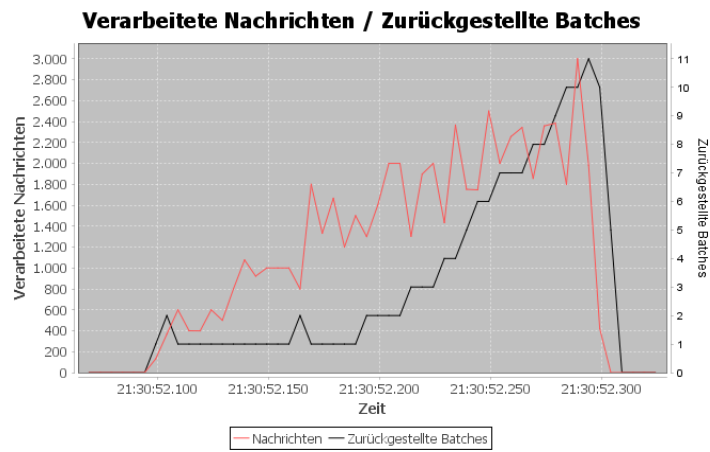

**Abbildung Anhang 19: 8 Worker - LS3 - Grafik 1**

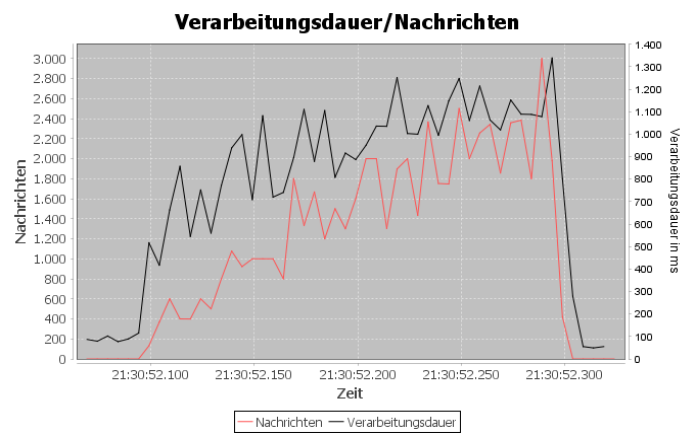

**Abbildung Anhang 20: 8 Worker - LS3 - Grafik 2**

In beiden Tests konnte durch die Skalierung des Systems eine Erhöhung des Durchsatzes erreicht werden.

**Tankstellensuche**:

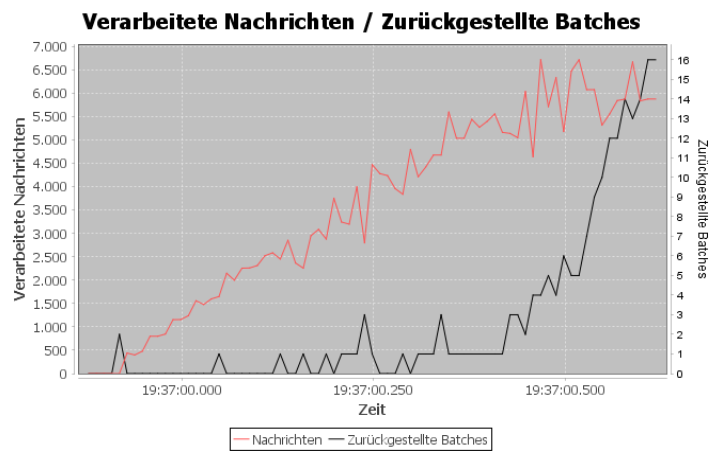

**Abbildung Anhang 21: 8 Worker - Tankstellensuche - Grafik 1**

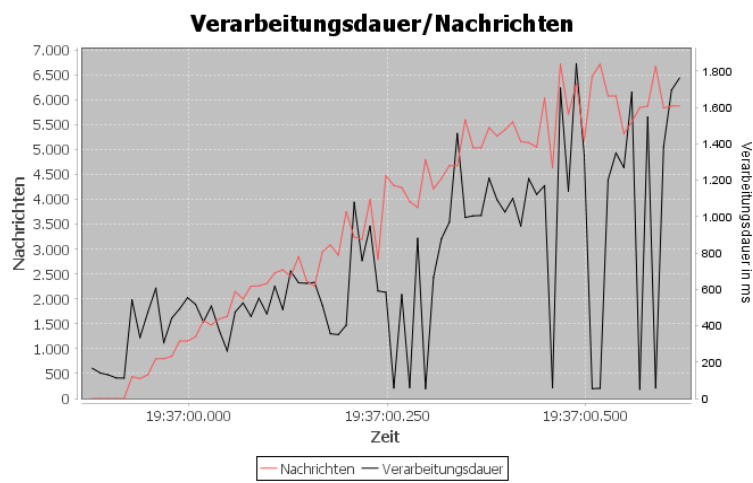

**Abbildung Anhang 22: 8 Worker - Tankstellensuche - Grafik 1**

In dieser Konfiguration können ca. 5500 Nachrichten verarbeitet werden.

## a.d 16 Worker

Im Folgenden wird ein Spark System mit 16 Worker Nodes betrachtet.

**Vorverarbeitungsschritt**:

- **LS2:**

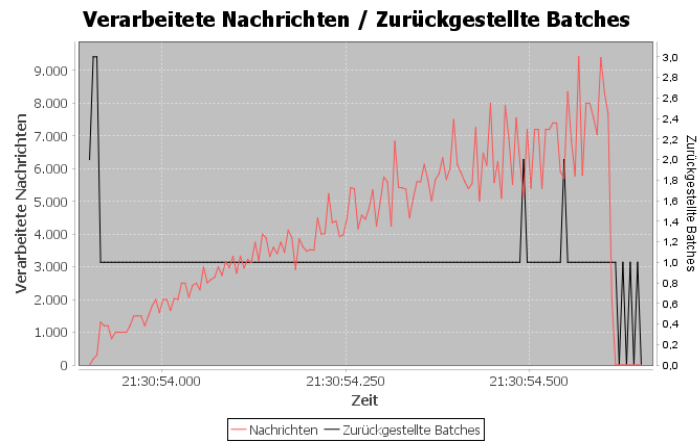

Abbildung Anhang 24: 16 Worker - LS2 - Grafik 1

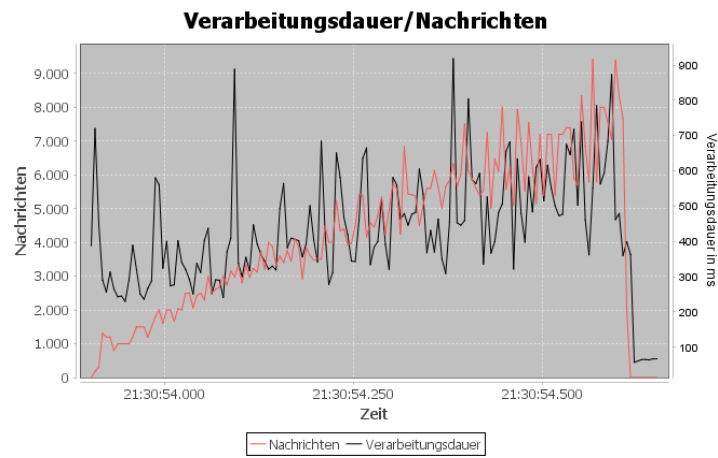

Abbildung Anhang 23: 16 Worker - LS2 - Grafik 2

Die Grafiken zeigen, dass bis zu der hier getesteten Last von 10.000 Nachrichten keine Batches verzögert werden.

- **LS3**

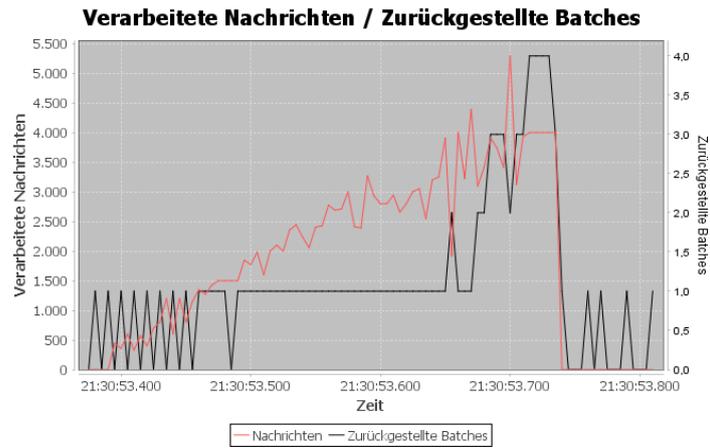

**Abbildung Anhang 25: 16 Worker - LS3 - Grafik 1**

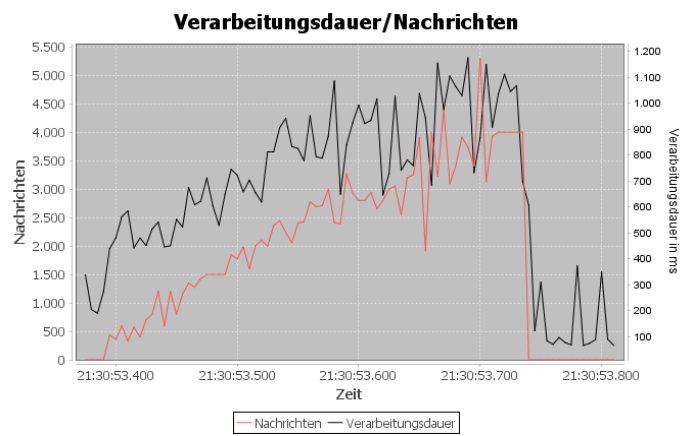

**Abbildung Anhang 26: 16 Worker - LS3 - Grafik 2**

In beiden Tests konnte durch die Skalierung des Systems eine Erhöhung des Durchsatzes erreicht werden.

**Tankstellensuche**:

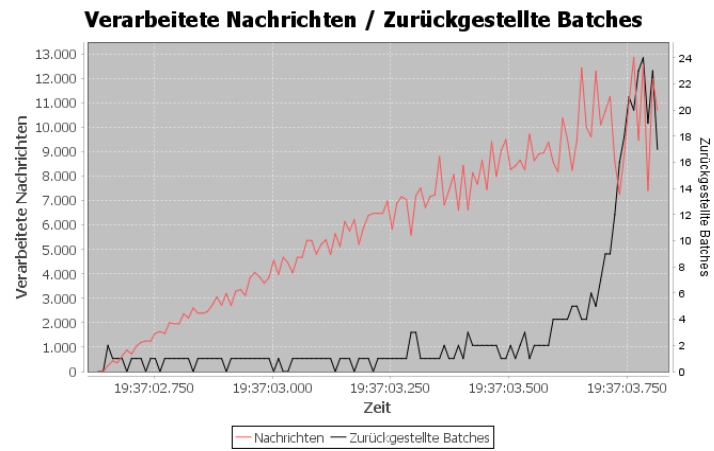

**Abbildung Anhang 27: 16 Worker - Tankstellensuche - Grafik 1**

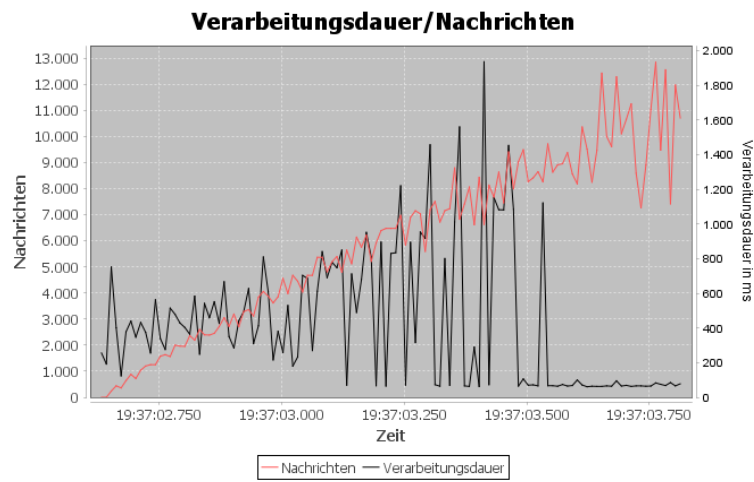

**Abbildung Anhang 28: 16 Worker - Tankstellensuche - Grafik 2**

In dieser Konfiguration können ca. 10.000 Nachrichten verarbeitet werden.

## b     AWS Instanzen

| Name | vCPU Cores | Ram (GB) |
|---|---|---|
| m3.large | 2 | 7,5 |
| m3.xlarge | 4 | 15 |
| m1.large | 2 | 7,5 |

## c  Spark Erlernbarkeit

Im Folgenden sind die Daten des Dokumentationsabgleiches angeführt. Hierbei wurde untersucht, welche Spark bzw. Spark Streaming Funktionalitäten in dem offiziellen Spark Guide beschrieben werden. Alle Daten wurden am 8.1.2016 erhoben.

### c.a  RDD Funktionen

| Transformationen | In Guide |
| --- | --- |
| ++(otherRDD) | alias |
| cartesian(otherRDD) | ja |
| collect(func) | alias |
| distinct() | ja |
| filter(func) | ja |
| flatMap(func) | ja |
| glom() | nein |
| groupBy(func) | nein |
| intersection(otherRDD) | ja |
| keyBy(func) | nein |
| map(func) | ja |
| mapPartitions(func) | ja |
| mapPartitionsWithIndex(func) | ja |
| pipe(command) | ja |
| sortBy(func) | ja |
| sample() | ja |
| subtract(otherRDD) | nein |
| union(other) | ja |
| zip(other) | nein |
| zipPartitions(other) | nein |
| zipWithIndex() | nein |
| zipWithUniqueId() | nein |

| Actions | In Guide |
| --- | --- |
| aggregate(func) | nein |
| collect() | ja |
| count() | ja |
| countApprox() | nein |
| countApproxDistinct() | nein |
| countByValue() | nein |
| countByValueApprox() | nein |
| first() | ja |
| fold() | nein |
| foreach(func): | ja |
| foreachPartition(func) | nein |
| isEmpty() | nein |
| max() | nein |
| min() | nein |
| saveAsObjectFile() | ja |
| saveAsTextFile() | ja |
| take() | ja |
| takeOrdered() | ja |
| takeSample() | ja |
| top() | alias |
| treeAggregate() | nein |
| treeReduce() | nein |

| Hilfsfunktionen | In Guide |
| --- | --- |
| cache() | ja |
| checkpoint() | nein |
| coalesce() | ja |
| localCheckpoint() | nein |
| randomSplit() | nein |
| repartition() | ja |
| persist() | ja |
| unpersist() | ja |

## c.b    Erweiterte RDD Funktionen

| PairRDDFunctions | In Guide |
|---|---|
| aggregateByKey | ja |
| cogroup | ja |
| collectAsMap | nein |
| combineByKey | nein |
| combineByKeyWithClassTag | nein |
| countApproxDistinctByKey | nein |
| countByKey | ja |
| countByKeyApprox | nein |
| flatMapValues | nein |
| foldByKey | nein |
| fullOuterJoin | ja |
| groupByKey | ja |
| groupWith | ja |
| join | ja |
| leftOuterJoin | ja |
| lookup | nein |
| mapValues | nein |
| reduceByKey | ja |
| reduceByKeyLocally | nein |
| rightOuterJoin | ja |
| sampleByKeyExact | nein |
| subtractByKey | nein |

| DoubleRDDFunctions | In Guide |
|---|---|
| histogram() | nein |
| mean(): | nein |
| meanApprox(): | nein |
| sampleStdev(): | nein |
| sampleVariance(): | nein |
| stats() | nein |
| stdev() | nein |
| sum(): | nein |
| sumApprox(): | nein |
| variance() | nein |

| AsyncRDDActions | In Guide |
|---|---|
| collectAsync() | nein |
| countAsync() | nein |
| foreachAsync(func): | nein |
| foreachPartition-Async(func) | nein |
| takeAsync(num): | nein |

## c.c DStream Funktionen

| Transformationen | In Guide |
|---|---|
| count | ja |
| countByValue | ja |
| countByValueAndWindow | ja |
| countByWindow | ja |
| filter | ja |
| flatMap | ja |
| glom | nein |
| map | ja |
| mapPartitions | nein |
| reduce | ja |
| reduceByWindow | ja |
| transform | ja |
| union | ja |
| window | ja |

| Actions | In Guide |
|---|---|
| foreachRDD | ja |
| print | ja |
| saveAsObjectFiles | ja |
| saveAsTextFiles | ja |
| slice | nein |

| Hilfsfunktionen | In Guide |
|---|---|
| cache() | alias |
| checkpoint() | ja |
| persist | ja |
| repartition | ja |

## c.d Erweiterte DStream Funktionen

| PairRDDFunctions | In Guide |
|---|---|
| cogroup | ja |
| combineByKey | nein |
| flatMapValues | nein |
| fullOuterJoin | ja |
| groupByKey | nein |
| groupByKeyAndWindow | nein |
| join | ja |
| leftOuterJoin | ja |
| mapValues | nein |
| mapWithState | nein |
| reduceByKey | ja |
| reduceByKeyAndWindow | ja |
| rightOuterJoin | ja |
| updateStateByKey | ja |

# Literaturverzeichnis

# Versicherung über Selbstständigkeit

*Hiermit versichere ich, dass ich die vorliegende Arbeit ohne fremde Hilfe selbststän-
dig verfasst und nur die angegebenen Hilfsmittel benutzt habe.*

*Hamburg, den \_\_\_\_\_\_\_\_\_\_\_\_\_\_\_\_     \_\_\_\_\_\_\_\_\_\_\_\_\_\_\_\_\_\_\_\_\_\_\_\_\_*